\documentclass[preprint,superscriptaddress,showkeys,showpacs,floatfix,eqsecnum,onecolumn]{revtex4}
\usepackage{graphics}
\usepackage{nicefrac}
\usepackage{graphicx}
\usepackage{dcolumn}
\usepackage{bm}
\usepackage{amssymb,amsmath,graphicx,graphics,color,epstopdf,mathtools}
\usepackage{caption}
\usepackage{hyperref}
\usepackage[capitalise]{cleveref}
\usepackage{rotfloat}
\usepackage{changes}

\hypersetup{
    colorlinks=true,
    linkcolor=blue,
    filecolor=magenta, citecolor=blue,    urlcolor=blue}

\begin{document}
\preprint{HEP/123-qed}
\title{Topical Review: Greybody Factors and Quasinormal Modes for Black Holes in Various Theories -- Fingerprints of Invisibles}

\author{\.{I}zzet Sakall{\i}}
\affiliation{Physics Department, Eastern Mediterranean
University, Famagusta, North Cyprus via Mersin 10, Turkey.}
\author{Sara Kanzi}
\affiliation{Faculty of Engineering, Final International University,
Kyrenia, Turkish Republic of Northern Cyprus, \\ 99370 via Mersin 10 Turkey.}


\begin{abstract}
We give a pedagogical introduction to black holes (BHs) greybody factors (GFs) and quasinormal modes (QNMs) and share the recent developments on those subjects. To this end, we present some particular analytical and approximation techniques for the computations of the GFs and QNMs. We first review the gravitational GFs and show how they are analytically calculated for static and spherically symmetric higher dimensional BHs, consisting the charged BHs and existence of cosmological constant (i.e., de Sitter (dS)/anti-de Sitter (AdS)AdS BHs). The computations performed involve both the low-energy (having real and small frequencies) and the asymptotic (having extremely high frequency of the scattered wave throughout the imaginary axis) cases. A generic method is discussed at low frequencies. This method can be used for all three types of spacetime asymptotics and it is unaffected by the BH's features. For asymptotically dS BHs, GF varies depending on whether the spacetime dimension is even or odd, and is proportional to the ratio of the event and cosmic horizon areas. At asymptotic frequencies, the GFs can be computed by using a matching technique inspired by the monodromy method. In the meantime, we also make a general literature review on the matching technique in a separate section. While the GFs for charged or asymptotically dS BHs are generated by non-trivial functions, the GF for asymptotically AdS BHs is precisely one: pure black-body emission. QNMs, which are solutions to the relevant perturbation equations that satisfy the boundary conditions for purely outgoing (gravitational) waves at spatial infinity and purely ingoing (gravitational) waves at the event horizon, are considered using some particular analytical (like the matching technique) and approximation methods. In this study, our primary focus will be on the bosonic and fermionic GFs and QNMs of various BH and brane geometries and reveal the fingerprints of the invisibles with the radiation spectra to be obtained by the WKB approximation and bounding the Bogoliubov coefficients (together with the Miller-Good transformation) methods.

\end{abstract}
\volumeyear{ }
\eid{ }
\date{\today}
\received{}

\maketitle
\tableofcontents

\section{Introduction} \label{sec1}

BH  thermal radiation or the so-called Hawking radiation (HR) exists at the intersection of classical general relativity (GR) and quantum field theory (QFT), and it may hold the key to unraveling the puzzles of a quantum gravity theory. In general relativity, classical macroscopic BHs obey rules that are similar to thermodynamic laws \cite{is1,bch73}. When it comes to describing quantum fields in a BH geometry, around the event horizon, the similarity to the ordinary thermodynamics becomes exact since the BH has a temperature and entropy \cite{h75,h76}. Namely, the Hawking radiation is nothing but the black-body radiation at the event horizon.
However, before reaching an observer (detected at spatial infinity in an asymptotically flat spacetime) and being observed, this thermal radiation must travel through a curved spacetime geometry. As a result, the surrounding spacetime acts as a potential radiation barrier, causing a variation from the black-body radiation spectrum as observed by the asymptotic observer. In summary, the GF is the difference between the spectrum of asymptotic and the black-body radiations.

The classic Hawking radiation calculation \cite{h75} employs a semi-classical approximation and thus it demonstrates that BHs have a thermal spectrum:
\begin{equation}  
\langle \Xi(\omega) \rangle = \frac{\Gamma(\omega)}{exp(\omega/T_H) \pm
1},\label{iz1}
\end{equation} 

in which $\langle \Xi(\omega) \rangle$ indicates the expectation value for the number of particles of a particular species released in a mode with frequency $\omega$, $T_H$ is the Hawking temperature, the minus/plus sign
describes bosons/fermions, and $\Gamma(\omega)$ represents the GF, which is the probability for an outgoing wave to reach spatial infinity or the observer \cite{is1}. This corresponds to the probability of an incoming wave to be absorbed by the BH (the absorption probability). The overall BH emission rate is obtained by integrating Eq. \eqref{iz1} across all spectra. Note that the BH emission spectrum would be identical to that of a black-body if $\Gamma(\omega)$ was a constant. 

The computations of the GFs date back to 1976s \cite{p76, u76}.
Although the computation setup is straightforward, yet retrieving accurate results might be problematic. The particle perturbation to the BH geometry is given by linearized wave equations, which explain the scattering of particles off BHs (the reader is referred to \cite{aj00} and references therein). The applications of linear perturbation theory to 4-dimensional and higher dimensional BH geometries were investigated in \cite{rw57, z70, z74} and \cite{ik03a, ik03b}, respectively. The resultant radial equation obtained for the perturbations in BH geometries is generally reduced to a 1-dimensional Schr\"odinger equation in which the 1-dimensional coordinate is created by the tortoise coordinate (representing the geometry of spacetime outside of the BH) and the effective potential identifies the BH geometry as well as the sort of perturbation that is being addressed. Normally, the potentials are enormously complicated and finding a definite solution to the Schr\"odinger equation (by required boundary conditions (BCs) for the scattering problem) could be impossible; thus one must rely on numerical works or approximation techniques like the Wentzel-Kramers-Brillouin (WKB) approximation described in \cite{p76, u76}. When studying the perturbation of BHs obtained in some particular theories like string, dilaton, axion etc., the exact solutions to the wave equations can be found in terms of Mathieu, hypergeometric, and Heun functions \cite{gh98, clpt99,Gursel:2019fyd,Sakalli:2016fif}.

The absorption probability $\Gamma(\omega)$ can be stated as the square of the transmission coefficient for the related Schr\"odinger problem of the examined BH geometry. In other words, the tunneling possibility for the barrier defined by the particular potential is known as the GF. One of the main purposes of this review article is to show the computation of $\Gamma(\omega)$ in a variety of BH spacetimes. Parameter $\sigma(\omega)$ denotes absorption cross-section whıch is strongly connected to the GF, which is derived from the optical theorem \cite{g97} as follows: $\sigma (\omega) = \Gamma(\omega) |\Psi(\omega)|^2$ in which $\Psi(\omega)$ denotes the asymptotic plane wave projection of the incoming spherical wave function, and thus $|\Psi(\omega)|^2$ can be normalized to unity. So, the answer to the question of why $\Gamma(\omega)$ and $\sigma (\omega)$ are considered as equivalent in many studies becomes clear for a reader. Moreover, it is worth noting that the above result is valid in the case of both incoming and outgoing asymptotic states are determined. The latter remark is mostly possible in asymptotic flat spacetimes, but the vice-versa cases are also possible, albeit to a lesser extent \cite{Sakalli:2016abx,Gursel:2018bts}.

In this paper, we shall review the computation of GFs \big($\Gamma(\omega)$ or $\sigma (\omega)$ and later on $T(\omega)$, whose details will be given in the next sections\big) for static, stationary, and spherically symmetric BHs (with and without charge/cosmological constant) even in higher dimensions. To this end, bosonic, fermionic, and gravitational\footnote{
The interest on gravitational perturbations has gained momentum since the first indirect observation of gravitational waves, which was made on 14 September 2015 and was announced by the LIGO and Virgo collaborations on 11 February 2016 \cite{LIGOScientific:2016aoc}.} perturbations are considered, and the transmission and reflection
coefficients for the regarded Schr\"odinger equatıon are then computed \cite{ik03a, ik03b}. When having analytical solutions are impossible, we consider specific approximation methods whose details will be given in the associated sections for the perturbed fields \cite{u76,n03}. One of the important information that we want to say at this point is that there is a general consequence for the asymptotically flat BHs in the low frequency approximation, $\omega \ll T_H$ and $\omega r_{H}  \ll 1$ where $r_{H} $ represents the radius of the event horizon: $\sigma (\omega)$ for the $s$-wave of a minimally coupled massless scalar field is equal to the area of the event horizon: $\sigma (\omega)= A_H$ \cite{dgm96}. After taking into account the scattering's sub-leading contributions (the higher partial-waves with
the angular momentum component $\ell > 0$), $\sigma (\omega)$ always modifies, although the universality is maintained \cite{dgm96}. 

GFs or absorption cross-sections, which are natural outcomes of the investigations made in the BH scattering theory, have a large literature (the reader is referred to \cite{aj00,gfthesis} for topical reviews). Let us take a quick look at some recent researches that are directly relevant to the current review on the GFs. At this point, it is useful to focus in particular on the most classical BHs belonging to the Schwarzschild family, whose GF investigations have been thoroughly conducted (mainly for bosonic emissions) \cite{is1}. In addition to the Schwarzschild family solutions, the so-called brane-world scenarios have also ignited the works on the GFs  \cite{Chen:2007ay,Maldacena:1996ix}. In this regard, the numerous researches have focused on the GFs of the brane BHs (see for example \cite{gf1,gf2}). Among them the notable works that we can show are Refs. \cite{km02a, km02b}, which use the matching technique of \cite{u76} in case of $r_{H}\omega \ll 1$ to investigate scalar, Dirac, and vector particles' emissions \cite{hk03, jp06a}. At this point, it is worth noting that spin-2 emission is an issue that has been extensively discussed in the current literature but has not yet been resolved. This issue is indeed very important in gravitational wave astronomy. A few of articles, which address the subject of gravitational emission from higher [$d(\geq4)$]-dimensional static BHs in the low frequency regime are Refs. \cite{cns05, ccg05a, ccg05b, cekt06, p06,
dkss06, jp06b,is1}. Let us also point out that Ref. \cite{jp05} studied the spin-0 emission of $d(\geq4)$-dimensional Reissner-Nordstr\"{o}m BH, again in the low frequency regime. However, GFs for those BHs were not explored in $d$-dimensional framework until \cite{kgb05}. Later on Harmark et al \cite{is1} worked on the GFs and the rates of energy-emission on a brane and bulk. What made their work unique comparing to \cite{kgb05} is the following: in \cite{kgb05}, the standard matching technique of \cite{u76} was not used and $\omega \to 0$ limit was imposed. However, Harmark et al made their computations further on the limitation of $\omega \rightarrow 0$ by applying the matching technique prescribed in \cite{u76} and they further extended their methodology to non-asymptotically flat (NAF) BHs. As an outcome, dissimilar to the Schwarzschild case, the Schwarzschild dS BH's absorption probability tends to a constant, not vanishing as the frequency. Meanwhile, there is significantly less literature on the GFs of BHs having NAF structure. Herein, we can state that Sakall{\i} and his co-authors can be considered as the forefront authors who contributed to the literature with the GFs of NAF BHs \cite{Sakalli:2016fif, Sakalli:2016abx, Gursel:2018bts,Kanzi:2020cyv,Sakalli:2022swm,Kanzi:2021jrl}. Besides, the GFs of $d$-dimensional asymptotically AdS spacetimes were  studied by Refs. \cite{is1} and \cite{hk00} using the ultrahigh energy and low energy regimes, respectively. While the geometrical optics approximation was employed in \cite{hk00}, the authors of \cite{is1} used the matching technique \cite{u76}, as we mentioned earlier. By getting inspired from the monodromy method \cite{cns04, ns04}, the first method for determining the GFs in asymptotically flat spacetimes was devised in \cite{n03}. Then, Harmark et al \cite{is1} generalized the method to the NAF spacetimes for both (low/high) frequency regimes. It was shown that the GFs approach the geometrical-optics limit at the ultra high real frequencies, and the results obtained do not depend on the emitted quanta spin \cite{is1}. One of the most important GF studies about the asymptotically flat BHs is undoubtedly the study of \cite{s76}. Moreover, Ref. \cite{ss06} demonstrated that the perturbation theory methods in quantum mechanics guide us to investigate the imaginary (asymptotic) frequency regimes. This methodology can also be applied to calculate the GFs of massless/massive bosonic, fermionic, and vector fields \cite{is1}. As a result, in-depth understanding of Hawking radiation from $d$-dimensional spherically symmetric BHs could be gained. Studies made in this direction for the BHs found in the string theory can be seen in \cite{gbk05, ms06}. GFs of the AdS BHs in the circumstance of the AdS/CFT correspondence \cite{agmoo99} are still attractive research area (see for example \cite{Chowdhury:2020bdi}). In summary, the findings of \cite{is1} revealed the generic behaviors of the thermal correlation functions, which are dual to the GFs in the dual gauge theory, at both low and high frequencies. On the other hand, one of the promising GF computation methods for obtaining the GF is the method described in Ref. \cite{GFmtd2}, which defines the general semi-analytic bounds for the GFs:
\begin{equation}
 \sigma_{\ell}(\omega) \geq \sec h^{2}\left(\int_{-\infty}^{+\infty} \wp d r_{\star}\right), \label{izs2}   
\end{equation}
where $\sigma_{\ell}(\omega)$ is the GF which relies on radiated particles' frequency $\omega$ and parameter $\ell$ (angular momentum quantum number) and $\wp$ is defined by
\begin{equation}
\wp=\frac{\sqrt{\left(h^{\prime}\right)^{2}+\left(\omega^{2}-V_{e f f}-h^{2}\right)^{2}}}{2 h},
\end{equation}
in which $V_{e f f}$ is the effective potential of the 1-dimensional Schr\"{o}dinger like wave equation and the prime symbol indicates the radial derivative: $\frac{d}{dr}$. There are two constrains for the explicit positive function $h$: 1) $ h\left(r_{\star}\right)>0$ and 2) $h(-\infty)=h(\infty)=\omega$. Without loss of generality, one can always set $h=\omega$  \cite{GFmtd2,SK27}. Besides, there are other methodologies to determine the GFs. The most used of these methods are the WKB approximation, which is without a doubt the most widely utilized method \cite{mtd1,mtd2,mtd3}, the
rigorous bound method \cite{mtd4,mtd6}, and the matching technique \cite{mtd7,mtd8}. Apart from those, the reader may also learn the additional approaches for computing the GFs by looking at those Refs. \cite{GFmtd,mtd9}.

The distinctive ringing of BHs are called QNMs, which are the damped oscillations with a discrete set of complex frequencies. More technically speaking, the eigenmodes in dissipative systems are known as QNMs. Those eigenmodes are naturally produced by perturbations of classical gravitational backgrounds involving BHs, branes, and even the wormholes \cite{Churilova:2021tgn}. QNMs decay with time for any stable system, and thus provide a useful description of the relaxation process. They are calculated by solving the wave equation of the considered BH perturbed by an external field.
A disrupted BH tries to re-establish its equilibrium by generating energy in the form of gravitational waves. As a result, QNMs play critical role on the gravitational investigations like LIGO detector, which is designed for detecting the gravitational waves \cite{ligo}. Additionally, in the gauge/gravity dualities theories, there is a relationship between the QNMs and the poles of a propagator in the dual field theory, which physicists use it as a tool to work on strong coupled gauge theories (or holography) \cite{qnm1,qnm2,qnm3}. The QNMs frequencies are defined by perturbing the spacetime of the BH with suitable boundary conditions: while the waves of the QNMs should be in the form of fully ingoing waves near the event horizon, they should be completely outgoing waves at spatial infinity. One of the most difficult aspects of understanding the QNMs is that they admit a non-self-adjoint eigenvalue problem: the system is not conserved, so that the energy is lost at both spatial infinity and the event horizon. Thus, unlike the ordinary or classical normal modes, the amplitude of oscillation decays in time for the QNMs \cite{Berti:2018vdi}. The QNMs can also be utilized to test the theorem of no-hair, which claims that BHs contain only mass, charge, and angular momentum and for the BH quantization \cite{Maggiore:2007nq,Bekenstein:1974ax,Corichi:2006wn,Wei:2009yj,Hod:2005dc,Sakalli:2018nug,Sakalli:2011zz,Sakalli:2013yha,Sakalli:2016jkf,Sakalli:2014wja,Sakalli:2016fif,Sakalli:2015uka,Tokgoz:2018irr,Sakalli:2014vca,Ovgun:2017dvs,Sakalli:2021dxd}. Investigating QNMs can therefore provide new insights into physics beyond Einstein's general theory of relativity, which is an important instrumental for the (incomplete) theory of quantum gravity \cite{Ashtekar:2021kfp}. References \cite{qnmr1,qnmr2,qnmr3} provide more detailed reviews of BH QNMs including their various computation methods, which are as follows: 1) Exact solutions \cite{Fernando:2008hb,Sakalli:2021dxd,Tokgoz:2018irr,Sakalli:2018nug} 2) WKB approximation \cite{Konoplya:2003ii,Konoplya:2019hlu,Hod:2012zzb,Chakrabarti:2006ei}  3) Monodromy technique for highly-damped modes \cite{Natario:2004jd} 4) Asymptotically AdS BHs: a series solution \cite{Mirbabayi:2018mdm} 5) Asymptotically AdS BHs: the resonance method \cite{Macedo:2013jja} 6) Continued fraction method \cite{Rostworowski:2006bp} 7) Asymptotic iteration method \cite{Cho:2009cj,qnmr3} 8) Feedforward neural network method \cite{Ovgun:2019yor,Ncube:2021jfu}. On the other hand, the method that we will mainly focus on in this review will be the WKB approximation. We would like to briefly summarize this method now.

As noted above, QNMs can be represented as the solutions to a perturbation (wave) equation satisfying the certain boundary conditions. After separating the wave equation into the radial and angular parts, the radial part can be reduced to a 1-dimensional Schr\"{o}dinger equation:

\begin{equation}
\frac{d^{2} \mathcal{Z}}{d r^{* 2}}+V(r) \mathcal{Z}=0, \label{is4}   
\end{equation}
\noindent
which $\mathcal{Z}$ represents the radial part of the wave function having a time-dependence component $exp(i \omega t)$ and $r^{*}$ denotes the tortoise coordinate. $V(r)$ is the effective potential. Considering a generic metric like 
\begin{equation}
ds^{2}=A(r) d t^{2}-H(r)^{-1} d r^{2}-B(r) d \theta^{2}-B(r) \sin ^{2} \theta d\phi^{2},    
\end{equation}
one can define the tortoise coordinate as follows
\begin{equation}
 \frac{d}{d r^{*}}=\sqrt{A H} \frac{d}{d r}. \label{is5}   
\end{equation}
\noindent
The potential represented in terms of $r^{*}$ coordinates behaves in such a manner that when $r^{*} \rightarrow -\infty$ (horizon) it vanishes and as $r^{*} \rightarrow+\infty$ (spatial infinity) it might be a non-zero. There is a particular location $r_{0}^{*}$ that the potential reaches its maximum. In fact, the region considered for the WKB analysis is examined in three parts: Part $I$ is from $-\infty$ to $r_{1}$, which is the initial turning point (where the potential vanishes), Part $II$ covers the region from $r_{1}$ to $r_{2}$, which is the second turning point, and Part $III$ starts from $r_{2}$ to $+\infty$. We apply the Taylor expansion around $r_{0}^{*}$ in part $I I$. In Parts $I$ and $III$, a special exponential function can be used to approximate the solution as
\begin{equation}
\mathcal{Z} \sim \exp \left[\sum_{n=0}^{\infty} \epsilon^{n-1} \mathcal{S}_{n}(x)\right], \quad \epsilon \rightarrow 0. \label{is6}
\end{equation}
After inserting Eq. \eqref{is6} into Eq. \eqref{is4}, one gets $\mathcal{S}_{j}$ as a function of $V(r)$ and its derivative $V^{'}(r)$. By considering the QNM boundary conditions:
\begin{equation}
\begin{array}{cc}
\mathcal{Z} \sim e^{-i \omega r^{*}}, & r^{*} \rightarrow-\infty, \\
\mathcal{Z} \sim e^{i \omega r^{*}}, & r^{*} \rightarrow+\infty,
\end{array} \label{is7} 
\end{equation}
\noindent
we obtain the solutions of Parts $I$ and $I I I$ by matching them to Part $I I$ solution at the turning points $r_{1}$ and $r_{2}$, respectively. In \cite{Konoplya:2003ii}, the WKB approach was effectively expanded to the sixth order. This permits the QNM frequencies to be calculated as a function of the potential and its derivatives at their maximum. Thus, one has the following QNM frequencies for the sixth order diagnosis:
\begin{equation}
\omega^{2}=V_{0}-i \sqrt{-2 V_{0}^{\prime \prime}}\left(\sum_{j=2}^{6} \Lambda_{j}+n+\frac{1}{2}\right). \label{is8} 
\end{equation}
$\Lambda_{j}$ expressions are also given in  \cite{Konoplya:2003ii}. While Ref. \cite{Hatsuda:2019eoj} has some interesting current ideas on the convergence of the WKB series, Ref. \cite{Hatsuda:2019eoj} provides more information on the expansion parameter.

This review article is constructed as follows. In Section \ref{sec2}, we supply a detailed review of GFs for the $d$-dimensional dS and AdS spacetimes. This section is nothing but a comprehensive overview of Ref. \cite{is1} and it is a useful starting point for learning the GFs in general. We discuss the black-body radiation, BH perturbation, reflection and transmission coefficients, GFs in the asympotic limit, flux, and universality. In Section \ref{sec3}, we summarize the matching technique for computing the GFs and QNMs. Section \ref{sec4} is devoted to using of the WKB approximation for computing the GFs and QNMs. To assist the reader in conducting these computations, we provide specific applications. In this regard, the WKB approximation method is considered to compute the bosonic GFs and QNMs of polytropic BHs and slowly rotating Kerr like BHs (KlBHs) in the bumblebee gravity model (BGM). Section \ref{sec4} ends with a discussion of the fermionic GFs and QNMs of the KlBHs. In Section \ref{sec5}, we consider the bounding Bogoliubov coefficients method including the Miller-Good transformation and discuss the computations of the bosonic and fermionic GFs and QNMs via that method. To this end, we consider various BH prototypes, which are the Schwarzschild-like BH in the BGM (SBHBGM), the Schwarzschild BH surrounded by quintessence field (SBHSQ), the brane-world BH solutions via a confining potential (BWBHSCP), and  the charged (Taub) NUT BHs (CTNBHs) to exemplify our viewpoint for computing the GFs and QNMs\footnote{Throughout this study, we shall usually refer to "QNMs" as \textit{frequencies of QNMs}, rather than the corresponding amplitudes. The units $\hbar=c=G=1$ will be used the most unless otherwise stated.}. Finally, we give our concluding remarks and comments on the GFs and QNMs in Section \ref{sec6}.

\section{\texorpdfstring{GF\MakeLowercase{s} for \MakeLowercase{d}S and A\MakeLowercase{d}S}{d} Spacetimes: Brief Review of Ref. \cite{is1}}
\label{sec2}

For spin-$0$ waves of asymptotically flat BHs that are static and spherically symmetric, Harmark et al \cite{is1} presented a simple but generic derivation of the cross-section $\sigma (\omega)$. Their approach extends the results presented in \cite{dgm96}, which are limited to Schwarzschild BHs. For dS and AdS BHs \cite{dgm96}, they \cite{is1} showed that the cross-section is $\sigma (\omega)= A_H$, which is the BH area. In particular, they considered foremost $s$-waves ($\ell=0$) to derive the GFs of asymptotically dS/AdS spacetimes. In fact, the literature concerning the GFs for BHs in NAF spacetimes, like dS or AdS, is preferably barren that there are less number of studies about this subject in the literature (see for example \cite{Sakalli:2016fif,g97,Sakalli:2016abx,Gursel:2018bts}). The most prominent feature of Ref. \cite{is1} is that the authors developed a general computation methodology that may be utilized to tackle different types of spacetime asymptotic.

For dS BHs, the GFs in the case of the low frequencies $\omega
\ll T_H$ and $\omega r_{H} \ll 1$ were analytically derived \cite{is1}. That was done in the case of small dS BHs, namely when $r_{H}$ is much shorter than
the distance-scale adjusted by the cosmological horizon. The separation of the near horizon area from the distant region is the basis for their methodology. After some straightforward computations, they got the following non-trivial result:
\begin{equation}  \label{uni1}
\Gamma ( \omega ) = 4 h(\hat{\omega}) A_HA_C^{-1},
\end{equation}
in which $A_C$ and $A_H$ represent the areas of the cosmological and event
horizons, respectively. Moreover, $h(\hat{\omega})$ implies a non-linear function of $\hat{\omega}$ (the frequency expressed in units of the cosmological constant's scale). For even and odd spacetime dimensions, the function $h(\hat{\omega})$ has a distinct formulation. However, within its region of validity, it is a monotonically growing function with $h(0)=1$ \cite{is1}. Besides, function $h(\hat{\omega})$ generalizes a finding derived in \cite{kgb05} for $d$-dimensional Schwarzschild dS BHs. They also derived the generic result for the GF \eqref{uni1} obtained for spin-0 waves, which is hold in all static and spherically symmetric asymptotically dS BHs. For AdS BHs, the GFs for two various regimes were obtained.
First regime considers the low frequencies and small BHs: $\omega \ll T_{H}$
and $\omega r_{H} \ll 1$, and with $r_{H}$ being far less than the cosmic constant's distance scale. The second regime has $\omega \ll T_{H}$ and $\omega$ is substantially lower than the cosmological constant's energy scale. To make the subject more clear, they defined a new area term $\hat{A}_{H}$, which is expressed in scale units determined by the cosmological constant. For
both regimes with $\hat{\omega}^{d-2} \ll \hat{A}_H$, it was found that  
\begin{equation}
\Gamma ( \hat{\omega} ) \propto \hat{\omega}^{d-2}\hat{A}_H^{-1}.
\end{equation}
\noindent
The above result is indeed valid for large AdS BHs having $\hat{A}_H \gg 1$ and $\hat{\omega} \ll 1$. On the other hand, for small AdS BHs possesing $\hat{A%
}_H \ll 1$, it was found that the GFs ($\Gamma(%
\hat{\omega})$ have a more richer structure. Moreover, $\hat{\omega}^{d-2} \sim \hat{A}_H$ admits a particular frequency, which corresponds to  no reflection or full transmission  \big($\Gamma(\hat{\omega})=1$\big). Conversely, in the case of $\hat{\omega} = 2n+d-1$ together with $n \in \{ 0, 1,
2, \cdots \}$, the pure
reflection for the radiation \big($\Gamma(\hat{\omega}) = 0$\big) was determined. On the other hand, those frequencies, interestingly enough, exactly match the usual frequencies of scalar wave perturbations in pure AdS spacetime \cite{ns04}. They also found other critical points, $(2n+d-1-\hat{%
\omega})^2 \sim \hat{A}_H$, that result in full transmission, $\Gamma(\hat{%
\omega}) = 1$. The AdS/CFT correspondence, which links AdS BH events to thermal gauge theory, is particularly noteworthy because of the multifarious structure revealed by the GF for the AdS BHs (see for instance \cite{agmoo99} and references therein). 

In the asymptotic limit, the frequency through the imaginary axis is extremely high \big($\omega \to + i \infty$ and thus $|r_{H} \omega | \gg 1$\big), there are no universal results for the GFs. Furthermore, it is notable that $\Gamma(\omega)$ for the Schwarzschild BH has the following expectation value for the radiated gravitons \cite{is1,n03}:
\begin{equation}
\langle \Xi(\omega) \rangle = \frac{1}{exp(\omega/T_H)+ 3}.
\end{equation}
\noindent
To comprehend the hypothesis outlined in \cite{n03}, one should check Refs. \cite{Maldacena:1996ix,k97} in which it was shown that, for a particular class of five-dimensional BHs, the GFs behave in such a way that at asymptotic infinity  BH spectroscopy resembles the microscopic string's excitation spectrum. This suggests that GFs include information on the quantum structure of BHs. One of the works made in \cite{n03} is  based on
the low-frequency results of \cite{Maldacena:1996ix}, which hypothesizes that by examining the aforementioned GFs one would be able to deduce on the BH's microscopical description at asymptotic frequencies. Whence, at asymptotic frequencies, BH microscopics would need new degrees of freedom (DoF) with somewhat unusual statistics, in particular, for the case of Schwarzschild BH (additional investigations along similar lines were later conducted in \cite{ks04}). On the other hand, the asymptotic calculations demonstrated that these additional microscopic DoF would have to include even more unusual statistics than in the Schwarzschild case for the charged or chargeless Reissner-Nordstr\"{o}m or asymptotically dS BHs (for the relevant equations, the reader is referred to \cite{is1}). Besides, for uncharged/charged AdS BHs and the asymptotically AdS case, $\Gamma(\omega)$ elicts the following expectation value for the number of radiated gravitons at asymptotic frequencies:

\begin{equation}
\langle \Xi (\omega) \rangle = \frac{1}{exp(\omega T_H^{-1}) - 1},
\end{equation}
which is obviously the well-known black-body radiation.\\
To explore the generality of GFs in the asymptotic limit, one must first study the BH scattering. However, such a study reveals that some transmission and reflection coefficients hide the generality in the asymptotic limit. The characterization of gravitational perturbations can be considered by a Schr\"{o}dinger equation in 1-dimension, with some potential $V(r^{*})$ related with the geometry of spacetime and characteristics of the perturbation, as previously stated. Let us deine the wave function $\Phi_\omega(r^{*})$ as solution to the 1-dimensional Schr\"odinger-like wave equation: 
\begin{equation}
- \frac{d^2 \Phi_{\omega}}{dr_{*}^2} + V(r^{*}) \Phi_{\omega} = \omega^2 \Phi_{\omega}, \label{my1}
\end{equation}
which describes the scattering of an ingoing wave begun to propagate from spatial infinity $r^{*}=+\infty$ (or the cosmological horizon for asymptotically dS BHs). Thus, we have
\begin{eqnarray}
\Phi_\omega &\sim& e^{i \omega r^{*}} + R e^{-i\omega r^{*}}, \qquad r^{*} \to + \infty,
\notag \\
\Phi_\omega &\sim& T e^{i \omega r^{*}}, \qquad r^{*} \to - \infty,
\end{eqnarray}
\noindent where the reflection and transmission coefficients are denoted by $R(\omega)$ and $T(\omega)$, respectively. Although the both $\Phi_{\pm\omega}$ admit the solution to the same wave equation, however one can consider the solution of $\Phi_{-\omega}$ 
\begin{eqnarray}
\Phi_{-\omega}&\sim&\widetilde{\Re} e^{i\omega r^{*}}+ e^{-i \omega r^{*}}, \qquad
r^{*} \to + \infty,  \notag \\
\Phi_{-\omega} &\sim&\mathcal{\widetilde{T}} e^{-i \omega r^{*}}, \qquad r^{*} \to - \infty,
\end{eqnarray}
which have other form of the reflection and transmission coefficients: $\widetilde{\Re}(\omega)$ and $\mathcal{\widetilde{T}}(\omega)$. Considering the following flux definition: 
\begin{equation}
J = \frac{1}{2i} \left( \Phi_{-\omega}\frac{d \Phi_\omega}{dr^{*}} -
\Phi_{\omega}\frac{d \Phi_{-\omega}}{dr^{*}} \right), \label{flx}
\end{equation}
one can infer that the flux is independent of $r^{*}$. For the entire region, $r^{*} \to \pm \infty$, after evaluating Eq. \eqref{flx}, we get
\begin{equation}
R \widetilde{\Re} + T \mathcal{\widetilde{T}} = 1.
\end{equation}
It is easy to check that when $\omega$ is real then $\Phi_{-\omega}=\Phi_{\omega}^*$ (here, $*$ symbol means the complex conjugation) and whence $\widetilde{\Re}=R^*$ and $\mathcal{\widetilde{T}}=T^*$.
As a result, we rediscover one of the main outcomes of the quantum mechanics: $|\widetilde{R}|^2 + |\widetilde{T}|^2 = 1$. On the other hand, considering $\Phi^{\prime }_\omega$ as the Schr\"{o}dinger solution with complex $\omega$, which represents the dispersing of the outgoing wave coming from $r^{*}=-\infty$  (the outer event horizon), we get
\begin{eqnarray}
\Phi^{\prime }_\omega &\sim& \mathcal{T}^{\prime -i\omega r^{*}}, \qquad r^{*} \to + \infty, 
\notag \\
\Phi^{\prime }_\omega &\sim& e^{-i \omega r^{*}} + \Re^{\prime i \omega r^{*}x}, \qquad
r^{*} \to - \infty,
\end{eqnarray}
in which $\mathcal{T}^{\prime }(\omega)$ and $\Re^{\prime }(\omega)$ have the same role as $T(\omega)$ and $R(\omega)$, respectively. As far as the space solutions of Eq. \eqref{my1} possesses dimension 2, $\Phi^{\prime}_\omega $ should be a linear superposition of $\Phi_{\omega}$ and $\Phi_{-\omega}$. So, one can set 
\begin{equation}
\Phi^{\prime }_\omega = - \frac{\widetilde{\Re}}{\mathcal{\widetilde{T}}} \Phi_\omega +
\frac1{\mathcal{\widetilde{T}}} \Phi_{-\omega}, \label{set1}
\end{equation}
and consequently, we have 
\begin{eqnarray}
\Re^{\prime }&=& - \frac{T}{\widetilde{\mathcal{T}}}\widetilde{\Re},  \notag \\
\mathcal{T}^{\prime }&=& T.
\end{eqnarray}
Again, for real $\omega$ we have $|\Re^{\prime }|=|R|$. However, it is not obligatory in the latter result to satisfy the complex frequencies $\omega$. But, $\mathcal{T}^{\prime }=T$ always holds. Lastly, we would like to point out that $\Phi^{\prime }_{-\omega}$ continues to solve Eq. \eqref{my1} by satisfying the following boundary conditions
\begin{eqnarray}
\Phi^{\prime }_{-\omega} &\sim& \mathcal{\widetilde{T}}^{\prime i \omega r^{*}}, \qquad r^{*}
\to + \infty,  \notag \\
\Phi^{\prime }_{-\omega} &\sim& e^{i \omega r^{*}} + \widetilde{\Re}^{\prime
-i\omega r^{*}}, \qquad r^{*} \to - \infty,\label{set11}
\end{eqnarray}
for the other scattering coefficients (transmission/reflection): $%
\widetilde{\Re}^{\prime }(\omega)$ and $\mathcal{\widetilde{T}}^{\prime }(\omega)$.
Similar to Eq. \eqref{set1}, $\Phi^{\prime }_{-\omega}$ is also a linear combination of $\Phi_{\omega}$ and $\Phi_{-\omega}$:
\begin{equation}
\Phi^{\prime }_{-\omega} = \frac1T \Phi_\omega - \frac{R}{T} \Phi_{-\omega},
\end{equation}
whence
\begin{eqnarray}
\widetilde{\Re}^{\prime }&=& - \frac{\mathcal{\widetilde{T}}}{T} R,  \notag \\
\mathcal{\widetilde{T}}^{\prime }&=& \mathcal{\widetilde{T}}.
\end{eqnarray}
So, we have double identities: $T\mathcal{\widetilde{T}} = \mathcal{T}^{\prime }\mathcal{\widetilde{T}}%
^{\prime }$ and $R\widetilde{\Re} = \Re^{\prime }\widetilde{\Re}^{\prime }$. While the GF with complex frequency $\omega$
is defined as $\Gamma(\omega) = T(\omega) \mathcal{\widetilde{T}}(\omega)$,
for the real frequencies it becomes $\Gamma(\omega) = | T(\omega) |^2$.
The GFs for the two scattering cases, in particular, turn out to be identical \cite{is1}.
Those findings are valid for either asymptotically flat or asymptotically dS
geometries. For asymptotically AdS BHs, the
$r^{*}$ coordinate spans the spacetime from $- \infty$ (\textit{i.e.}, from event horizon) to a fix point, which can be chosen as zero, at spatial infinity. As shown in Ref. \cite{ns04}, in the case of $r^{*} \sim
0 $ the wave function behaves as 
\begin{equation}
\Phi (r^{*}) \sim C_+ \sqrt{2\pi\omega r^{*}}\ \mathcal{J}_{\frac{j_\infty}{2}} \left( \omega
r^{*} \right) + C_- \sqrt{2\pi\omega r^{*}}\ \mathcal{J}_{-\frac{j_\infty}{2}} \left( \omega r^{*}
\right), \label{sol1}
\end{equation}
in which $\mathcal{J}_\nu$ is a first kind of Bessel function and $j_\infty = d-1, d-3, d-5$ stand for the perturbations of tensor, vector, and
scalar types, respectively. $C_\pm$ are complex integration constants \cite{ns04}. Therefore, when $Re(\omega)>0$ with $r^{*} \ll -1$, Eq. \eqref{sol1} admits the following asymptotic behavior
\begin{equation}  
\Phi (r^{*}) \sim \left( C_{+} e^{i\beta_+} + C_{-} e^{i\beta_{-}}\right) e^{i \omega
r^{*}} + \left( C_{+} e^{-i\beta_{+}} + C_{-} e^{-i\beta_{-}}\right) e^{-i \omega r^{*}}, \label{sol2}
\end{equation}
where $\beta_\pm = \frac{\pi}4 (1 \pm j_\infty)$ \cite{is1}. The transmission and reflection coefficients at spatial infinity may therefore be defined in terms of the coefficients of the Bessel functions $\mathcal{J}_\nu$ with the aid of Eq. \eqref{sol2}. For instance, the incoming waves obey the following matrix equation:
\begin{equation}
\left( 
\begin{matrix}
R \\ 
1%
\end{matrix}
\right)=\left( 
\begin{matrix}
e^{i\beta_+} & e^{i\beta_-} \\ 
e^{-i\beta_+} & e^{-i\beta_-}%
\end{matrix}
\right) \left( 
\begin{matrix}
C_+ \\ 
C_-%
\end{matrix}
\right). 
\end{equation}
In fact, we should consider the solution at spatial infinity as if a plane-wave for $r^{*} \ll -1$ \cite{is1}. As previously stated, some of the $T(\omega)$ and $R(\omega)$ conceal generality of the GFs in the asymptotic limit. For all considered BH spacetimes, if $\mathcal{\widetilde{T}}=1$, then $\Gamma(\omega) = T(\omega)$. Plus,
for all the asymptotically flat spacetimes \cite{is1} 
\begin{equation}
\frac{i\widetilde{\Re}}{2} = \cos \left( \frac{\pi j}{2} \right),
\end{equation}
where the parameter $j$ could be $j=0\left(j=\frac{d-3}{2 d-5}\right)$ for tensor or scalar-type gravitational perturbations of neutral (charged) BHs and $j=2\left(j=\frac{3 d-7}{2 d-5}\right)$ represents the vector-type gravitational perturbations of neutral (charged) BHs. On the other hand, it is found that for all asymptotically flat cases, we have
\begin{equation}
\frac{T(\omega)-1}{2iR(\omega)}=\cos \left( \frac{\pi j}{2} \right),
\end{equation}
which yields the generic asymptotic GF:
\begin{equation}
\Gamma(\omega) = T(\omega) \mathcal{\widetilde{T}}(\omega) = 1+2i \cos \left( \frac{%
\pi j}{2} \right) R(\omega).
\end{equation}
Although the asymptotically dS spacetimes do not reveal any generality for the GFs aside from the $\mathcal{\widetilde{T}}=1$, asymptotically AdS geometries do. Namely, for all AdS
BHs, it was shown that when $\mathcal{\widetilde{T}}=1=T$, it leads to a generic result for the GF: $\Gamma = 1$ \cite{is1}. Moreover, in the case of $\widetilde{\Re}=0$, one gets
\begin{equation}
R = 2i \cos \left( \frac{\pi j}{2} \right).
\end{equation}
The resemblance to the asymptotically flat case has now become even more obvious. Future investigations to be made on those topics will definitely help the researchers to get a better understanding of the observed signs of the generality/universality.

\section{\texorpdfstring{Brief Review of Matching Technique for Computing GF\MakeLowercase{s} and QNM\MakeLowercase{s}}{d}}
\label{sec3}
In this section, we shall address the investigations on the matching technique for computing the GFs. This technique is based on solution of the field equations in the near-horizon and far-field regions, and then comparing the solutions, via a intermediate region, to set up a complete and straightforward solution for the radial part of the field. Getting inspired from his PhD thesis, Unruh \cite{u76} did, in fact, use this procedure in 1976 for the Schwarzschild BH. The transmission/reflection coefficients for not only the bosons but also for the massive fermions from small BHs were considered \cite{u76}, and the same results were independently obtained by Ford \cite{SK2}.

In Ref. \cite{u76}, the matching technique was first employed for the scalar field equation to determine the solution of the Schr\"{o}dinger equation if the particle wavelength is greater than the Schwarzschild radius of the BH. To overlap the solutions of Part $I$ (near-horizon region) and Part $III$ (distant region) in Part $II$ (intermediate region), one must first get approximate solutions of the wave equation in the relevant regions, and then determine the rate of the incoming wave transmitted through the BH. Furthermore, in this method, the ratio of the total number of particles absorbed by the BH to the entering flux of particles is known as absorption cross-section, which can be determined for bosons for all velocities under the approximation.

Another useful and insightful work about this technique was also done in \cite{is1}, which calculates the GFs in the static and spherically symmetric BHs in $d$-dimensions, as indicated earlier. In fact, the monodromy matching techniques were initially designed to calculate the QNMs \cite{is1} and then were developed by different studies \cite{Natario:2004jd, SK6}. In \cite{SK4}, the monodromy matching technique was applied to derive the QNM frequencies of the $d>4$ Schwarzschild and $d=4$ Reissner-Nordstr\"{o}m BHs for the ultra-high damping case. The equations are calculated using the monodromy of the perturbation technique, which is then carried out analytically to the complex plane. The potential in the non-physical locality appeared around the BH singularity \cite{SK4} is heavily weighted in the evaluation. This research includes the low-energy form of the scattered wave with real and low frequencies, and the asymptotic case, in which the scattered wave's frequencies are extremely high along the imaginary axis. The gravitational perturbations marked out with the master equations of Ishibashi-Kodama were explored in both cases \cite{ik03a, ik03b}. For further approaches in this framework, we refer the reader to \cite{SK9,SK10}, in which the evaporation of a $(d\geq4)$-dimensional rotating BH possessing scalar DoF on the brane was considered. The GFs (and thus the absorption probabilities and cross-sections) for those BH spacetimes were analytically obtained for both high and low-energy regimes \cite{SK10}. Essentially, the procedure performed was to consider the scalar fields propagating in the gravitational background persuade on the brane and to analytically solve the scalar field equation by applying the matching technique on the low-energy regime and matching up the far-field and near-horizon parts of the solution. In general, the analytic expression obtained for the absorption probability has dependency on the angular quantum numbers $(l,m)$. The analytical results obtained in \cite{SK9,SK10} are hold just for low values of the energy parameter $(\omega r_h)$ and rotation parameter $(a)$, and they were in harmony with the numerical results \cite{SK10}. In addition to this, for the large values of $\omega r_h$ and $a$, the agreement on both qualitative and quantitative levels was also confirmed.\\
Using the technique of matched asymptotic expansions in order to analytically compute the GFs was considered in \cite{SK11}, where the minimally-coupled scalar fields propagating on the background of rotating BHs were used in the higher (odd) dimensions. Besides, the non-existence of the superradiant was shown to be a result of reflecting or transparent border conditions since a global timelike Killing vector exists $(\Omega_{h}<l^{-1})$, (here $\Omega_{h}$ denotes the horizon angular velocity) \cite{SK11}. It is worth noting that for this purpose only the positive frequencies modes are considered, namely the frequencies having $\omega-m\Omega_{h}>0$ \cite{SK11}.\\
As being stated above, the first significant study on matching technique \cite{Aliev:2014aba,Aliev:2008yk,Aliev:2015wla} concerning the GFs of the bosons and fermions was dexterously made by \cite{u76}. In that study, the Dirac equation (for fermions), after the operation of separating the angular and radial equations, was studied in the various regions and the solutions to the field equations were obtained via the approximation method. In the sequel, by matching the solutions in the overlap regions, one finds an appropriate intermediate approximate solution for the wave equation and determine the ratio of the incident signal which is transmitted from the BH. Another interesting result appeared, based on this work \cite{u76}, is that all (quantum and/or classical) particles with a velocity $v=1$ (see \cite{u76} for details) observe the BH as having roughly the same size.\\
In Ref. \cite{SK9}, the emission of fermions from  the aforementioned projected BH background were also studied. It was considered that the emitted particle modes couple at best minimally to the gravitational background, thus, they hold the free equations of motion \cite{SK9}. When the Newman-Penrose (NP) formalism \cite{chandra} is applied, by further assuming the factorized ansatz with spin-weighted spheroidal harmonics, the free equations of motion for particles with spin $s=0,\frac{1}{2}$ can form a master equation on the brane. Then, the analytical results to the radial general equation are obtained for the two asymptotic regions of the BH horizon $(r\approx r_{h})$ and in the far-field $(r>>r_{h})$, these two solutions are related at a middle radial region for constructing a complete solution, which is valid at all radial regimes, for a field with arbitrary spin $s$. The derived expressions of the absorption probabilities (valid for $\omega\rightarrow 0$) for gauge bosons show that thermal radiation of spin-$0$ particles vanishes faster than the fermionic ones, spin-$\frac{1}{2}$. Furthermore, it is proven that the total absorption cross-section behaves differently for gauge bosons and fermions in the low-energy limit than it does in the non-rotating case, while in the high-energy regime it behaves universally, achieving a constant, spin-independent asymptotic value.
    
    \section{\texorpdfstring{Computing GF\MakeLowercase{s} and QNM\MakeLowercase{s} via WKB Approximation}{d}} \label{sec4}
      This section is addressed to the calculation of GFs and QNMs by applying the WKB approximation. This approximation method is nothing but a semi-analytic technique, which is mostly used for determining the mode frequencies of the perturbed systems. This method is also used to derive analytic expressions for the QNMs with real and imaginary frequency components in terms of the BH parameters and numbers $(n=0,1,2,..)$, which label the modes; for instance $n=0$ is the fundamental mode, $n=1$ denotes the first overtone mode, and so on. The WKB approach has found itself many different application areas in the literature including the computations of the GFs. It is indeed an accurate approach, which can be carried to the higher orders. In this context, Schutz and Will conducted one of the foundation investigations for the Schwarzschild BH in 1985 \cite{SK12}. Moreover, it was found that the agreement of WKB approach with other methods is excellent especially for low lying modes \cite{SK13,Alsup:2008fr}, which are carried the results to third order beyond (within the eikonal approximation) \cite{SK14}. In the following, we will go over the detailed applications for both GFs and QNMs based on the WKB approximation for both bosons and fermions.
 
\subsection{Bosonic GFs and QNMs}

The WKB approximation for the GFs is mostly used to calculate the reflection and transmission probabilities. In fact, the transmission coefficient can also be interpreted as GF, which determines the spectrum of Hawking radiation \cite{n03}. There are several valuable studies along this line. Its applications do not only take place in quantum mechanics, but also in higher dimensional gravity, astrophysics, string theory \cite{SK16}, wormhole physics \cite{SKL16n}, magnetized BHs \cite{SKL17},  and $5$-dimensional charged Bardeen BH spacetimes (by involving the perturbations of scalar and electromagnetic fields) \cite{SK17} etc. The use of the WKB approximation in BH physics has gained momentum when extended to the sixth order. The pioneering works on this subject undoubtedly belong to Konoplya and his co-authors (see \cite{Konoplya:2003ii,Konoplya:2019hlu,SK16,SKL16n} and references therein). Besides, the third order WKB approximation is also in the agenda literature for studying the GFs; for example, the QNMs of the Hayward BH surrounded by quintessence have recently been studied in \cite{SKW17} for some special cases of the quintessence parameter.

\begin{figure}[h]
\centering\includegraphics[scale=0.5]{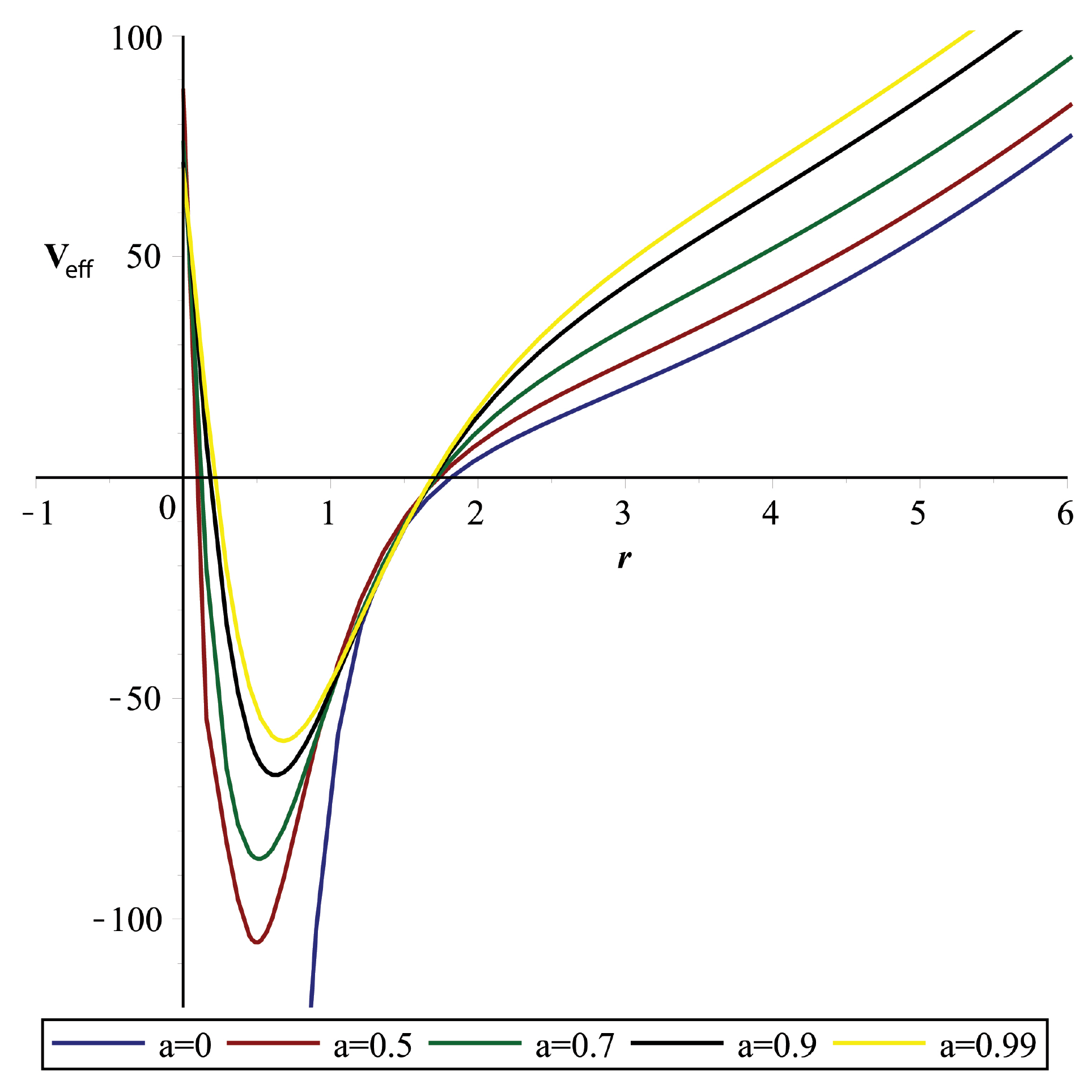}\caption{
$V_{eff}$ \eqref{32} plots for the scalar particles of the rotating polytropic BH. Physical quantities are chosen as $M=\omega=3, m=1, \lambda=6$ and $L=1,$.} \label{fig1}
\end{figure}

At this stage, let us look at our first practical application. To this end, we consider Ref. \cite{Kanzi:2021jrl}, which is one of our (Kanzi and Sakall{\i}) most recent works. Thus, we shall employ the WKB approximation for computing the transmission and reflection coefficients for scalar perturbation of the polytropic BH \cite{Kanzi:2021jrl} whose line element is%

\begin{equation}
ds^{2}=\left(  1-\frac{2f}{\rho^{2}}\right)  dt^{2}-\frac{\rho^{2}}{\Delta
}dr^{2}+\frac{4af\sin^{2}\theta}{\rho^{2}}dtd\phi-\rho^{2}d\theta^{2}%
-\frac{\Sigma\sin^{2}\theta}{\rho^{2}}d\phi^{2}, \label{1}%
\end{equation}
where 
\begin{equation}
 \begin{aligned}
2f=r^{2}\left(  1-F\right),\\
\rho^{2}=r^{2}+a^{2}\cos^{2}\theta, \quad \\
\Sigma=\left(  r^{2}+a^{2}\right)  ^{2}-a^{2}\Delta\sin^{2}\theta,
\end{aligned}
\end{equation}
in which%
\begin{equation}
\Delta=a^{2}+r^{2}F, \label{2}%
\end{equation}
and%
\begin{equation}
F=\frac{r^{2}}{L^{2}}-\frac{2M}{r}. \label{3}%
\end{equation}
In Eq. (\ref{3}), $L^{2}=-\frac{3}{\Lambda}$ in which $\Lambda$ is the cosmological
constant. The case of negative $\Lambda$ will be considered throughout this section. In other words, we assume that the empty space has positive pressure but negative energy density, such as the AdS space. To determine the scattering coefficients, one needs to derive first the effective potential from the 1-dimensional Schr\"{o}dinger-like wave equation. The following massive ($m_{0}$: mass of boson) Klein-Gordon (KG) equation can be used to determine the wave equation of the associated scalar field:
\begin{equation}
\left(  \nabla^{\nu}\nabla_{\nu}-m_{0}^{2}\right)  \Phi=0, \label{s22}%
\end{equation}
where $\nabla_{\nu}$ denotes the covariant derivative. Equation \eqref{s22} can be alternatively expressed as
\begin{equation}
\left[  \frac{1}{\sqrt{-g}}\partial_{\mu}\left(  g^{\mu\nu}\sqrt{-g}%
\partial_{\nu}\right)  \right]  \Phi-m_{0}^{2}\Phi=0. \label{23}%
\end{equation}
\noindent
Since the line-element under consideration is Eq. \eqref{1}, one computes the square root of determinant of the metric as $\sqrt{-g}=\rho^{2}\sin\theta$. In the sequel, considering the symmetries of the metric, we can define the following ansatz for the wave function:
\begin{equation}
\Phi\left(  r,t\right)  =R\left(  r\right)  S\left(  \theta\right)
e^{im\varphi}e^{-i\omega t}, \label{26}%
\end{equation}
\noindent
where, needless to say, $\omega$ is the frequency of wave and $m$ denotes the well-known azimuthal quantum number. Thus, one can get the radial equation as follows%
\begin{multline}
\Delta\frac{d}{dr}\left(  \Delta\frac{dR\left(  r\right)  }{dr}\right)  +\\
\left[  m^{2}a^{2}-4afm\omega+\omega^{2}\left(  r^{2}+a^{2}\right)
^{2}-\left(  \omega^{2}a^{2}+m_{0}^{2}r^{2}+\lambda_{lm}\right)  \Delta\right]
R\left(  r\right)  =0, \label{s27}%
\end{multline}
\noindent
which $\lambda_{lm}$ represents the angular solution eigenvalue whose differential equation is given by%
    \begin{equation}
\frac{1}{S\left(  \theta\right)  }\frac{1}{\sin\theta}\frac{d}{d\theta}\left(
\sin\theta\frac{dS\left(  \theta\right)  }{d\theta}\right)  -\frac{m^{2}}%
{\sin^{2}\theta}+c^{2}\cos^{2}\theta=-\lambda_{lm}, \label{28}%
\end{equation}
\noindent
here $c^{2}=\left(  \omega^{2}-\mu_{0}^{2}\right)  a^{2}.$ On the other hand, we have%
\begin{equation}
\frac{dr_{\ast}}{dr}=\frac{r^{2}+a^{2}}{\Delta}. \label{29}%
\end{equation}
\noindent
In addition, let us utilize a transformation as%
\begin{equation}
R=\frac{U}{\sqrt{r^{2}+a^{2}}}, \label{30}%
\end{equation}
\noindent
to obtain the 1-dimensional Schr\"{o}dinger equation:
\begin{equation}
\frac{d^{2}U}{dr_{\ast}^{2}}=\left(  V_{eff}-\omega^{2}\right)U,
\label{31}%
\end{equation}
in which the effective potential $V_{eff}$ is obtained as
\begin{equation}
V_{eff}=\Delta\left(  r^{2}+a^{2}\right)  ^{-2}\left[  \frac
{\Delta+\Delta^{\prime}r}{r^{2}+a^{2}}-\frac{3r^{2}\Delta}{\left(  r^{2}%
+a^{2}\right)  ^{2}}+\frac{4afm\omega-m^{2}a^{2}}{\Delta}+\omega^{2}a^{2}%
+\mu_{0}^{2}r^{2}+\lambda_{lm}\right]  , \label{32}%
\end{equation}
where $^\prime=\frac{d}{dr}$. In Fig. \eqref{fig1}, we have demonstrated the influence of the rotational parameter $a$ on $V_{eff}$, \textit{i.e.}, Eq. \eqref{32}. As it is shown in Fig. \eqref{fig1}, the height of the $V_{eff}$ barrier rises as the rotation $a$ increases. Now, it is worth recalling a well-known scientific fact: the chance of a particle tunneling relies on the proceeding particle energy in relation with the height of the barrier \cite{qmbook}.\\
Let us consider a moving wave that approaches the BH ($r_{\ast}\rightarrow-\infty$) from spatial infinity $r_{\ast}\rightarrow\infty$ to determine the scattering coefficients. Naturally, while some of the waves are transmitted, the others will be reflected, due to the effective potential. Considering the scattering boundary conditions \eqref{set11}
with the following normalization condition
\begin{equation}
\left\vert T\right\vert ^{2}+\left\vert R\right\vert ^{2}=1, \label{34}%
\end{equation}
\noindent
one can define the GF as (see \cite{SK27} and references therein) 
\begin{equation}
\Gamma_{l}\left(  \omega\right)  =\left\vert T\left(  \omega\right)
\right\vert ^{2}.\label{35}%
\end{equation}
\noindent
The definitions of $R\left(  \omega\right)  $ and $T\left(  \omega\right)$ are given by \cite{Konoplya:2003ii,Kanzi:2021jrl}
\begin{equation}
T(\omega)=\big(1+\exp\left(2\pi i\mathfrak{K}\right)\big)^{-2}, \label{36}%
\end{equation}
\begin{equation}
R(\omega)=\big(1+\exp\left(  -2\pi i\mathfrak{K}\right)\big)^{-2}, \label{37}%
\end{equation}
where
\begin{equation}
\mathfrak{K}=i\frac{\omega_{n}^{2}-V\left(  r_{0}\right)  }{\sqrt{-2V^{\prime\prime
}\left(  r_{0}\right)  }}-\Lambda_{2}-\Lambda_{3}, \label{e38}%
\end{equation}
in which $r_{0}$ is the special radial location which lead the potential $V(r)$ to reach its maximum ($V_{0}$), \textit{i.e.}, to its peak. Let us set $V(r_{0})=V_{0}$. Moreover,
\begin{equation}
\Lambda_{2}=\frac{1}{\sqrt{-2V_{0}^{\prime\prime}}}\left[  \frac{1}{8}\left(
\frac{V_{0}^{\left(  4\right)  }}{V_{0}^{\prime\prime}}\right)  \left(
\frac{1}{4}+\alpha^{2}\right)  -\frac{1}{288}\left(  \frac{V_{0}^{\left(
3\right)  }}{V_{0}^{\prime\prime}}\right)  ^{2}\left(  7+60\alpha^{2}\right)
\right]  , \label{e39}%
\end{equation}
and%
\begin{multline}
\Lambda_{3}=\frac{1}{\sqrt{-2V_{0}^{\prime\prime}}}\left[  \frac{5}%
{6912}\left(  \frac{V_{0}^{\left(  3\right)  }}{V_{0}^{\prime\prime}}\right)
^{4}\left(  77+188\alpha^{2}\right)  -\frac{1}{384}\left(  \frac{V_{0}%
^{\prime\prime\prime2}V_{0}^{\left(  4\right)  }}{V_{0}^{\prime\prime3}%
}\right)  \left(  51+100\alpha^{2}\right)  +\right. \\
\left.  \frac{1}{2304}\left(  \frac{V_{0}^{\left(  4\right)  }}{V_{0}%
^{\prime\prime}}\right)  ^{2}\left(  67+68\alpha^{2}\right)  -\frac{1}%
{288}\left(  \frac{V_{0}^{\prime\prime\prime}V_{0}^{\left(  5\right)  }}%
{V_{0}^{\prime\prime2}}\right)  \left(  19+28\alpha^{2}\right)  -\frac{1}%
{288}\left(  \frac{V_{0}^{\left(  6\right)  }}{V_{0}^{\prime\prime}}\right)
\left(  5+4\alpha^{2}\right)  \right]. \label{e40}%
\end{multline}

\begin{figure}[h]
\centering\includegraphics[scale=0.5]{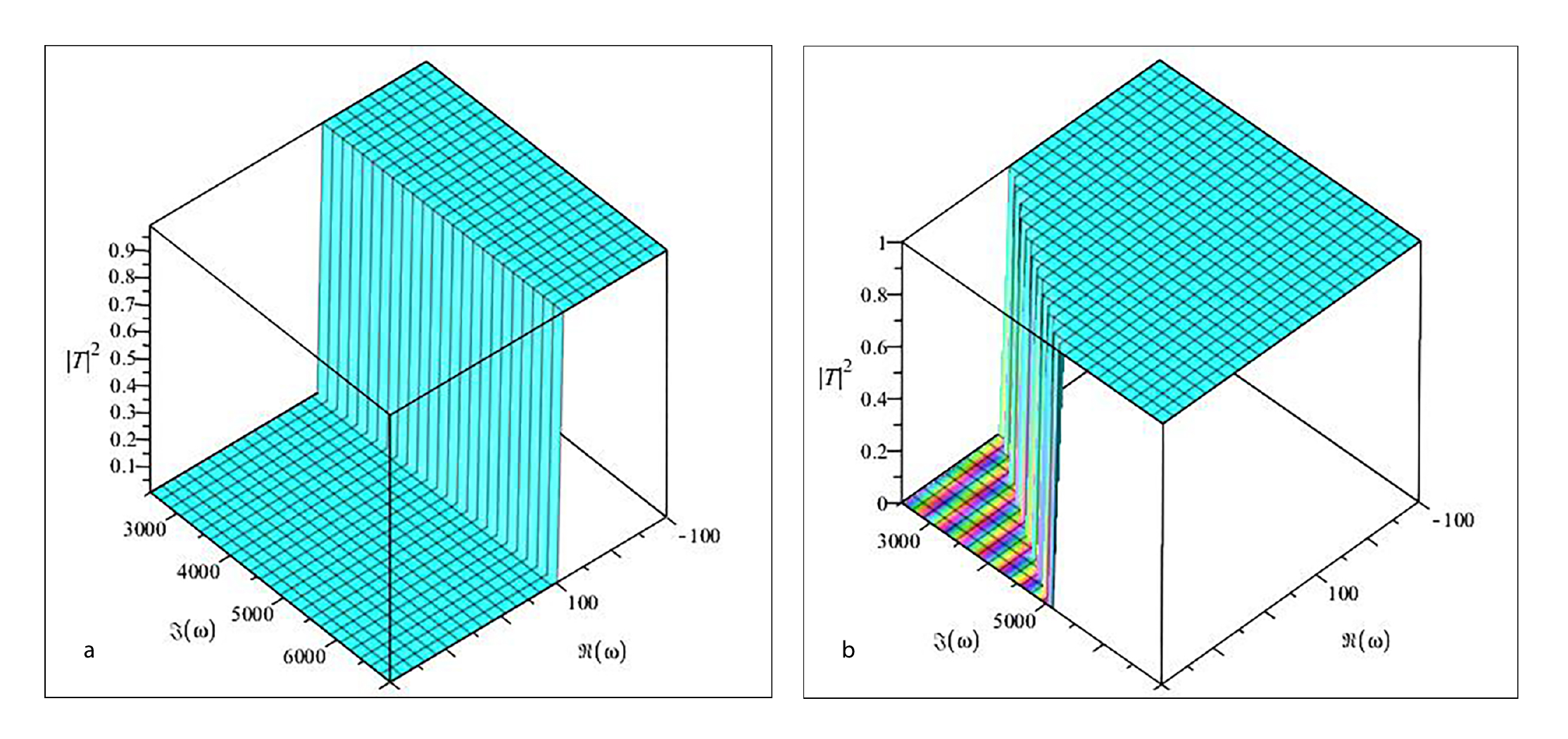}\caption{Transmission coefficient \eqref{36} plots for the rotating polytropic BH for $a=0.011$ (a) and $a=0.50$ (b). The physical quantities are chosen as $m=L=1, M=3, \lambda=6$. } \label{figeS1}%
\end{figure}

\begin{figure}[h]
\centering\includegraphics[scale=0.5]{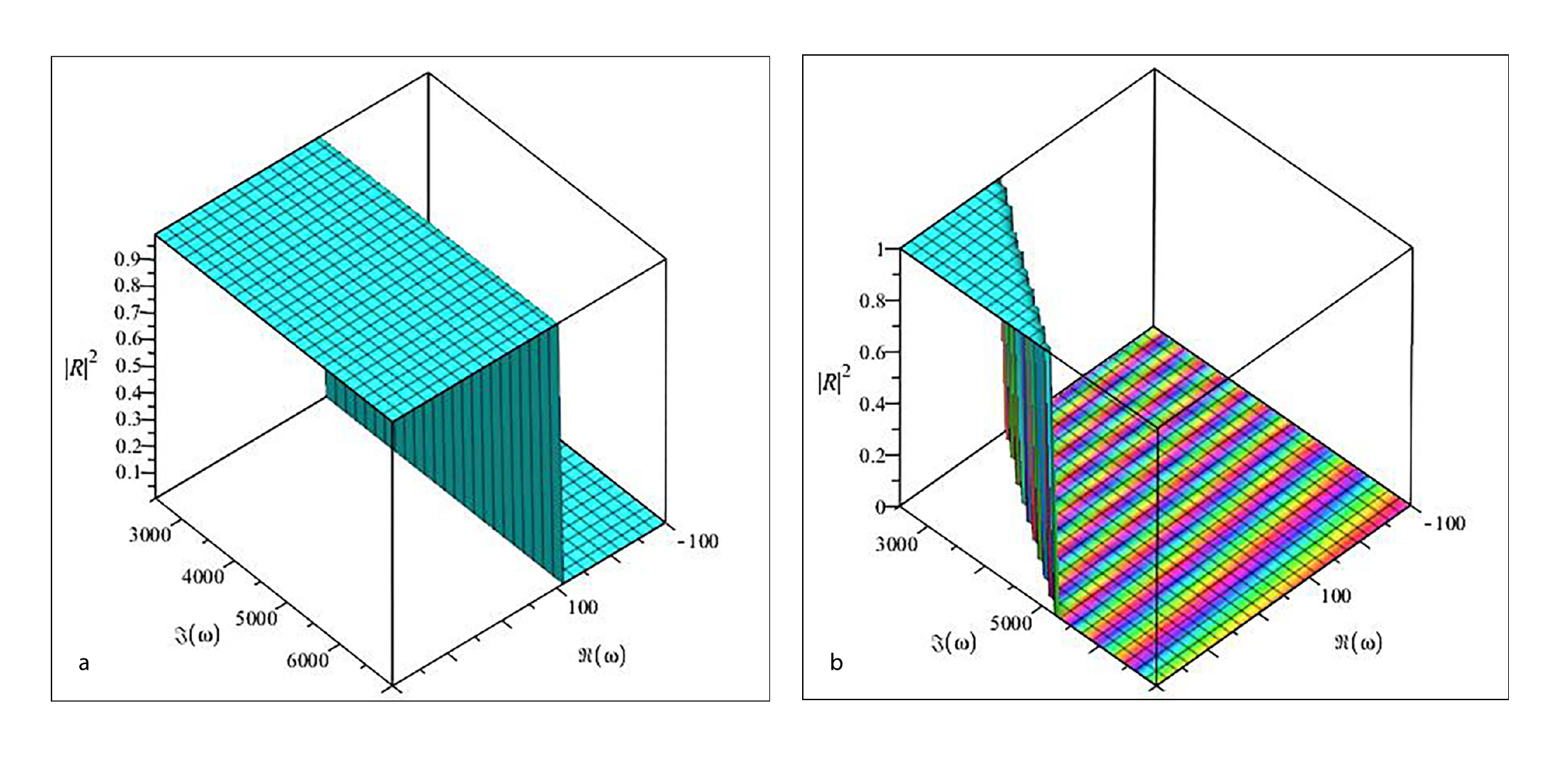}\caption{Reflection coefficient \eqref{37} for the rotating polytropic BH for $a=0.011$ (a) and $a=0.50$ (b). The Physical quantities are chosen as $m=L=1, M=3, \lambda=6$ } \label{figeS2}%
\end{figure}
In Eqs. \eqref{e38}-\eqref{e40}, both prime symbol and  superscript numbers $\left(3-6\right)$ denote the order of derivative with respect to $r_{\ast}$, and $2\alpha=2n+1$ in which $n$ represents the overtone number.

The behaviors of $R\left(  \omega\right)  $ and $T\left(  \omega\right)$ for $ a=0.011$ and $a=0.5$ are illustrated in Figs. \eqref{figeS1} and \eqref{figeS2}, respectively. The evaluations under varying $a$, in particular, are performed for both real and imaginary parts. By varying $a$ parameter, the damping modes change more than the oscillatory ones. \\
As second application, we now study the influence of the Lorentz symmetry breaking (LSB) parameter in the scalar and Dirac perturbations, and study the GFs and QNMs for the bosonic and fermionic particles. In fact, we studied this problem before by considering the Kerr-like BH (KlBH) in bumblebee gravity model (BGM) \cite{Ding:2019mal} and conducted a thorough investigation of their GFs and QNMs \cite{Kanzi:2021cbg}. But, after being discovered by \cite{Maluf:2022knd}, it was understood that the metric of \cite{Ding:2019mal} that was already used by us in \cite{Kanzi:2021cbg} is wrong. Then, we revised all of our studies and demonstrated \cite{Kanzi:2022vhp} that problem can be corrected when the slowly rotating KlBH \cite{Ding:2020kfr} is considered. In this context, the information that we shall serve in this section is based on Ref. \cite{Kanzi:2022vhp}.

We now give the computation details of the bosonic QNMs for slowly rotating KlBH spacetime via the six order WKB approximation \cite{Kanzi:2022vhp}, in which the scalar QNMs are elaborated. The slowly rotating KlBH metric in the BGM is given by \cite{Ding:2020kfr}%
\begin{equation}
d s^{2}	\approx-\left(1-\frac{2 M}{r}\right) d t^{2}-\frac{4 M \tilde{a} \sin ^{2} \theta}{r} d t d \varphi+\frac{(1+L) r}{r-2 M} d r^{2}+r^{2} d \theta^{2}+r^{2} \sin ^{2} \theta d \phi^{2}, \label{W1}
\end{equation}
\noindent
where $\tilde{a}=\sqrt{1+L}a$.
Following the same process as previously prescribed in Eqs. \eqref{23}-\eqref{32}, one gets the effective potential of the scalar waves propagating in the BH geometry \eqref{W1} as follows
\begin{equation}
V_{\text {eff }}=\frac{\left(1-\frac{2M}{r}\right)}{r^2}\left[\frac{2M}{r\left(1+L\right)}+\mu^2r^2+\lambda\right]+\frac{4Mm\omega a}{r^3}.  \label{iz8}
\end{equation}
\noindent
\begin{figure}[h]
\centering\includegraphics[scale=0.5]{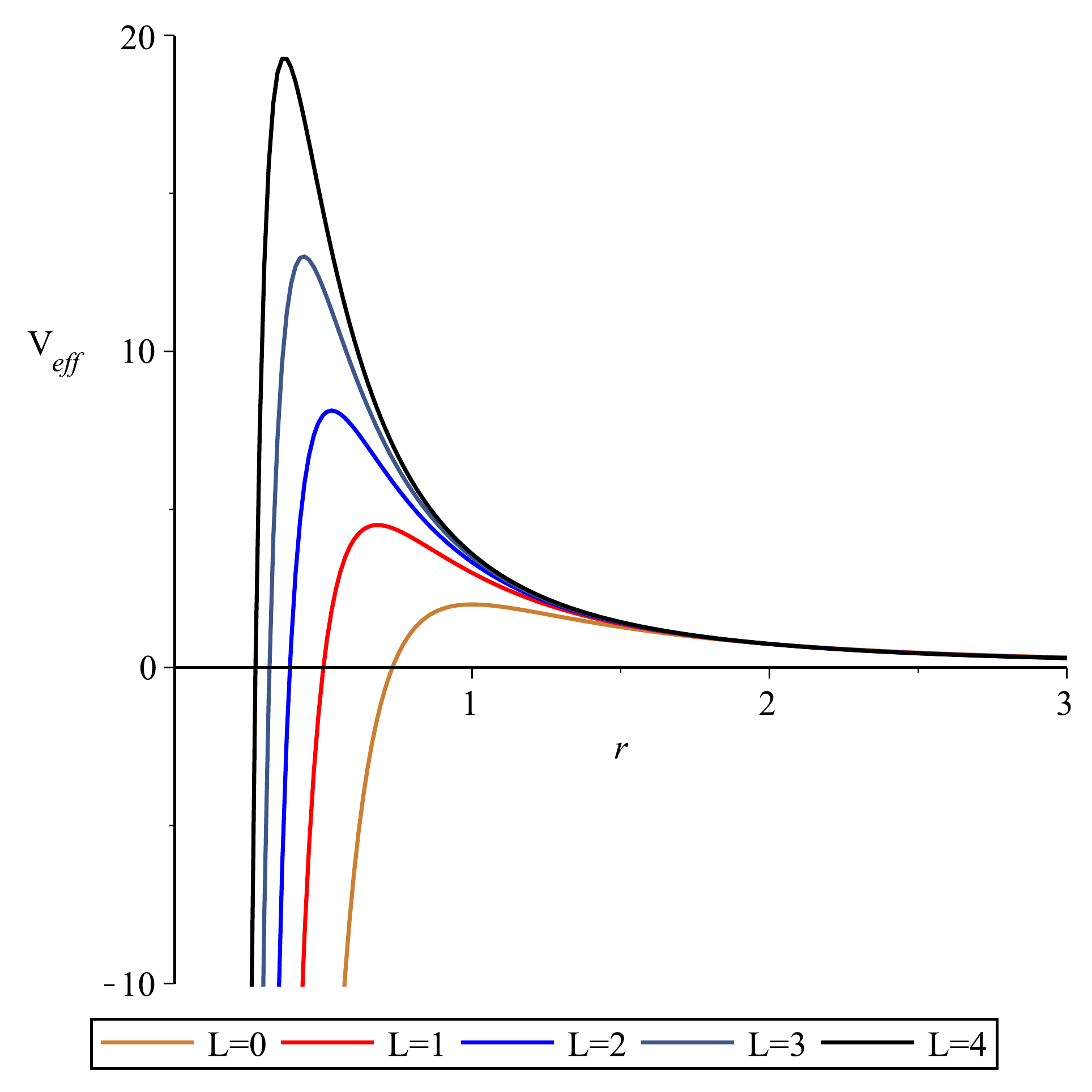}\caption{Plots of $V_{eff}$ \eqref{iz8} for the bosonic waves of the slowly rotating bumblebee BH. The physical quantities are chosen as $M=m=1, a=0.1, \omega=15,$ and $\lambda=2$.}%
\label{myfig1}%
\end{figure}
The effective potential behavior \eqref{iz8} under the influence of LSB quantity is plotted in Fig. \ref{myfig1}. It is obvious from  Fig. \ref{myfig1} that there exists a significant reduction in the peak when the LSB parameter is increased. For evaluating the QNMs via the WKB approach, one should transform the obtained 1-dimensional Schr\"odinger-like equation \eqref{31} to the wave equation \eqref{is4} (see also \cite{Konoplya:2003ii}). Thus, one can compute the QNMs as:

\begin{equation}
\omega^{2}=\left[  V_{0}+\sqrt{-2V_{0}^{\prime\prime}}\Lambda\left(  n\right)  -i\left(  n+\frac{1}{2}\right)  \sqrt{-2V_{0}^{\prime\prime }}\left(  1+\Omega\left(  n\right)  \right)  \right]  ,\label{W26}%
\end{equation}
\noindent
where $\Lambda$ and $\Omega$ are the same as Eqs. \eqref{e39} and \eqref{e40}. The results for bosonic QNMs are tabulated in Table \ref{etab1} for $m=0$, $l=2$, and the initial overtone $n=0$. It can be inferred from Table \ref{etab1} that the oscillations which are governed by the real values reduce if the LSB quantity rises. However for the imaginary part, there is no coherent facts about the LSB attitude. For example, in the case of $L\approx0-1.2$, we have a reduction with the increasing LSB effect in the damping mode first and then this harmonious trend vanishes and a random behaviors appear, which can be deduced from Fig. \ref{myfig1}.

\begin{table}
  \centering
    \begin{tabular}{ |c|c| }
\hline
$L$ & $\omega_{bosons}$ \\
\hline
 0 &  0.6193445868-0.4259671652i\\
 1 &  0.7688968549-0.6024536531i \\
1.1 & 0.7791058962-0.6159336514i \\
1.2 & 0.7887835943-0.6289091139i \\
1.3 & 0.7979767859-0.6414186083i \\
1.4 & 0.8067264683-0.6534964325i \\
1.5 & 0.8075365232-0.6587911664i \\
1.6 & 0.8230350287-0.6764758541i \\
1.7 & 0.8306539984-0.6874294407i \\
1.8 & 0.8379506594-0.6980559713i \\
1.9 & 0.8449475817-0.7083755957i\\
2   & 0.8516649690-0.7184065252i\\
\hline 
    \end{tabular}
  \caption{QNMs of scalar waves in the slowly rotating KlBH background.} \label{etab1}
\end{table}

\subsection{Fermionic GFs and QNMs}
\begin{figure}[h]
\centering\includegraphics[scale=0.5]{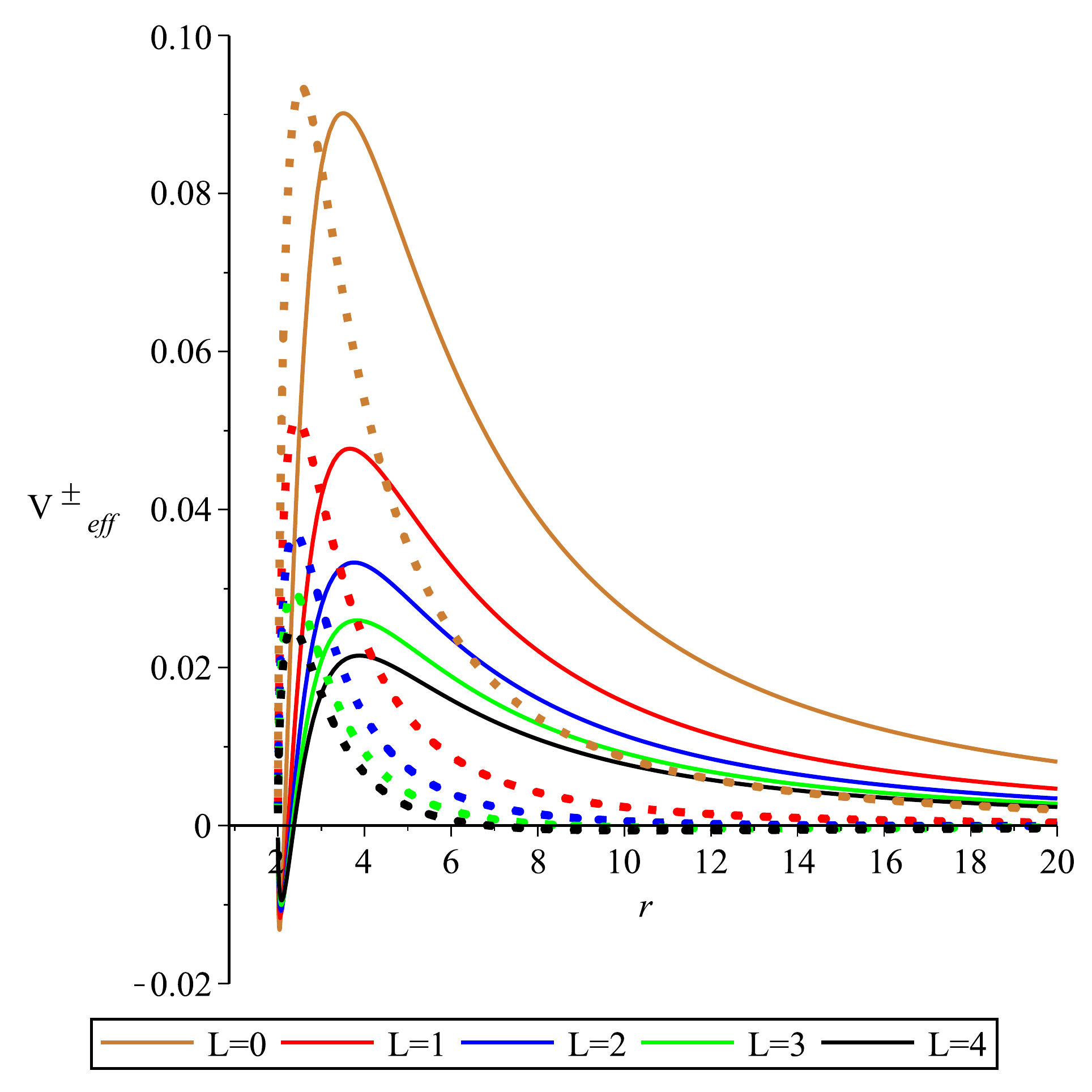}\caption{$V_{eff}^{\pm}$ \eqref{S43} plots for the fermionic waves. While solid lines belong to $V_{eff}^{+}$, the dotted lines represent $V_{eff}^{-}$. The physical quantities are chosen as $\lambda=-1.5$ and $M=1$.}%
\label{myfig3}
\end{figure}

Before we get into the details of the Dirac perturbation (see Section \ref{5b}) to derive the QNMs and GFs by using the WKB method, it might be worthy to mention some more investigations about the subject. In Ref. \cite{SK19}, to analyze the behavior of Dirac perturbations when vacuum fluctuations are taken into consideration, the QNMs of a quantum corrected Schwarzschild BH were explored. For $4$-dimensional Kerr BH, the numerical analyses of the fermionic QNMs, which were produced by using the third and sixth order semi-analytic WKB(J) and advanced iteration methods, can be found in \cite{WSK19,SKW19,SKWR19}. Similar works were made for the $f(R)$ global monopole \cite{SKW20} and for the massive Dirac field in the Kerr background \cite{SK20}.\\
Furthermore, in Refs. \cite{Kanzi:2021cbg,Kanzi:2022vhp}, we  also investigated the Dirac QNMs of KlBH in the BGM. To proceed the evaluations of the Dirac fields in the KlBH geometry, we used the $4$-dimensional Dirac equation formulated in the NP formalism
\cite{SK21} in order to compute the fermionic $V_{eff}$ for the KlBH \cite{Kanzi:2021cbg,Kanzi:2022vhp}:

\begin{equation}
V_{\text {eff }}^{\pm}=\frac{\lambda\left(1-\frac{2M}{r}\right)}{\sqrt{1+L}}\left[\frac{\lambda}{r^2\sqrt{1+L}}\pm\left(\frac{r-M}{r^2\sqrt{r^2-2Mr}}-\frac{2\sqrt{r^2-2Mr}}{r^3}\right)\right], \label{S43}
\end{equation}
\noindent
where $\pm$ symbols represent the spin-up/down fermions. Fig. \ref{myfig3} shows the behaviors of the effective potentials (\ref{S43}).

To derive the fermionic QNMs, one can employ the method for computing the bosonic QNMs prescribed in the previous section. Therefore, we analyze the potentials seen in Eq. (\ref{S43}) and use them in Eqs. \eqref{e39}-\eqref{e40}, and \eqref{W26}. The numerically computed QNM frequencies are displayed in Table \ref{tab2} for different values of the LSB quantity together for $a=0.4$ . As it is shown in the Table \ref{tab2}, both damping and oscillatory terms of the fermionic QNMs tend to reduce with the rising LSB parameter, and inversely rise with the growing $l$ and $n$ parameters. In summary, in comparison to bosonic QNMs, fermionic QNMs have a distinct character.

 \section{\texorpdfstring{Calculating GF\MakeLowercase{s} Via Bounding Bogoliubov Coefficients Method}{d}} \label{sec5}

The general bounds on the Bogoliubov coefficients approach to determining the transmission probability or GFs, proposed in \cite{SK24}, are discussed in this section. For 1-dimensional potential scattering, some generic bounds for reflection and transmission coefficients were established and also shown how the bounds are extended with various approaches \cite{,SKl25,mtd4,SK26}. Regardless of the various successful methods applied for getting the GFs and QNMs, what matters at the end of the day is to have the effective potential and analyze its behavior under the physical changes. However, based on the particles character \big(spin-0, spin-$(\pm1/2)$\big), the paths for deriving the effective potentials could be disparate.
\begin{table}
  \centering
    \begin{tabular}{ |c|c|c|c| }
\hline
$l$ & $n$ & $L$ & $\omega_{fermions}$ \\
\hline
 1 & 0 &  0  &  0.1615755163-0.1986761458i\\
   &   & 0.1 &  0.1057644220-0.1612179485i \\
   &   & 0.2 &  0.0530479984-0.1300388664i \\
   &   & 0.3 &  0.0019930161-0.1113914550i \\
   & 1 &  0  &  0.2408767549+0.5575170642i \\
   &   & 0.1 &  0.3028756279+0.5575297813i \\
   &   & 0.2 &  0.3477807164+0.5531523256i \\
   &   & 0.3 &  0.3780138537+0.5433679210i \\
2  & 0 &  0  &  1.4891900030-0.8296137229i \\
   &   & 0.1 &  1.0377655280-0.3211236139i \\
   &   & 0.2 &  0.8543487310-0.3189534898i\\
   &   & 0.3 &  0.7108515298-0.3159381339i\\
   &   & 0.4 &  0.5961995167-0.3122465225i\\
\hline 
    \end{tabular}
  \caption{QNMs of Dirac waves in the slowly rotating KIBH background.} \label{etab2}
\end{table}

\subsection{Bosonic GFs and QNMs}
Before going into the details of the subject seen in the title of this subsection, it would be better if we continue to make more literature reviews about the GFs. Calculation of HR in the semi-analytic approach for the $5$-dimensional BH conformally coupled with scalar field yielded that the GF can ‘suppress’ the total Hawking radiation power \cite{WSK27}. In addition, in the impacts of Hawking radiation, both conformally coupled and charge parameters play an opposing role. For a $d-$dimensional Schwarzschild-Tangherlini BH spacetime \cite{LSK27}, it was shown that the GFs bound in bulk and brane-localized scalar field reduces as the
spacetime dimension $d$ rises.\\
Let us now consider the SBHBGM spacetime \cite{SK27,Casana:2017jkc}, which is given by \cite{Casana:2017jkc}

\begin{equation}
ds^{2}=-\left(1-\frac{2M}{r}\right)dt^{2}+\left(1+L\right)\left(1-\frac{2M}{r}\right)^{-1}dr^{2}+r^{2}\left(d\theta^{2}+sin^{2}\theta d\phi^{2}\right). \label{ww1}%
\end{equation}
Recall that $L$ denotes the LSB parameter. In order to derive the effective potential, one can invoke the following ansatz for the massless scalar field equation \eqref{23}  \cite{SK27}
\begin{equation}
\psi=\frac{U\left(r\right)}{r}A\left(\theta\right)e^{-i\omega t}e^{im\phi}, \label{ww3}%
\end{equation}
and $r^{*}$ coordinate for the SBHBGM is found by $r_{\star}=\sqrt{1+L}\int\frac{dr}{f}$.
Thus, after some straightforward computations, the scalar effective potential of the SBHBGM can be derived as 
\begin{equation}
V_{eff}=f\left[\frac{f^{\prime}}{\left(1+L\right)r}+\frac{\lambda}{r^{2}}\right], \label{ww2}%
\end{equation}
where $\lambda$ represents the angular eigenvalue, $f$ is the metric function, and $^{\prime}=\frac{d}{dr}$.

\begin{figure}[h]
\centering\includegraphics[scale=0.6]{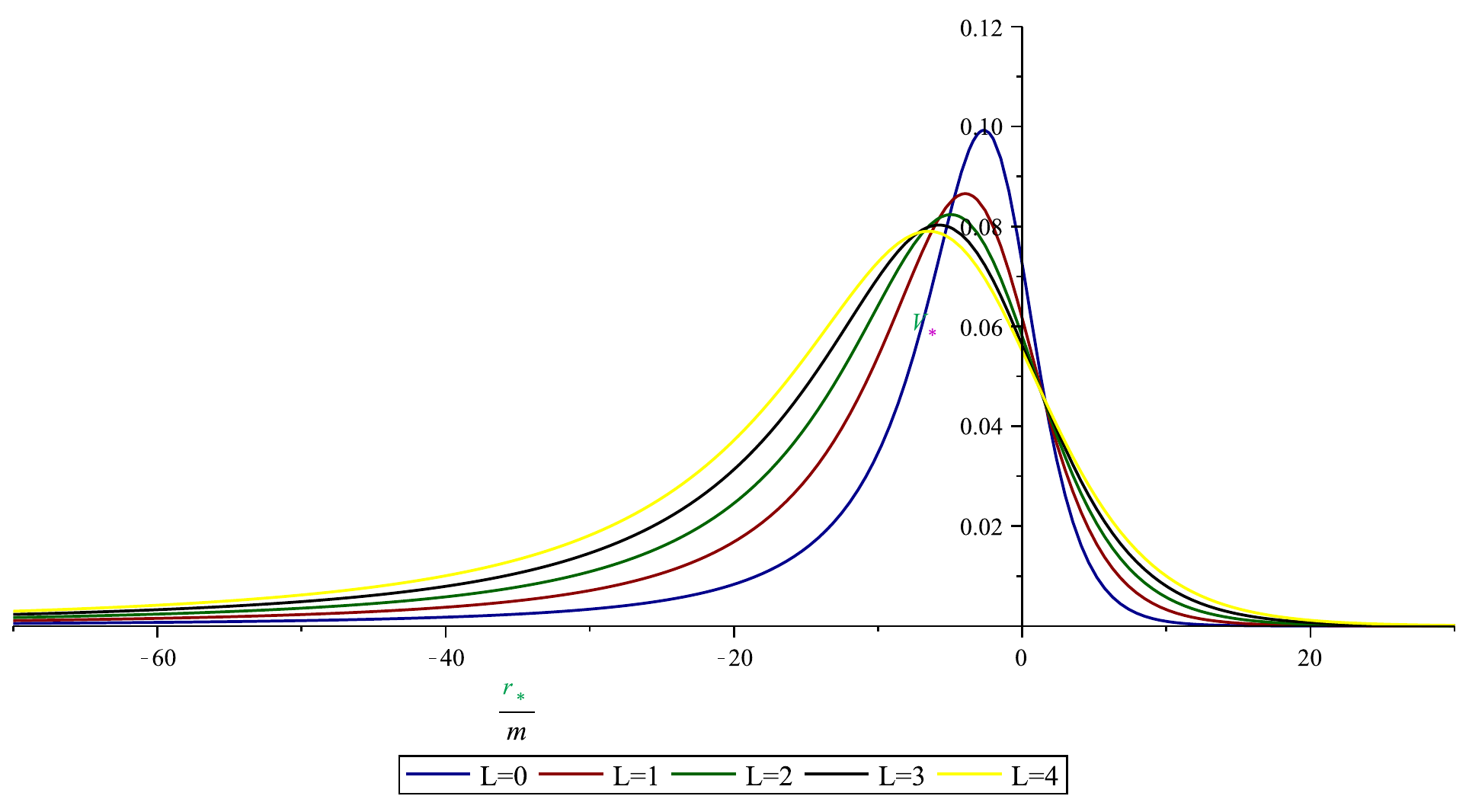}\caption{
$V_{eff}$ \eqref{ww2} plots for the scalar waves propagating in the geometry of the SBHBGM.} \label{Myfig2}
\end{figure}
\begin{figure}[h]
\centering\includegraphics[scale=0.5]{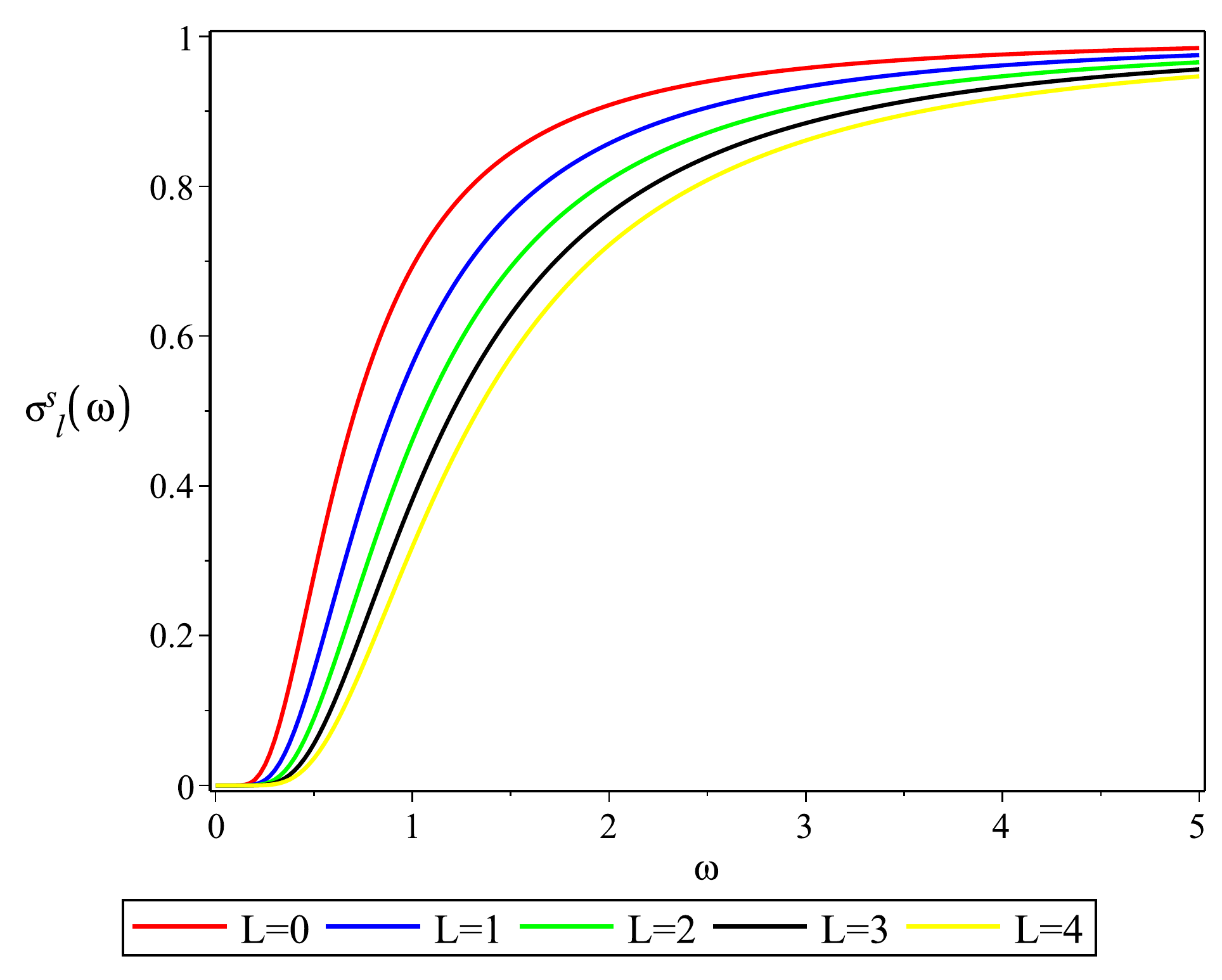}\caption{$\sigma_{l}^{s}\left(\omega\right)$ plots for the scalar particles of SBHBGM. The graph is governed by Eq. \eqref{gb2} with $M=1$.} \label{Myfig3}
\end{figure}

Figure \eqref{Myfig2} reveals that $V_{eff}$ disappears both at the event horizon and spatial infinity of the SBHBGM geometry. This picture assists us to analytically compute the GFs of the scalar field emitted from the SBHBGM.
For obtaining the GFs, we employ the following general semi-analytic bound formula (see \cite{SK27} and references therein)
\begin{equation}
\sigma _{l}\left( \omega \right) \geq \sec h^{2}\left( \int_{-\infty
}^{+\infty }\wp dr_{\ast }\right) ,  \label{is8}
\end{equation}
\noindent
where 
\begin{equation}
\wp =\frac{1}{2h}\sqrt{\left( \frac{dh\left( r_{\ast }\right) }{dr_{\ast }}%
\right) ^{2}+(\omega ^{2}-V_{eff}-h^{2}\left( r_{\ast }\right) )^{2}},
\label{is9}
\end{equation}
\noindent
in which $h(r_{\ast} )$ represents a positive function with condition $h\left( -\infty \right)
=h\left( -\infty \right) =\omega $, and, in general, one sets $%
h=\omega $. Therefore,%
\begin{equation}
\sigma_{l} \left( \omega \right) \geq \sec h^{2}\left( \int_{r_{h}}^{+\infty }%
\frac{V_{eff}}{2\omega }dr_{\ast }\right) .  \label{gb1}
\end{equation}%
We then use the potential (\ref{ww2}) to obtain the scalar GFs in the SBHBGM background. Thus, scalar ($s$) GFs are found to be
\begin{equation}
\sigma_{l}^{s}\left(\omega\right) \geq \sec h^{2}\left[\frac{\sqrt{1+L}}{2\omega r_{h}} \left(\lambda+\frac{1}{2\left(1+L\right)}\right)\right].  \label{gb2}
\end{equation}%
GFs of the SBHBGM arising from the scalar potential are depicted in Fig. \eqref{Myfig3}. Since a BH is not a perfect black-body, therefore, the GF of the HR must be $<1$. Thus our findings, as shown in Fig. \eqref{Myfig3}, are fully in agreement with the latter remark. Moreover, the peak of the GFs reduce with the $L$ parameter. To sum up, LSB has a GF-reducing effect for the bosonic fields.

\begin{figure}[h]
\centering\includegraphics[scale=0.5]{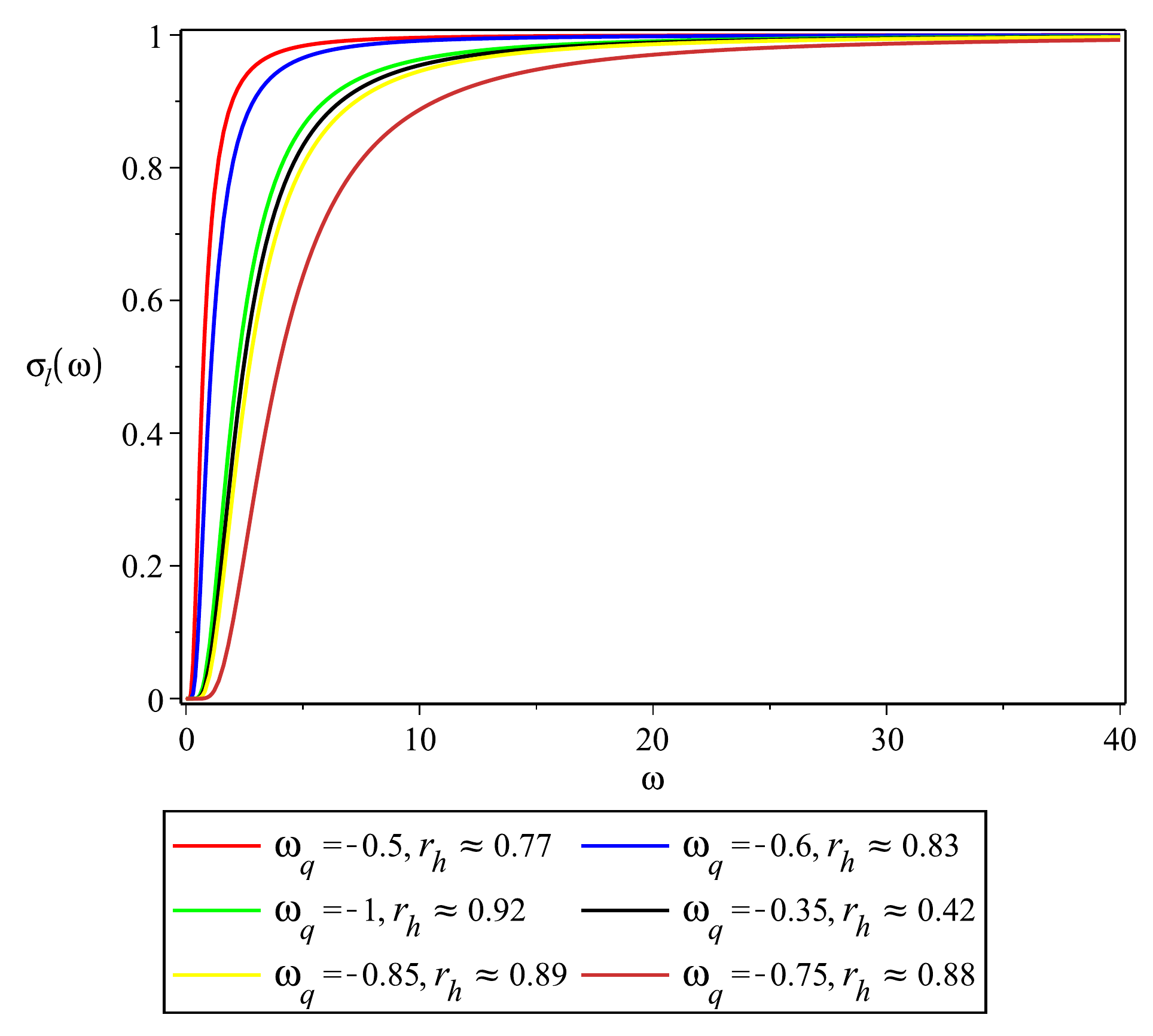}\caption{$\sigma_{l}\left(\omega\right)$ plots for the scalar particles of the BBHSQ. The graph is governed by Eq. \eqref{gb3} with $M=1$.} \label{Myfig4}
\end{figure}

We also investigated the same approach for regular Bardeen BH surrounded by quintessence (BBHSQ) in \cite{mtd8}, in which the scalar GFs are given by
\begin{equation}
\sigma_{l} \left( \omega \right) \geq \sec h^{2}\left[\frac{1}{2\omega}\left(\frac{\lambda}{r_{h}}+\frac{c\left(3\omega_{q}+1\right)}{\left(3\omega_{q}+2\right)r_{h}^{3\omega_{q}+2}}-\frac{2M}{\beta^{2}}+\frac{2M}{\beta^{2}\sqrt{1+\frac{\beta^{2}}{r_{h}^{2}}}}+\frac{2M}{r_{2}\left(1+\frac{\beta^{2}}{r_{h}^2}\right)^{3/2}}\right)\right],  \label{gb3}
\end{equation}%
where $\omega_{q}$ denotes the parameter of state, $c$ represents a positive normalization factor, and $r_{h}$ stands for the event horizon. We depicted the GFs of the BBHSQ in Fig. \eqref{Myfig4}. Meanwhile, the state parameter $\omega_{q}$ and event horizon are related with each other \cite{mtd8}. It is clear from Fig. \eqref{Myfig4} that GF characteristics strongly depend on the state parameter $\omega_{q}$. Furthermore, according to Fig. \eqref{Myfig4}, very similar emissions occur near the critical $\omega_{q}$ values (-$\frac{1}{3}$ and -$1$). 

In addition, the above methodology can also be used in the slowly rotating KlBH \eqref{W1}. Thus, one finds out \cite{Kanzi:2022vhp}%
\begin{equation}
\sigma_{l}\left(\omega\right)\geq\sec h^{2}\left\{ \frac{1}{2\omega} \int_{r_{h}}^{\infty}\left(\frac{2M}{r^3\sqrt{1+L}}+\frac{\lambda\sqrt{1+L}}{r^2}+\frac{4Mma\sqrt{1+L}}{r^2\left(r-2M\right)}\right)\right\}, \label{D24}
\end{equation}
which yields
\begin{equation}
\sigma_{l}\left(\omega\right)\geq\sec h^{2}\left\{ \frac{1}{2\omega} \left(\frac{M}{r_{h}^3\sqrt{1+L}}+\frac{\lambda\sqrt{1+L}}{r_{h}}+4Mma\sqrt{1+L}\left(\frac{1}{2r_{h}^2}+\frac{M}{4r_{h}^4}\right)\right)\right\}.
\label{D25}
\end{equation}

\begin{figure}[h]
\centering\includegraphics[scale=0.5]{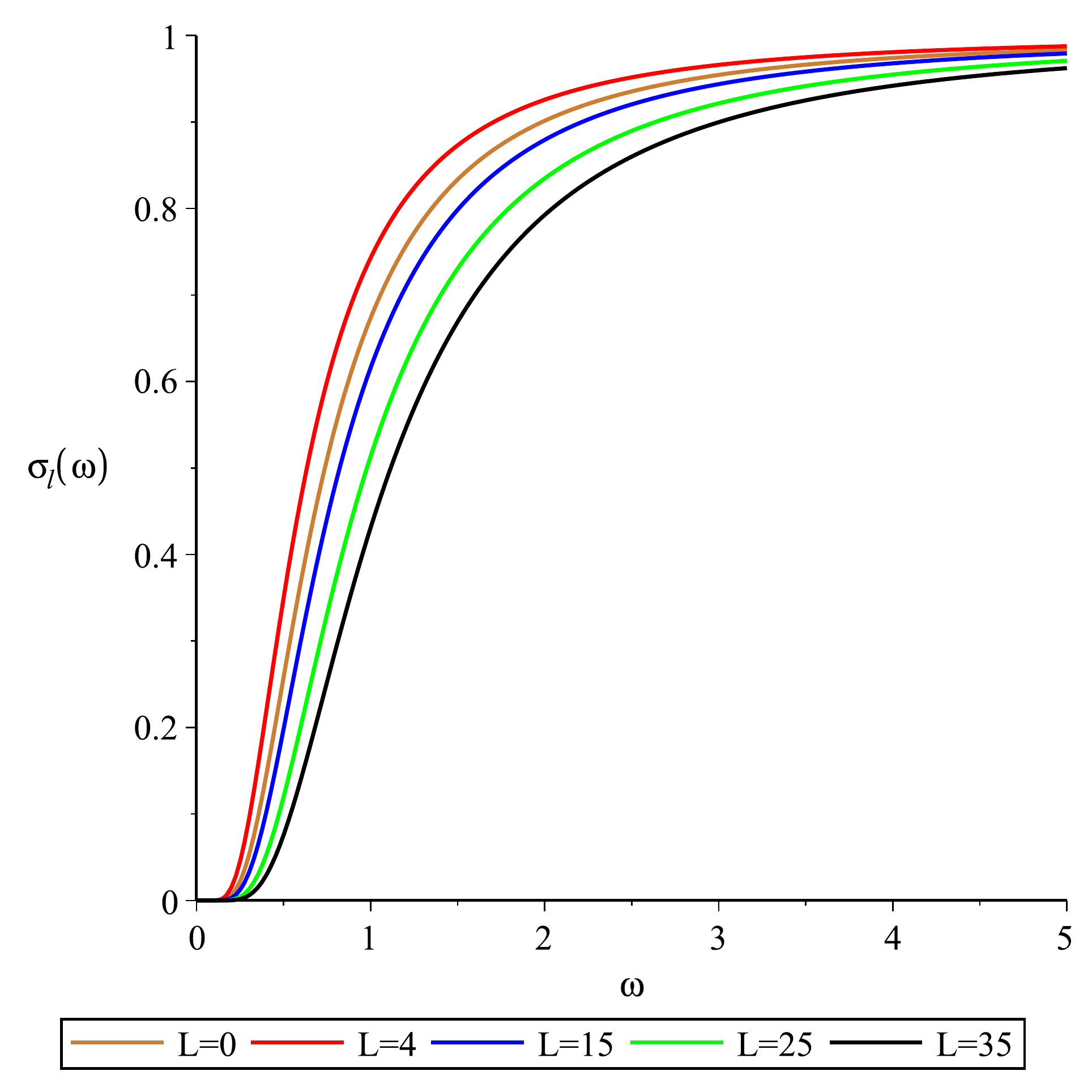}\caption{$\sigma_{l}\left(  \omega\right)  $ plots for the scalar particles of the KlBH. The graph is govermed by Eq. \eqref{D25}. The physical quantities are chosen as $ a=0.1, M=m=r=1$ and $l=0$.} \label{myfig4}
\end{figure}

In Fig. \ref{myfig4}, the obtained bosonic GFs of the
KlBH are demonstrated. As it is shown in the plots, $\sigma_{\ell}$ clearly drops with the growth in the LSB parameter. Namely, the LSB parameter acts as a fortifier for GFs of the spin-$0$ particles.\\
However, it is worth noting that employing a single semi-analytical method in combination with certain other methods can occasionally be necessary to complement and enrich the GFs. Therefore, we use supplementary methods like series, Miller-Good transformation, and rigorous bound to determine and improve the accuracy of the GFs \cite{Rincon:2020cos,mtd9,gf2,Boonserm:2008dk}. In 
study \cite{mtd9}, we concentrate on the Miller–Good transformation approach \cite{Boonserm:2008dk} to obtain the GF of the SBHSQ. This method establishes a generic bound on the probability of quantum transmission. Namely, with a particular transformation employed, one gets the 1-dimensional Schr\"{o}dinger equation in such a way that $V_{eff}$ is being modified to obtain an appropriate transmission probability for the Hawking quanta.\\
Substituting $U\left(r\right)=\frac{\psi\left(r\right)}{\sqrt{f}}$ into  the Schr\"{o}dinger equation \eqref{31}, we obtain

\begin{equation}
\frac{d^{2}\psi\left(r\right)}{dr^{2}}+k^{2}\left(r\right)\psi\left(r\right)=0,  \label{MG1}
\end{equation}%
where
\begin{equation}
k^{2}\left(r\right)=\left[\frac{\omega^{2}}{f^{2}}-\frac{f^{\prime}}{rf}-\frac{\lambda}{r^{2}f}-\frac{f^{\prime\prime}}{2f}+\frac{f^{\prime 2}}{4f^{2}}\right].\label{MG2}
\end{equation}%
Further, we set \cite{mtd9}
\begin{equation}
\psi\left(r\right)=\frac{1}{\sqrt{\rho^{\prime}\left(r\right)}}\Psi\left[\rho\left(r\right)\right],\label{MG3}
\end{equation}%
where $\rho$ denotes a new radial function and $\rho^{\prime}=\frac{d\rho\left(r\right)}{dr}\neq 0$.
Without loss of generality, it is possible to declare $\frac{d\rho\left(r\right)}{dr}>0$ and thus $\frac{dr}{d\rho\left(r\right)}>0$. The derivatives of the wave function \eqref{MG3} with respect to $r$ result in

\begin{equation}
\psi^{\prime}\left(r\right)=\sqrt{\rho^{\prime}\left(r\right)}\Psi_{\rho}\left(\rho\left(r\right)\right)-\frac{1}{2}\frac{\rho^{\prime\prime}\left(r\right)}{\rho^{\prime}\left(r\right)\sqrt{\rho^{\prime}\left(r\right)}}\Psi\left(\rho\left(r\right)\right),\label{MG4}
\end{equation}%
and
\begin{equation}
\psi^{\prime\prime}\left(r\right)=\rho^{\prime}\left(r\right)\sqrt{\rho^{\prime}\left(r\right)}\Psi_{\rho\rho}\left(\rho\left(r\right)\right)-\frac{1}{2}\frac{\rho^{\prime\prime\prime}\left(r\right)}{\rho^{\prime}\left(r\right)\sqrt{\rho^{\prime}\left(r\right)}}\Psi\left(\rho\left(r\right)\right)+\frac{3\rho^{\prime\prime 2}\left(r\right)}{4\rho^{\prime 2}\left(r\right)\sqrt{\rho^{\prime}\left(r\right)}}\Psi\left(\rho\left(r\right)\right),\label{MG5}
\end{equation}%
where $\Psi_{\rho}$ denotes the derivative with respect to $\rho$. Thus, the wave equation \eqref{MG1} recasts in

\begin{equation}
\Psi_{\rho\rho}\left(\rho\left(r\right)\right)+\left[\frac{k^{2}}{\rho^{\prime 2}\left(r\right)}+\frac{3\rho^{\prime\prime 2}\left(r\right)}{4\rho^{\prime 4}\left(r\right)}-\frac{\rho^{\prime\prime\prime}\left(r\right)}{2\rho^{\prime 3}\left(r\right)}\right]\Psi\left(\rho\left(r\right)\right)=0,\label{MG6}
\end{equation}%
\noindent
which can be simplified to the following form by defining a new effective potential $K^2$:
\begin{equation}
\Psi_{\rho\rho}\left(\rho\left(r\right)\right)+K^{2}\Psi\left(\rho\left(r\right)\right)=0.\label{MG7}
\end{equation}
\noindent
Thus, the Schr\"{o}dinger equation \eqref{MG1} which was previously represented in terms of $\psi\left(r\right)$ and $k\left(r\right)$ is now turned into anew 1-dimensional wave equation in terms of $\rho\left(r\right)$ and $K(\rho\left(r\right))$. Using the so-called Schwarzian derivative:

\begin{equation}
\sqrt{\rho^{\prime}\left(r\right)}\left(\frac{1}{\sqrt{\rho^{\prime}\left(r\right)}}\right)^{\prime\prime}=-\frac{\rho^{\prime\prime\prime}}{2\rho^{\prime}\left(r\right)}+\frac{3\rho^{\prime\prime 2}\left(r\right)}{4\rho^{\prime 2}\left(r\right)},\label{MG8}
\end{equation}
\noindent
$K^{2}$ can be rewritten as

\begin{equation}
K^{2}=\frac{1}{\rho^{\prime 2}\left(r\right)}\left[\frac{\omega^{2}}{f^{2}}-\frac{f^{\prime}}{rf}-\frac{\lambda}{r^{2}f}-\frac{f^{\prime\prime}}{2f}+\frac{f^{\prime 2}}{4f^{2}}+\sqrt{\rho^{\prime}\left(r\right)}\left(\frac{1}{\sqrt{\rho^{\prime}\left(r\right)}}\right)^{\prime\prime}\right].\label{MG9}
\end{equation}%
In order to have positive $\rho^{\prime}\left(r\right)$, one can set another parameter as $j\left(r\right)=\rho^{\prime}\left(r\right)$ where $j\left(r\right)>0$. Furthermore, by setting $j\left(r\right)=J(r)^{-2}$ with $J\left(r\right)>0$, one can obtain

\begin{equation}
K^{2}=J^{4}\left(r\right)\left[\frac{\omega^{2}}{f^2}-\frac{f^{\prime}}{rf}-\frac{\lambda}{r^{2}f}-\frac{f^{\prime\prime}}{2f}+\frac{f^{\prime 2}}{4f^{2}}+\frac{J^{\prime\prime}}{J}\right].\label{MG10}
\end{equation}%
 
\begin{figure}[h]
\centering\includegraphics[scale=0.5]{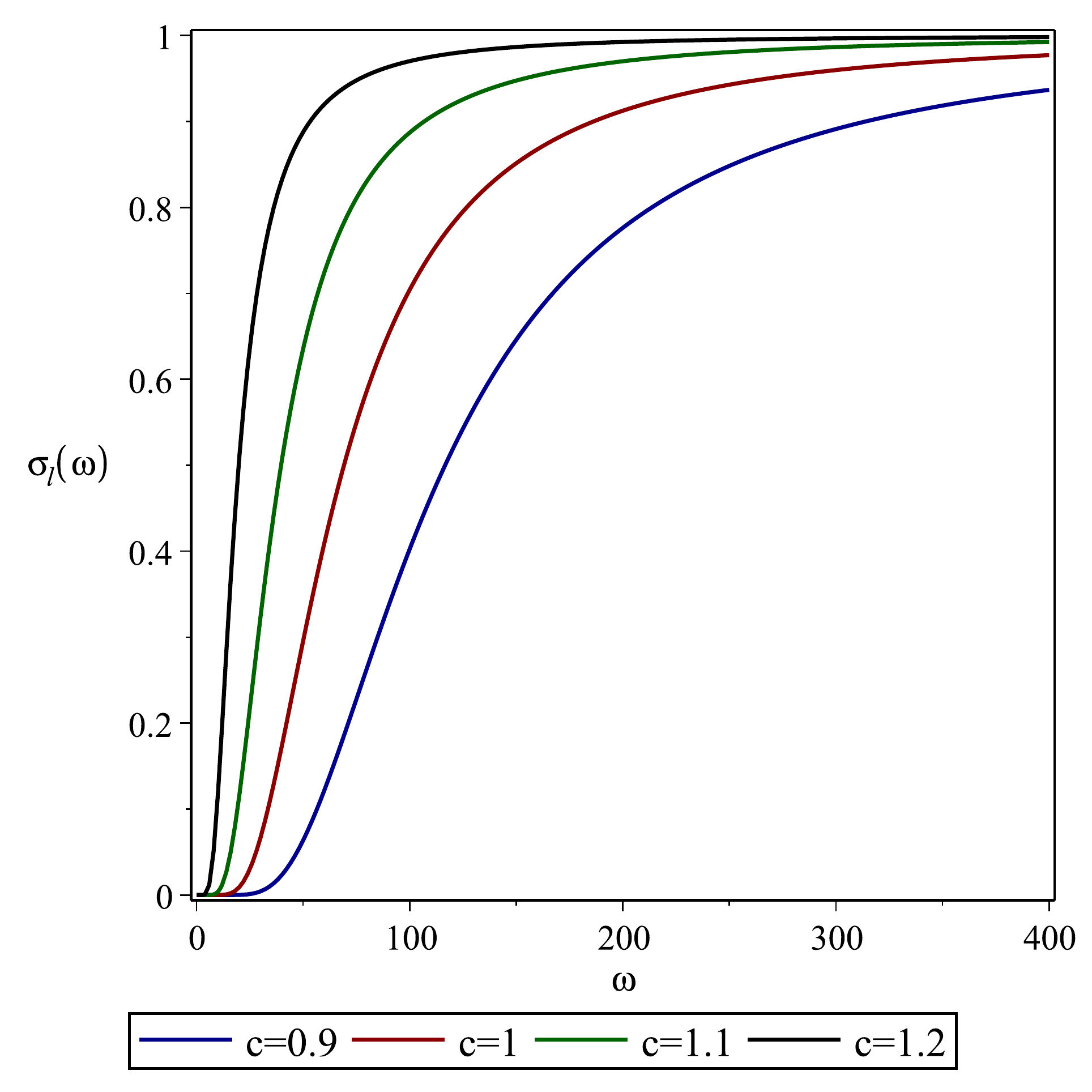}\caption{$T=\sigma_{l}\left(  \omega\right)$ plots for the scalar particles of the SBHSQ. The graph is governed by Eq. \eqref{MG15}. The physical quantities are chosen as $M=l=1$.} \label{myfig8}
\end{figure}

For $J = 1$, it is clear that $K^{2}\left(\rho\right)= k^{2}\left(r\right)$. As a result, the transmission amplitudes of both potentials are the same, and hence the transmission probability is the same. The second option is to discover a relationship between two parameters $J$ and $f$ that results in distinct potentials and hence the different transmission probabilities. To this end, Eq. \eqref{is8} is redefined as (see \cite{mtd9} and references therein)

\begin{equation}
T\geq \sec h^{2}\left( \int_{-\infty
}^{+\infty }\Tilde{\wp} d\rho\right) ,  \label{MG11}
\end{equation}
where $\Tilde{\wp}$ corresponds to a new function having new transformation parameters and $d\rho=\rho^{\prime}dr=jdr$. Using $h_{\rho}=\frac{dh}{dr}\frac{dr}{d\rho}$ in Eq. \eqref{MG11}, we get

\begin{equation}
T\geq \sec h^{2}\left( \int_{-\infty
}^{+\infty }\frac{1}{2h}\sqrt{\left(h^{\prime}\right)^{2}+\left[\frac{1}{j}\left(\frac{\omega^{2}}{f^{2}}-\frac{f^{\prime}}{rf}-\frac{\lambda}{r^{2}f}-\frac{f^{\prime\prime}}{2f}+\frac{f^{\prime 2}}{4f^{2}}-\frac{j^{\prime\prime}}{2j}+\frac{3j^{\prime 2}}{4j^{2}}\right)-jh^{2}\right]^{2}}d\rho\right).  \label{MG12}
\end{equation}
\noindent
We now have the initial form of the enhanced bound with the constraint of $h (r ) > 0$; so that
$j (r ) > 0$, as well. On the other hand, the bound can be improved by transforming $j$ to $J$ as $d\rho=\rho^{\prime}dr=J^{-2}dr$. When $J_{\pm\infty}\neq 1$ is considered with $h\left(+\infty\right)=h\left(-\infty\right)=\omega$, and hence $h^{\prime}=0$, we can write Eq. \eqref{MG12} as

\begin{equation}
T\geq \sec h^{2}\left( \int_{-\infty
}^{+\infty }\frac{1}{2\omega}{\left[J^{2}\left(\frac{\omega^{2}}{f^{2}}-\frac{f^{\prime}}{rf}-\frac{\lambda}{r^{2}f}+\frac{f^{\prime 2}}{4f^{2}}+\frac{J^{\prime\prime}}{J}\right)-\frac{\omega^{2}}{J^{2}}\right]}dr\right).  \label{MG13}
\end{equation}
\noindent
Moreover, by tuning $J$ as being $J^{2}=f$, we get a rather simpler integral and thus the transmission probability of the SBHSQ is obtained as follows:

\begin{equation}
T\geq \sec h^{2}\left(\frac{1}{2\omega} \int_{-\infty
}^{+\infty }\left(\frac{2JJ^{\prime}}{r}-\frac{\lambda}{r^{2}}\right)dr\right),  \label{MG14}
\end{equation}
which yields \cite{mtd9}

\begin{multline}
T\geq \sec h^{2}\left(\frac{1}{2\omega}\left[\frac{2M}{3cr_{h}^{3}}+\frac{2M}{4c^{2}r_{h}^{4}}+\frac{2M\left(\frac{1}{c^2}-\frac{2M}{c}\right)}{5cr_{h^{5}}}-\frac{1}{r_{h}}-\frac{1}{2cr_{h}^{2}}-\right. \right. \\
\left.  \left.
\frac{-\frac{2M}{c}+\frac{1}{c^{2}}}{3r_{h}^{3}}+\frac{\lambda}{2cr_{h}^{2}}+\frac{\lambda}{3c^{2}r_{h}^{3}}\frac{\lambda\left(\frac{-2M}{c}+\frac{1}{c^{2}}\right)}{4cr_{h}^{4}}\right]\right).  \label{MG15}
\end{multline}
In summary, the obtained expression yields the GFs of the SBHSQ spacetime, which is obtained by the
Miller–Good transformation. The result served in Eq. \eqref{MG15} is depicted in Fig. \eqref{myfig8}.

Another efficient method for calculating GFs is to use the rigorous bound on the GFs, which was applied first by Boonserm and Visser \cite{gf2} for the Schwarzschild BH. The Regge-Wheeler equation that describes the motion of any sort of particle in a non-rotational spacetime can be used to calculate these bounds, which are valid for all frequencies. In the literature, so far, the rigorous bounds are considered for the GFs of the $4$-dimensional Reissner-Nordstr\"{o}m BH, the higher dimensional Schwarzschild-Tangherlini BH, the charged dilatonic BH \big(in $(2 + 1)$ dimensions\big), the charged dilatonic BHs, the Kerr-Newman BH, and also for the BH and black string solutions in the dRGT massive gravity (see \cite{SK33,SK34,Barman:2019vst,SK35, SK37,Kanzi:2020cyv,SK38} and references therein). \\

Now, we would like to draw the reader's attention to our yet another study \cite{Kanzi:2020cyv} as an example of the use of rigorous bound on the GFs. In the framework of massive gravity theory coupled with non-linear electrodynamics ($nem$), the metric of the corresponding BH is given by \cite{Kanzi:2020cyv}
\begin{equation}
ds^{2}=-f\left(  r\right)  dt^{2}+\frac{dr^{2}}{f\left(  r\right)  }+
r^{2}\left(  d\theta^{2}+\sin^{2}\theta d\varphi^{2}\right),
\label{H13}%
\end{equation}
which represents two metric function solutions of the charged dRGT BHs in the $nem$. The first metric function solution reads \cite{Kanzi:2020cyv}

\begin{multline}
f\left(  r\right)  =1-\frac{2M}{r}+\frac{8r^{2}}{3b^{2}}\left(  1+\frac
{qb}{r^{2}}\right)  ^{3/2}-\frac{4q}{b}\left(  1+\frac{2r^{2}}{3qb}\right)
+\\
m_{g}^{2}\left(  \left(  1+\alpha+\beta\right)  r^{2}-\left(  1+2\alpha
+3\beta\right)  h_{0}r+h_{0}^{2}\left(  \alpha+3\beta\right)  \right)  \label{mym1},
\end{multline} 
\noindent
where $M$ is a constant of integration (mass parameter), $b$ is a positive parameter, $h_{0}$ is a constant parameter, $\alpha=1+3\alpha_{3}$, and $\beta=\alpha_{3}+4\alpha_{4}$. The second metric function solution is given by \cite{Kanzi:2020cyv}

\begin{equation}
f\left(  r\right)  =1-\frac{2M}{r}-\left(  m_{g}^{2}\frac{\left(  1+\alpha
^{2}+\alpha-3\beta\right)  }{3\left(  \alpha+3\beta\right)  }+\frac{8}{3b^{2}%
}\right)  r^{2}-\\
\frac{4q}{b}+\frac{8r^{2}}{3b^{2}}\left(  1+\frac{qb}{r^{2}}\right)  ^{3/2}. \label{mym2}
\end{equation} 
\noindent
To investigate the thermal radiation of the charged dRGT BHs in the $nem$, we consider the massless and charged KG equation:

\begin{equation}
\frac{1}{\sqrt{-g}}D_{\mu}\left[  \sqrt{-g}g^{\mu\nu}D_{\nu}\right]  \Psi=0,
\label{S27}%
\end{equation}
\noindent
where%
\begin{equation}
D_{\mu}=\partial_{\mu}-iqA_{\mu}, \label{S28}%
\end{equation}
and the electromagnetic potential is defined as%
\begin{equation}
A_{t}=-\frac{2}{b}\left(  r-\sqrt{r^{2}+qb}\right)  ,\text{ }A_{r}=A_{\theta
}=A_{\varphi}=0. \label{S29}%
\end{equation}
\noindent
Let us consider the line-element (\ref{H13}) of the charged dRGT BH in the KG equation (\ref{S27}) with the following ansatz:
\begin{equation}
\Psi\left(  t,r,\Omega\right)  =e^{i\omega t}\frac{\varphi\left(  r\right)
}{r}Y_{lm}\left(  \Omega\right)  , \label{S31}%
\end{equation}
in which $e^{i\omega t}$ indicates the oscillating function and $Y_{lm}\left(
\Omega\right)  $ denotes the spherical harmonics.
The tortoise coordinate can be easily obtained via the expression of $\frac{dr_{\ast}}{dr}=\frac{1}{f\left(
r\right)  },$ which assists us to permute the radial wave equation that can be cast in 1-dimensional Schr\"{o}dinger equation:%
\begin{equation}
\frac{d^{2}\varphi\left(  r\right)  }{dr_{\ast}^{2}}+\left[  \omega
^{2}-V_{eff}\right]  \varphi\left(  r\right)  =0, \label{S35}%
\end{equation}
where%

\begin{equation}
V_{eff}=2\omega qA_{t}-q^{2}A_{t}^{2}+\frac{\lambda f}{r^{2}}+\frac{f}{r}f^{\prime}
, \label{SS36}%
\end{equation}
where $f^{\prime}=\frac{df}{dr}$. Hereafter, we shall name the solutions by indices of $1$ and $2$. 
The $V_{eff}$ of the first solution is found as
\begin{multline}
V_{eff\left(  1\right)  }=2\omega q(-\frac{2}{b}\left(  r-\sqrt{r^{2}%
+qb}\right)  )-(-\frac{2q}{b}\left(  r-\sqrt{r^{2}+qb}\right)  )^{2}%
+\\
\frac{\lambda}{r^{2}}\left(  1-\frac{2M}{r}+\frac{8r^{2}}{3b^{2}}\left(
1+\frac{qb}{r}\right)  ^{3/2}-\frac{4q}{b}\left(  1+\frac{2r^{2}}{3qb}\right)
+\left(  Ar^{2}-Br+C\right)  \right)  +\\
\frac{1}{r}\left(  1-\frac{2M}{r}+\frac{8r^{2}}{3b^{2}}\left(  1+\frac{qb}%
{r}\right)  ^{3/2}-\frac{4q}{b}\left(  1+\frac{2r^{2}}{3qb}\right)  +\left(
Ar^{2}-Br+C\right)  \right)  \times\\
\left(  \frac{2M}{r^{2}}+\sqrt{1+\frac{qb}{r}}\left(  \frac{16r}{3b^{2}}%
+\frac{4q}{3b}\right)  -\frac{16qr}{3qb^{2}}+2Ar-B\right), \label{SW36}
\end{multline}
\noindent
where 
\begin{align}
A  &  =m_{g}^{2}\left(  1+\alpha+\beta\right),\nonumber\\
B  &  =m_{g}^{2}\left(  1+2\alpha+3\beta\right)  h_{0},\nonumber\\
C  &  =m_{g}^{2}\left(  \alpha+3\beta\right)  h_{0}^{2}.\label{SSW38}
\end{align}
\noindent
The $V_{eff}$ of the second solution can be found out to be as

\begin{multline}
V_{eff\left(  2\right)  }=2\omega q(-\frac{2}{b}\left(  r-\sqrt{r^{2}%
+qb}\right)  )-(-\frac{2q}{b}\left(  r-\sqrt{r^{2}+qb}\right)  )^{2}%
+\\
\frac{\lambda}{r^{2}}\left(  1-\frac{2M}{r}-\left(  D+\frac{8}{3b^{2}}\right)
r^{2}-\frac{4q}{b}+\frac{8r^{2}}{3b^{2}}\left(  1+\frac{qb}{r^{2}}\right)
^{3/2}\right)  +\\
\frac{1}{r}\left(  1-\frac{2M}{r}-\left(  D+\frac{8}{3b^{2}}\right)
r^{2}-\frac{4q}{b}+\frac{8r^{2}}{3b^{2}}\left(  1+\frac{qb}{r^{2}}\right)
^{3/2}\right)  \times\\
\left(  \frac{2M}{r^{2}}-2r\left(  D+\frac{8}{3b^{2}}\right)  +\sqrt
{1+\frac{qb}{r^{2}}}\left(  \frac{16r}{3b^{2}}-\frac{8q}{3br} \right)  \right), \label{SW37}
\end{multline}
\noindent
where%
\begin{equation}
D=m_{g}^{2}\frac{\left(  1+\alpha^{2}+\alpha-3\beta\right)  }{3\left(
\alpha+3\beta\right)  }.\label{SK2}%
\end{equation}

\begin{figure}[h]
\centering
\includegraphics[scale=0.6]{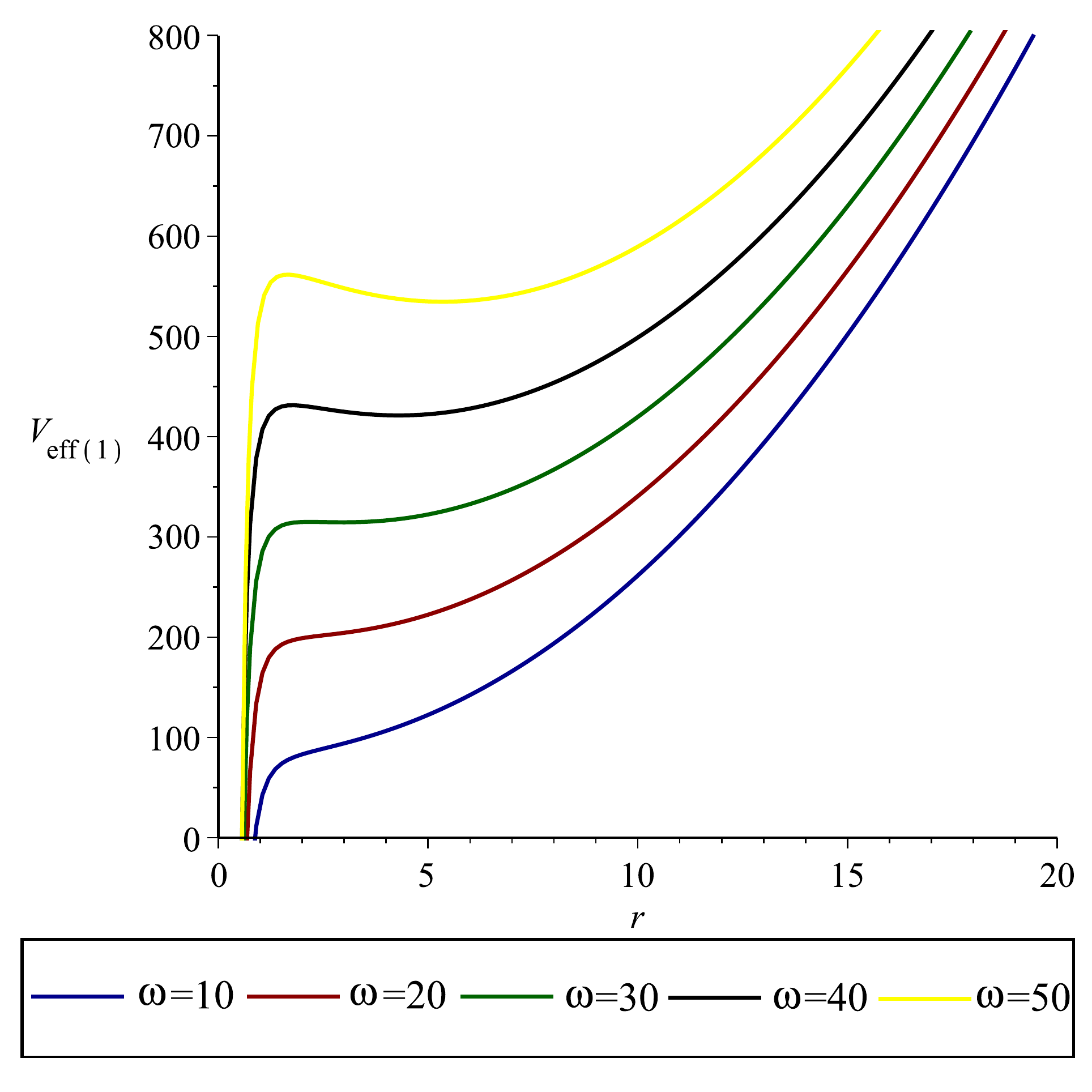}\caption{$V_{eff \left(1\right)}$ \eqref{SW36} plots for the scalar particles
of the first solution of the charged dRGT BHs. The graphics are governed by Eq.
(\ref{SW36}). The physical quantities are chosen as $M=1, q=8, b=50,$
and $\lambda=0$.}%
\label{myfigS1}%
\end{figure}

\begin{figure}[h]
\centering
\includegraphics[scale=0.6]{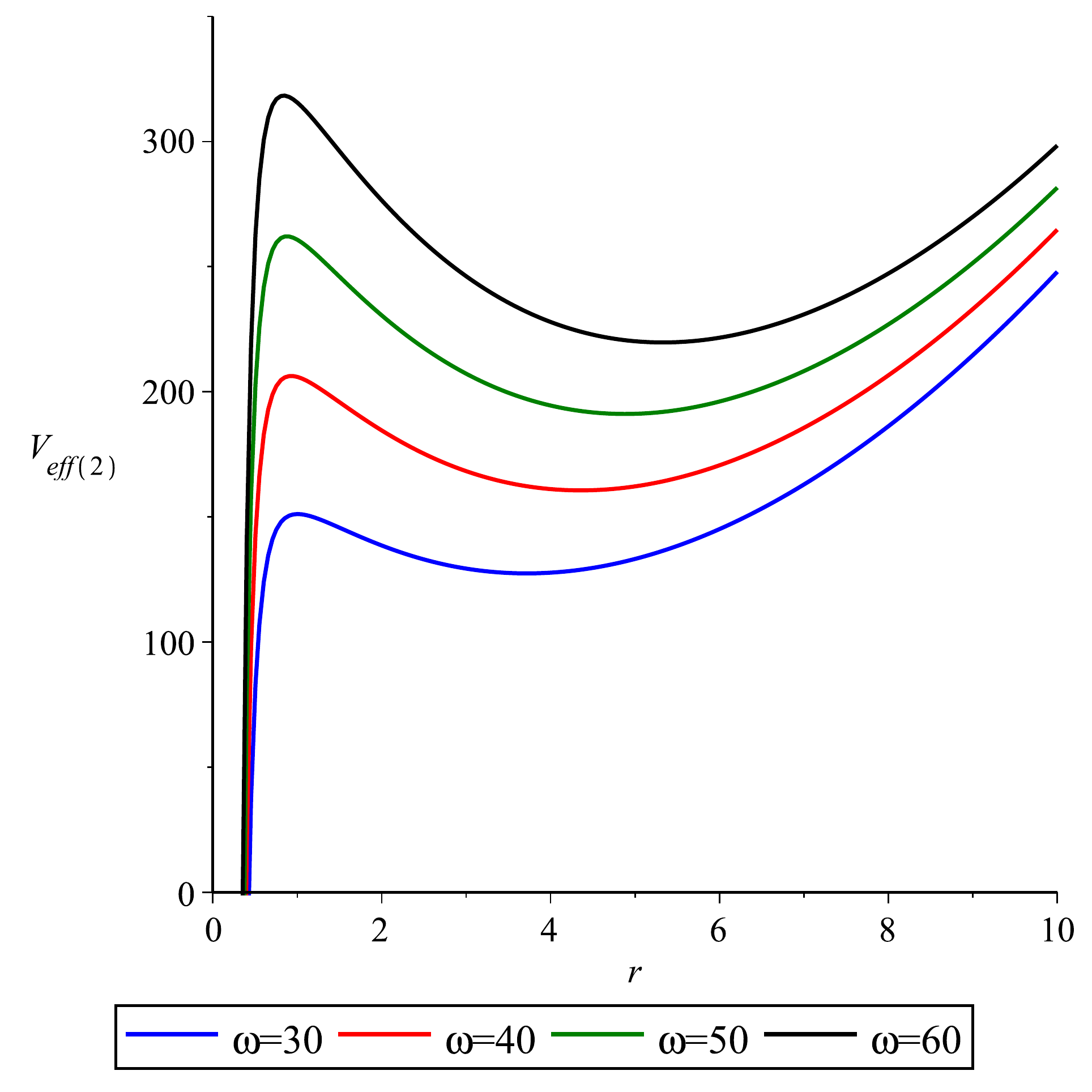}\caption{$V_{eff \left(2\right)}$ \eqref{SW37} plots for the scalar particles
of the second solution of the charged dRGT BHs. The physical quantities are chosen as $M=1, q=3, b=10,$
and $\lambda=0$.}%
\label{myfigS2}%
\end{figure}

Figures (\ref{myfigS1}) and (\ref{myfigS2}) represent the behaviors of the effective potentials for both metrics \textit{i.e.,} Eqs. (\ref{mym1}) and (\ref{mym2}) are plotted for different values of $\omega$. Without loss of generality, the quantities $B$ and $C$ can be set to zero, and $A=-1$. As can be seen from the figures, the effective potentials disappear at the horizon and make peak right after the horizon. By rising the frequency, the potential barriers' peaks increase as well. Moreover, by growing the scalar waves energy, the value of the potential barrier near the event horizon rises as well. This also gives rise to the caging of waves. Since the essential contribution of the transmission amplitude originates from the $l=0$ mode
($s$-waves \cite{chandra}), it is sufficient to qualitatively investigate the potential (\ref{SW36}) for the $s$-waves. Normally, the behavior \eqref{SW37} of the potential in the second metric function is much smoother, in the same period, in comparison with the first one, which indicates the role of constant parameter $b$ is much more notable.

Now, as yet another illustrative example, we utilize the rigorous bounds constrained to the $4$-dimensional BH for investigating the GF in\ dRGT massive gravity background which is
coupled with $nem$. For this purpose, we shall employ the
equation of the GF, i.e., Eq. \eqref{is8}. Considering the $V_{eff}$
of the first solution (\ref{SW36}), one can obtain the expression of GF \eqref{MG11} with the following expression 

\begin{multline}
T_{1}\geq\sec h^{2}\frac{1}{2\omega}\left\{  \int_{r_{h}}^{R_{h}}\left(
\frac{\lambda}{r^{2}}+\frac{2M}{r^{3}}-\frac{2q^{2}}{r^{4}}+2A-\frac{B}%
{r}\right)  dr+\right. \label{S44}\\
\left.  \int_{r_{h}}^{R_{h}}\frac{2\omega q^{2}r}{Ar^{4}-Br^{3}+\left(
1+c\right)  r^{2}-2Mr+q^{2}}dr-\right. \\
\left.  \int_{r_{h}}^{R_{h}}\frac{\omega q^{3}b}{2\left(  Ar^{5}%
-Br^{4}+\left(  1+c\right)  r^{3}-2Mr^{2}+q^{2}r\right)  }dr\right. \\
\left.  -\int_{r_{h}}^{R_{h}}\frac{q^{4}}{Ar^{4}-Br^{3}+\left(  1+c\right)
r^{2}-2Mr+q^{2}}\right\}  ,
\end{multline}
By applying the Taylor expansion, we can derive the GF as
\begin{multline}
T_{1}\geq\sec h^{2}\frac{1}{2\omega}\left\{  -\frac{\lambda}{R_{h}-r_{h}%
}-\frac{M}{R_{h}^{2}-r_{h}^{2}}+\frac{2q^{2}}{3\left(  R_{h}^{3}-r_{h}%
^{3}\right)  }\right. \\
\left.  -(B+\frac{1}{2}\omega qb)\ln\left(  R_{h}-r_{h}\right)  +W_{1}\left(
R_{h}-r_{h}\right)  +X_{1}\left(  R_{h}^{2}-r_{h}^{2}\right)  \right. \\
\left.  +Y_{1}\left(  R_{h}^{3}-r_{h}^{3}\right)  +Z_{1}\left(  R_{h}%
^{4}-r_{h}^{4}\right)  -P_{1}\left(  R^{5}-r^{5}\right)  \right\} \label{SW38}  ,
\end{multline}
where%
\begin{equation}
W_{1}=2A-\frac{\omega bM}{q}-q^{2}, \label{S46}%
\end{equation}

\begin{equation}
X_{1}=(\omega+\frac{\omega b}{4q^{2}}\left(  q\left(  1+c\right)
-\frac{4M^{2}}{q}\right)  -M), \label{S47}%
\end{equation}

\begin{equation}
Y_{1}=-\frac{\omega b}{6q^{2}}\left(  qB-\frac{4M\left(  q^{2}\left(
1+c\right)  -2M^{2}\right)  }{q^{3}}\right)  +\frac{q^{2}\left(  1+c\right)
-4M^{2}+4\omega M}{3q^{2}}, \label{SS48}%
\end{equation}
and%

\begin{multline}
Z_{1}=\left.  \frac{\omega}{2q^{2}}\left(  -\left(  1+c\right)  +\frac{4M^{2}%
}{q^{2}}\right)  \right.  -\label{z49}\\
\left.  \frac{1}{8q^{2}}\left(  -\omega qbA+\frac{\omega b\left(  1+c\right)
^{2}}{q}+\frac{4\omega bM\left(  Bq^{4}-3Mq^{2}\left(  1+c\right)
+4M^{3}\right)  }{q^{5}}\right)  \right. \\
\left.  -\frac{1}{4q^{2}}\left(  q^{2}B+\frac{4M\left(  -q^{2}\left(
1+c\right)  +2M^{2}\right)  }{q^{2}}\right)  \right.  ,
\end{multline}
and 
\begin{multline}
P_{1}=\left.  \frac{\omega b}{5q^{5}}\left(  \left(  -q^{2}\left(  1+c\right)
+6M^{2}\right)  B-2MAq^{2}\right)  \right.  -\label{S50}\\
\left.  \frac{2\omega bM(1+c)}{5q^{7}}\left(  -q^{2}\left(  1+c\right)
+2M^{2}\right)  \right.  -\\
\left.  \frac{\omega bM}{5q^{9}}\left(  -q^{4}\left(  1+c^{2}\right)
+12M^{2}q^{2}\left(  1+c\right)  -2cq^{4}-16M^{4}\right)  \right.  +\\
\left.  \frac{1}{5q^{2}}\left(  -q^{2}A+4BM+\left(  1+c\right)  ^{2}%
-\frac{4M^{2}\left(  3q^{2}\left(  1+c\right)  -4M^{2}\right)  }{q^{4}%
}\right)  \right.  -\\
\left.  \frac{2}{q^{2}}\left(  \omega B+\frac{4\omega M\left(  -q^{2}\left(
1+c\right)  +2M^{2}\right)  }{q^{4}}\right)  \right.  ,
\end{multline}
two parameters $R_{h}$/$r_{h}$ denote upper/lower rigorous bounds, respectively, which are given by%
\begin{equation}
R_{h}=\frac{2}{\left(  -2A\right)  ^{1/3}}\left[  \sqrt{\frac{2\sqrt{3}}%
{\beta}+4}\cos\left(  \frac{1}{3}\sec^{-1}\left(  -\frac{\sqrt{\frac{\sqrt{3}%
}{\beta}+2}\left(  2\sqrt{2}\beta+\sqrt{6}\right)  }{5\beta+3\sqrt{3}}\right)
\right)  -1\right]  , \label{S51}%
\end{equation}
and 
\begin{equation}
r_{h}=\frac{-2}{\left(  -2A\right)  ^{1/3}}\left[  \sqrt{\frac{2\sqrt{3}%
}{\beta}+4}\cos\left(  \frac{1}{3}\sec^{-1}\left(  -\frac{\sqrt{\frac{\sqrt
{3}}{\beta}+2}\left(  2\sqrt{2}\beta+\sqrt{6}\right)  }{5\beta+3\sqrt{3}%
}\right)  +\frac{\pi}{3}\right)  +1\right]  . \label{S52}%
\end{equation}

\begin{figure}[h]
\centering
\includegraphics[scale=0.6]{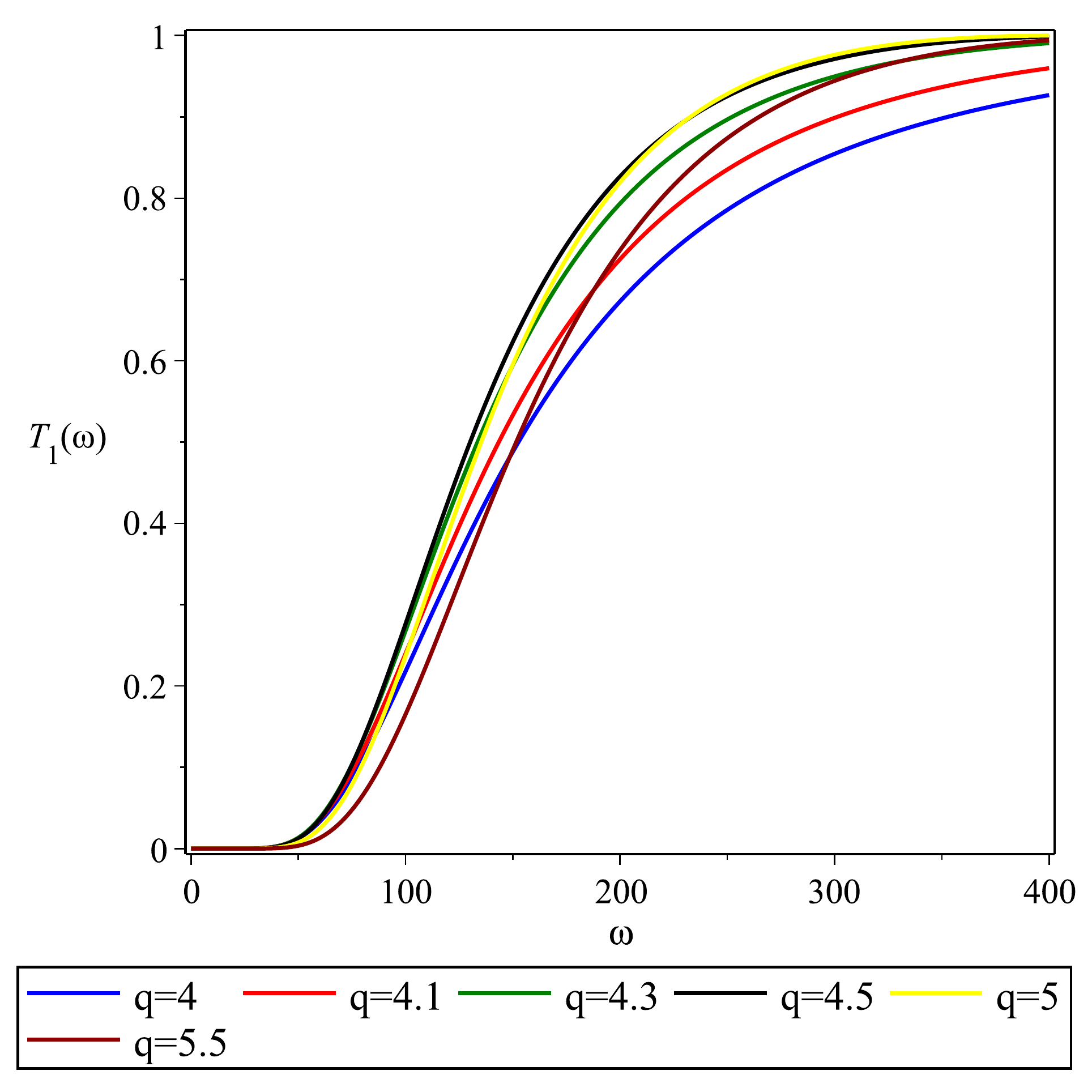}\caption{Transmission probability \eqref{SW38} plots for the metric function $f_{1}$. The physical quantities are chosen as $\lambda=0, M=1, A=-1, b=0.1,$ and $B=C=0$.}%
\label{myfigS3}%
\end{figure}

One can elucidate the results obtained by plotting the GFs (with $b=0.1$) for different charge values. The
notable point in the obtained figure, Fig. \eqref{myfigS3}, is that GF for $h_{0}=0$ acts as if the AdS/dS black string \cite{SK35}. Furthermore, the GF increases only for a limited range of small charge values, but it then reverses its behavior.

Substituting the $V_{eff}$ of second
solution, i.e., Eq. (\ref{SW37}), in Eq. (\ref{is9}) to get%
\begin{multline}
T_{2}\geq\sec h^{2}\frac{1}{2\omega}\left\{  \int_{r_{h}}^{R_{h}}\left(
\frac{\lambda}{r^{2}}+\frac{2M}{r^{3}}-\frac{2q^{2}}{r^{4}}-2D\right)
dr+\right. \\
\left.  \int_{r_{h}}^{R_{h}}\frac{2\omega q^{2}r}{-Dr^{4}+r^{2}-2Mr+q^{2}%
}dr-\int_{r_{h}}^{R_{h}}\frac{\omega q^{3}b}{2\left(  -Dr^{5}+r^{3}%
-2Mr^{2}+q^{2}r\right)  }dr\right. \\
\left.  -\int_{r_{h}}^{R_{h}}\frac{q^{4}}{-Dr^{4}+r^{2}-2Mr+q^{2}}dr\right\},
\end{multline}
by using the Taylor expansion and integration, one can obtain%
\begin{multline}
T_{2}\geq\sec h^{2}\frac{1}{2\omega}\left\{  -\frac{\lambda}{R_{h}-r_{h}%
}-\frac{M}{R_{h}^{2}-r_{h}^{2}}+\frac{2q^{2}}{3\left(  R_{h}^{3}-r_{h}%
^{3}\right)  }-\right. \label{S56}\\
\left.  \frac{1}{2}\omega qb\ln(R_{h}-r_{h})-W_{2}(R_{h}-r_{h})+X_{2}\left(
R_{h}^{2}-r_{h}^{2}\right)  +\right. \\
\left.  Y_{2}\left(  R_{h}^{3}-r_{h}^{3}\right)  +Z_{2}\left(  R_{h}^{4}%
-r_{h}^{4}\right)  +P_{2}\left(  R_{h}^{5}-r_{h}^{5}\right)  \right\}  ,
\end{multline}
\noindent
where%

\begin{equation}
W_{2}=2D+q^{2}+\frac{\omega bM}{q}, \label{S57}%
\end{equation}

\begin{equation}
X_{2}=\omega-\left(  \frac{-\omega b}{4q}+\frac{\omega bM^{2}}{q^{3}}\right)
-M, \label{S58}%
\end{equation}

\begin{equation}
Y_{2}=\frac{q^{2}-4M^{2}+4\omega M}{3q^{2}}+\frac{2\omega bM\left(
q^{2}-2M^{2}\right)  }{3q^{5}}, \label{S59}%
\end{equation}

\begin{equation}
Z_{2}=\frac{\omega}{2q^{2}}\left(  \frac{4M^{2}}{q^{2}}-1\right)
-\frac{\omega b}{8q^{2}}\left(  -qD+\frac{\left(  q^{2}-12M^{2}\right)
}{q^{3}}+\frac{16M^{4}}{q^{5}}\right)  , \label{Sd60}%
\end{equation}
and%

\begin{multline}
P_{2}=-\frac{1}{10q^{3}}\left(  2\omega bM\left(  2D+\frac{3}{q^{2}}%
-\frac{16M^{2}}{q^{4}}+\frac{16M^{4}}{q^{6}}\right)  \right)  -\label{Sd61}\\
\frac{1}{5q^{2}}\left(  1-q^{2}D+\frac{4M}{q^{4}}\left(  M\left(
4M^{2}-3q^{2}\right)  +2\omega\left(  -q^{2}+2M^{2}\right)  \right)  \right)
.
\end{multline}

\begin{figure}[h]
\centering
\includegraphics[scale=0.6]{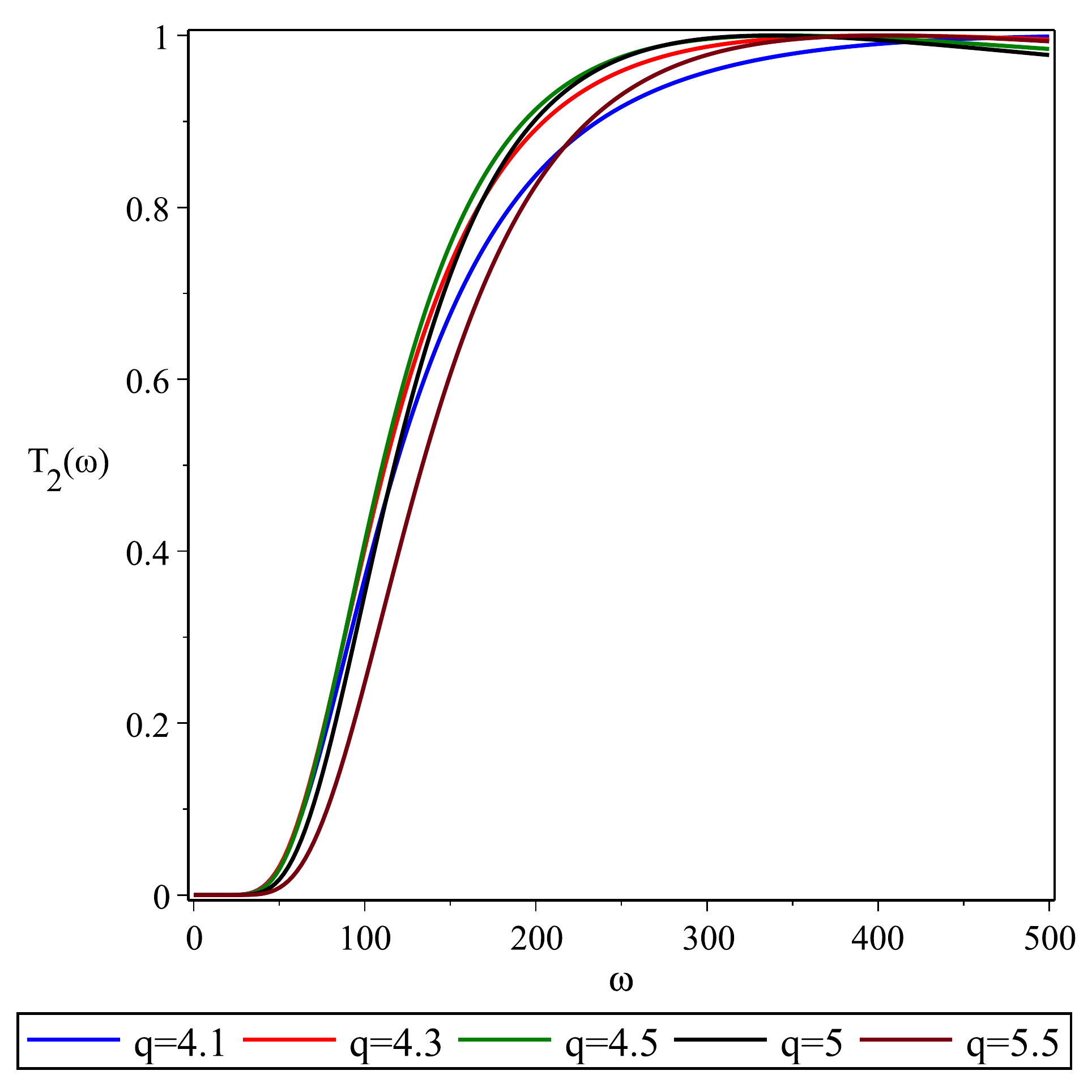}\caption{Transmission probability plots for the metric function $f_{2}$. The graph is governed by Eq.
(\ref{S56}). The physical quantities are chosen as $M=1, \lambda
=0, b=0.1,$ and $D=0.8$.}%
\label{myfigS4}%
\end{figure}

Using Eq. (\ref{S56}), the rigorous bounds on the GFs for the second solution of the charged dRGT BH in the $nem$ are plotted in Fig. \eqref{myfigS4}. For both GF solutions, the constant parameter $b$ is set to be minimal. The charge parameter's increase first aids the GF in reaching its maximum value, but later the GF begins to decline. In both solutions, the aforementioned change can be seen after $q=4.5$. As a result, we claim that in the $nem$, the charged GF cannot have a monotonous behavior. 

After this stage, it will be appropriate to serve another interesting application for the GFs, which is about the BWBHSCP \cite{Sakalli:2022swm}. Before moving on to the GF computation details, let us take a look at the BWBHSCP metric structure.  BWBHSCP has a static spherically symmetric line-element that can be described by the following metric ansatz:
\begin{equation}
ds^{2}=-e^{\mu\left(  r\right)  }dt^{2}+e^{\upsilon\left(  r\right)  }%
dr^{2}+r^{2}\left(  d\theta^{2}+\sin^{2}\theta d\varphi^{2}\right).
\label{Ssk3}%
\end{equation}
 There are two types of solutions \cite{Heydar-Fard:2007ahl}, the first solution reads

\begin{equation}
e^{\mu\left(  r\right)  }=e^{-\upsilon\left(  r\right)  }=1-\frac{B}%
{r}-\alpha^{2}r^{2}-2\alpha\beta r-\beta^{2}, \label{4}%
\end{equation}
\noindent
in which $B$, $\alpha$, and $\beta$ indicate the constants of integration. Moreover, $\alpha$ is investigated as the energy scale on the brane. By
considering $\beta=0$, the second type of solution is defined as follows
\begin{equation}
e^{\mu\left(  r\right)  }=e^{-\upsilon\left(  r\right)  }=1-\frac{B}%
{r}-\alpha^{2}r^{2}. \label{8}%
\end{equation}
Since the metric \eqref{Ssk3} represents an ordinary $4$-dimensional BH, its effective potential should be in the form of Eq. \eqref{ww2}. Even so, when considering Eq. (\ref{is8}), in both BWBHSCP background solutions the GF computations are found to be indecisive. The integral is set for the lower bound, which causes this problem. In order to overcome this issue, one can replace the applied method with the one, which was prescribed in \cite{Sakalli:2022swm,gf2}. To this end, we first set $h=\sqrt{\omega^{2}-V_{eff}}$ in Eq. (\ref{gb1}). Since there are no restricted regions in this method \cite{gf2}, one has%
\begin{equation}
\sigma_{l}(\omega)\geq\sec h^{2}\left\{  \frac{1}{2}\int_{-\infty}^{+\infty
}\left\vert \frac{h^{\prime}}{h}\right\vert dr_{\ast}\right\}  .\label{40}%
\end{equation}
\noindent
\begin{figure}[h]
\centering\includegraphics[scale=0.6]{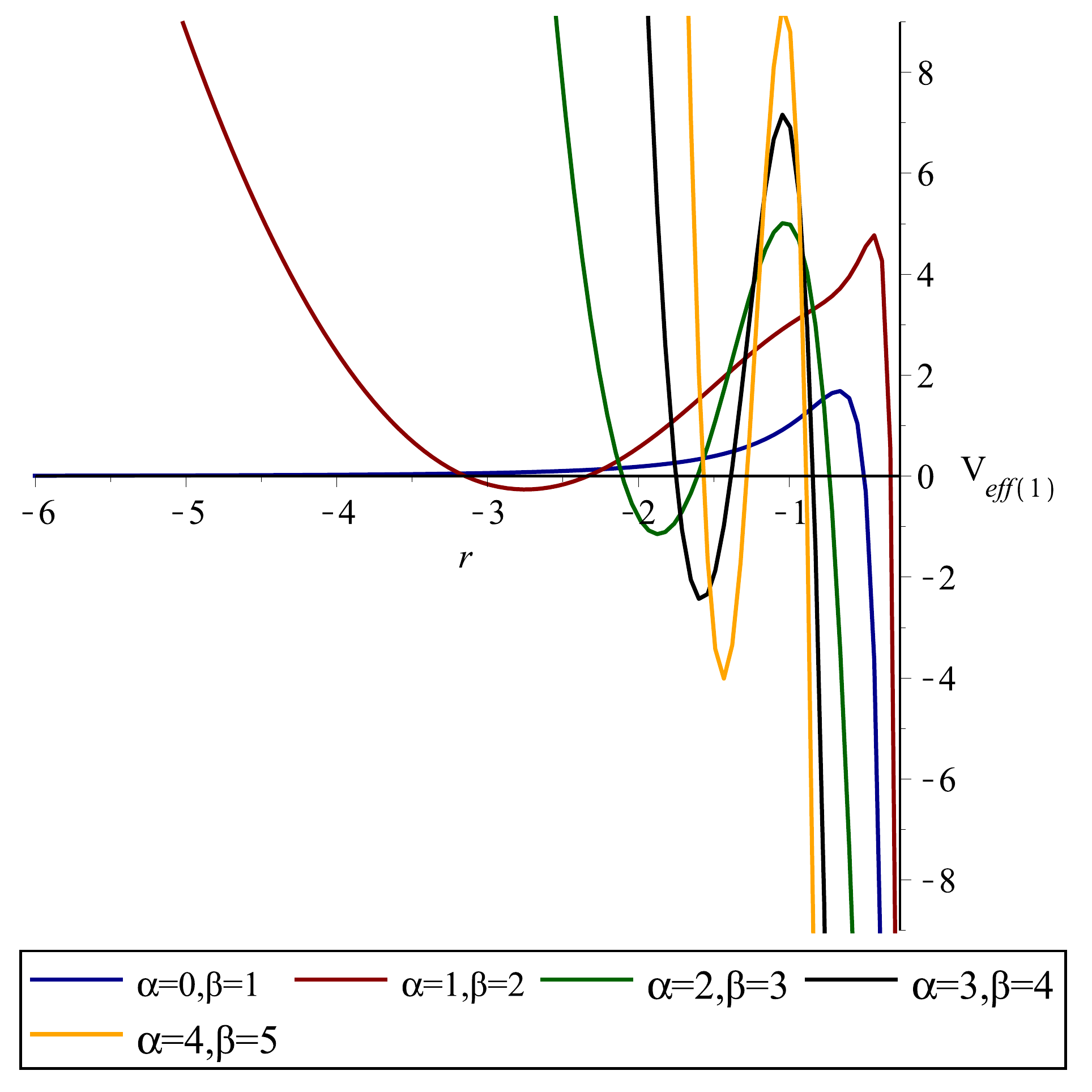}\caption{Effective potential plots for the first solution having different values of parameters $\alpha$ and $\beta$ (see Ref. \cite{Sakalli:2022swm}). The physical quantities are chosen as  $B=1$ and $\lambda=2$.} \label{figsk1}
\end{figure}

\begin{figure}[h]
\centering\includegraphics[scale=0.6]{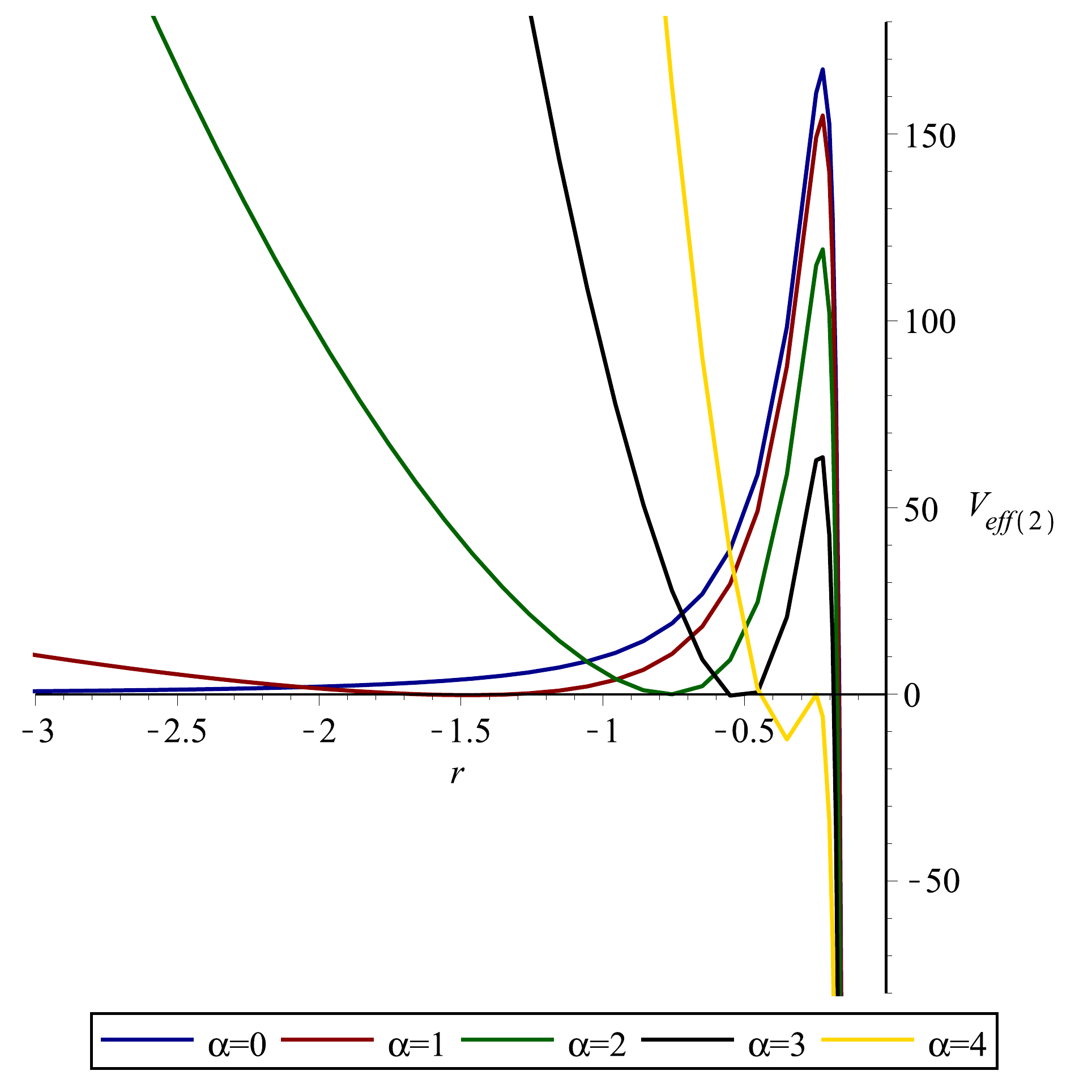}\caption{Effective potential plots for the second solution \cite{Sakalli:2022swm}. The physical quantities are chosen as $B=1$ and $\lambda=2$.} \label{figsk2}
\end{figure}

\begin{figure}[h]
\centering\includegraphics[scale=0.6]{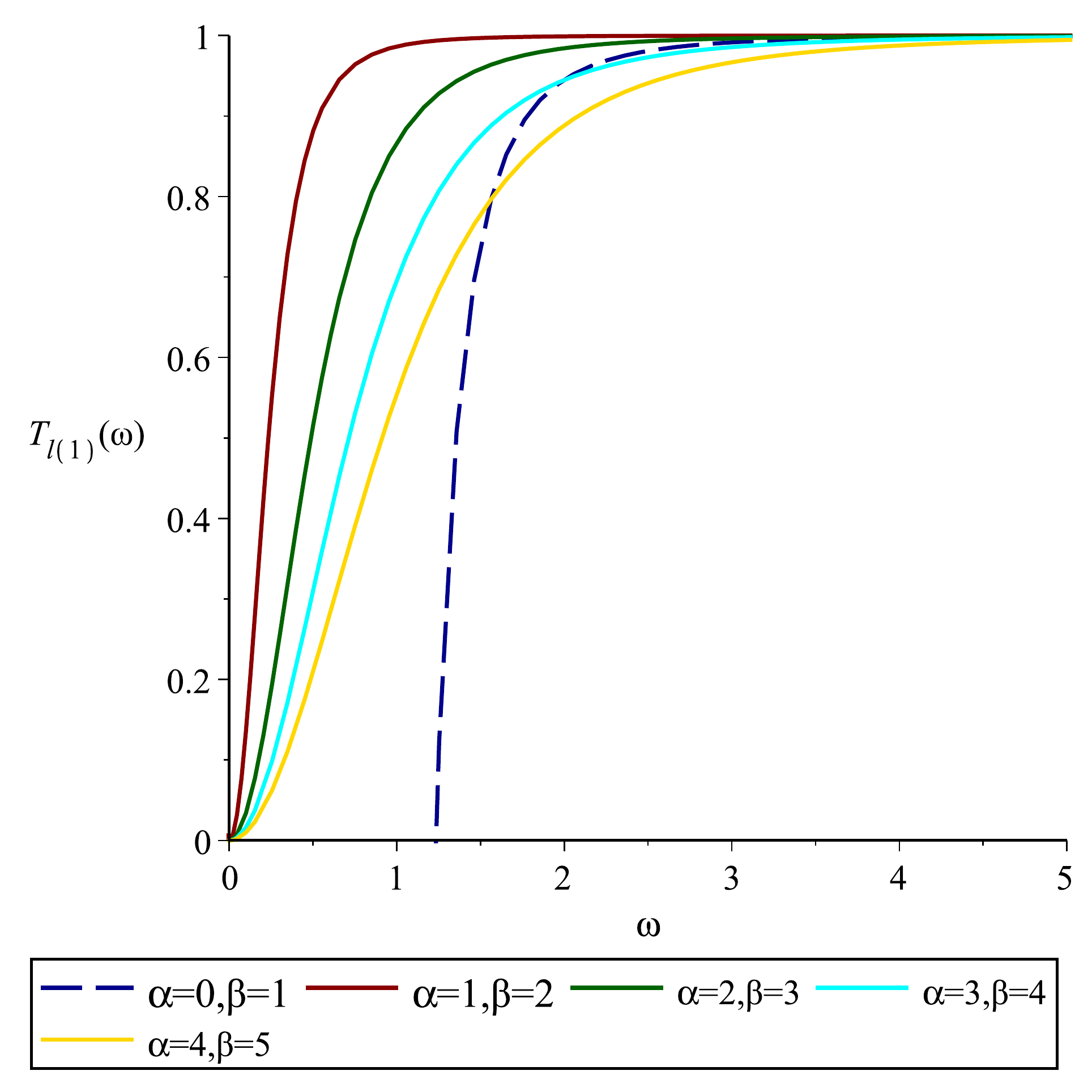}\caption{Transmission probability plots scalar particles in the first solution \cite{Sakalli:2022swm}. The physical quantities are chosen as $B=1$ and
$\lambda=2$.} \label{fig3}
\end{figure}

\begin{figure}[h]
\centering\includegraphics[scale=0.6]{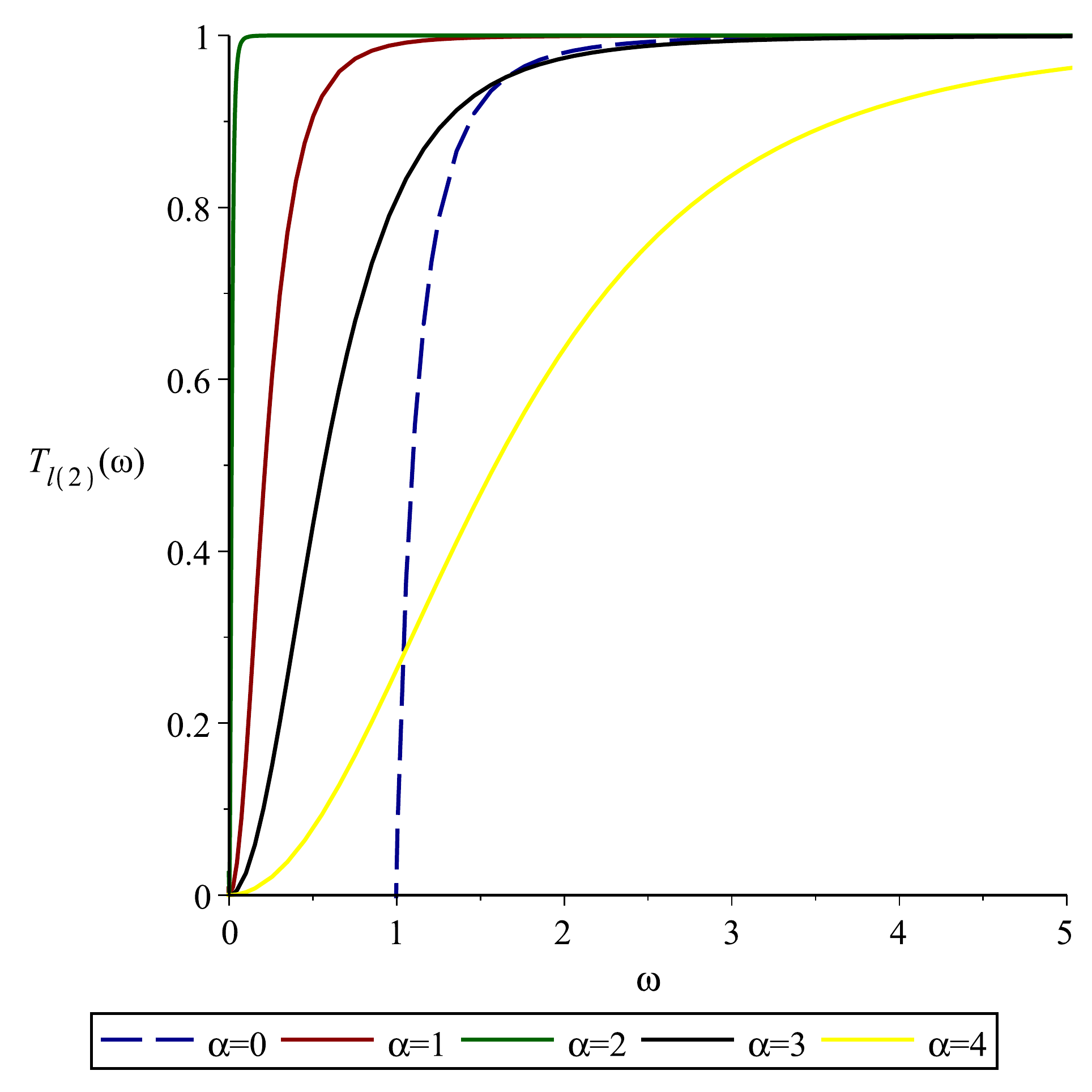}\caption{Plots of the transmission probability for the bosons in the second solution \cite{Sakalli:2022swm}. The physical quantities are chosen as $B=1$ and $\lambda=6$.} \label{fig4}
\end{figure}

Figures \eqref{figsk1} and \eqref{figsk2} represent the peaks of effective potential for first and second solutions, respectively, which are related to
$h(+\infty)=h_{peak}=\sqrt{\omega^{2}-V_{peak}}$. Thus, by characterizing $h(-\infty)=\omega$, Eq. \eqref{40} yields
\begin{equation}
\sigma_{l}(\omega)\geq\sec h^{2}\left\{  \ln\left(  \frac{h_{peak}}{h_{\infty
}}\right)  \right\}  =\sec h^{2}\left\{  \ln\left(  \frac{\sqrt{\omega
^{2}-V_{peak}}}{\omega}\right)  \right\}  . \label{41}%
\end{equation}
This bound becomes pointless if $\omega^{2}<
V_{peak}$. Furthermore, we can compute the GF by using the transmission coefficient $\big(T_{l}(\omega)\big)$ with the Miller-Good
transformation technique \cite{Boonserm:2008dk}. Thus, Eq. (\ref{41}) is given by
\begin{equation}
\sigma_{l}\left(  \omega\right)  \equiv T_{l}(\omega)\geq\frac{4\omega^{2}(\omega^{2}-V_{peak})}{(2\omega
^{2}-V_{peak})^{2}}.\label{42}%
\end{equation}
To compute $V_{peak}$, one should first define $r_{peak}$. For instance, for an illustrative demonstration, if one sets $B=1, \lambda=2$, and $\alpha=1$, in the first solution, $r_{peak}$ can be computed as 
\begin{equation}
r_{peak}=\frac{3\beta^{2}+3\pm\sqrt{9\beta^4-46\beta^2+73}}{8\left(\beta^{2}-1\right)}.\label{dd42}%
\end{equation}
By plugging  Eq. \eqref{dd42} into Eq. (\ref{35}), the regarding $V_{peak}$ is determined. One can follow the the same series of steps for the other solution to derive the correspoing $V_{peak}$. Figures (\ref{fig3}) and (\ref{fig4}) show the behaviors of the BWBHSCP GFs for some fixed quantities and various $\alpha$ and $\beta$ parameters of the first and second solutions. It should be noted that, by growing the $\alpha$ and $\beta$ parameters, the GFs in the first solution reduce, and in the second one by rising $\alpha$ the GFs grow up to $\alpha=2$ then start to decline.

The QNMs, within the framework of WKB approximation (see Sec. \eqref{sec4}), are also determined via the WKB approximation method. The results obtained are tabulated in Tables \eqref{tabs1} and \eqref{tabs2} for the first and second types of BWBHSCP solutions, respectively.\\
\begin{table}
\begin{center}
  \begin{tabular}{ |c|c|c|c|c|c|c|c|c| }
  \hline
$l$ & $n$ & $\alpha$ & $\beta$ & $\omega_{solution(1)}$& $n$ & $\alpha$ & $\beta$ & $\omega_{solution(1)}$\\
\hline
1 & 0 & 0 & 0.1 & 0.286589623-0.096033341i& 1 & 0 & 0.1 & 0.258327264-0.301168481i\\
&  & 0.1 & 0.2 & 0.033656151-0.197605927i&  & 0.1 & 0.2 & 0.023903202-0.025845636i\\
&  & 0.2 & 0.3 & 0.033454334-0.280248891i&  & 0.2 & 0.3 & 0.220700417-0.414178179i\\
&  & 0.3 & 0.4 & 0.034255997+0.501632599i&  & 0.3 & 0.4 & 0.387300679-0.818680303i\\
&  & 0.4 & 0.5 & 0.152033452+0.713727366i&  & 0.4 & 0.5 & 0.512701323-1.148126858i\\
&  & 0.5 & 0.6 & 0.311345572+0.918153498i&  & 0.5 & 0.6 & 0.652993941-1.349528912i\\
&  & 0.6 & 0.7 & 0.507559367+1.112884744i&  & 0.6 & 0.7 & 0.873546922-1.356239820i\\
&  & 0.7 & 0.8 & 0.737220788+1.295439260i&  & 0.7 & 0.8 & 1.310381508-1.124252140i\\

\hline 
\end{tabular}
\end{center}
\captionof{table}{QNMs of scalar waves of the first kind BWBHSCP spacetime.}\label{tabs1}
\end{table}

\begin{table}
\begin{center}
  \begin{tabular}{ |c|c|c|c|c|c|c|c| }
\hline
$l$ & $n$ & $\alpha$ & $\omega_{solution(2)}$ & $n$ & $\alpha$ & $\omega_{solution(2)}$\\
\hline
1 & 0 & 0 & 0.291114116-0.098001363i& 1 & 0 & 0.262211870-0.3074323461i\\
&  & 0.1 & 0.029982323-0.406591230i&  & 0.1 & 0.068560465-0.619918791i\\
&  & 0.2 & 0.106971882-0.838143510i&  & 0.2 & 0.231082654-1.302481917i\\
&  & 0.3 & 0.217170609-1.295283573i&  & 0.3 & 0.455973672-2.038317313i\\
&  & 0.4 & 0.352035324-1.773332950i&  & 0.4 & 0.725807111-2.816329955i\\
&  & 0.5 & 0.504836567-2.269166503i&  & 0.5 & 1.026894923-3.631446894i\\
&  & 0.6 & 0.670912075-2.781643191i&  & 0,6 & 1.351964324-4.482378509i\\
&  & 0.7 & 0.847564070-3.310501544i&  & 0.7 & 1.698488478-5.368078250i\\
\hline 
\end{tabular}
\end{center}
\captionof{table}{QNMs of scalar waves of the second kind BWBHSCP spacetime.}\label{tabs2}
\end{table}

The behaviors of the QNMs under the influence of the brane-world parameters $\alpha$ and $\beta$ are shown in Table \eqref{tabs1} for the first solution. The oscillation frequencies (Re$(\omega)$) initially reduce with the rising $\alpha$ and $\beta$ parameters then start to monotonically rise; the same characteristics can easily be observed for the damping rates (Im$(\omega)$). As $\alpha$ parameter is increased, the QNMs of the second solution exhibit more regular behaviors: both Re$(\omega)$ and Im$(\omega)$ grow with the increasing $\alpha$ values.

\subsection{Fermionic GFs and QNMs} \label{5b}
\begin{figure}[h]
\centering
\includegraphics[width=13cm,height=11cm]{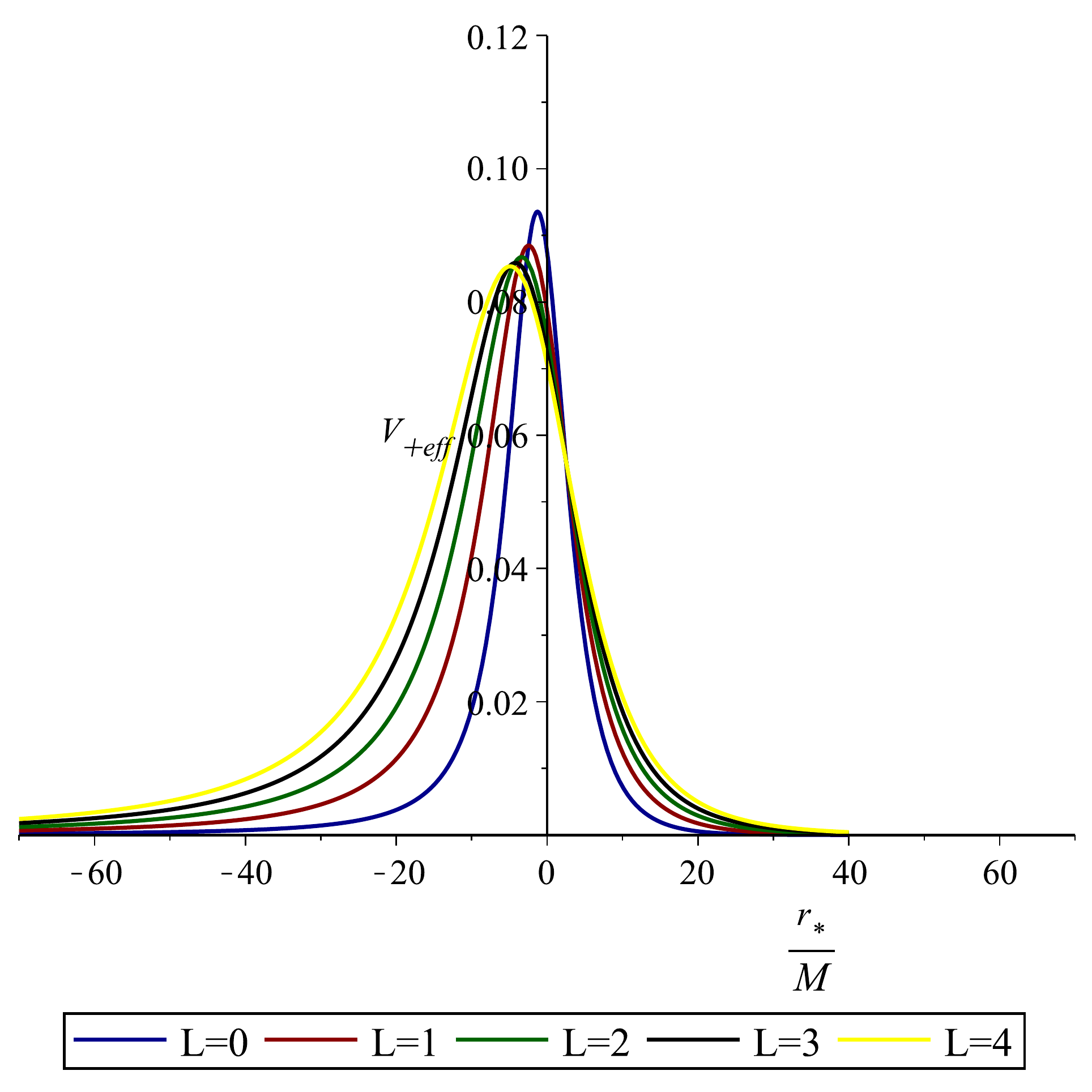}\caption{$V_{+}$ versus $\frac{r_{\star}}{M}$ (see Ref. \cite{SK27}).}%
\label{EffSBHBGM1}%
\end{figure}

\begin{figure}[h]
\centering
\includegraphics[scale=0.6]{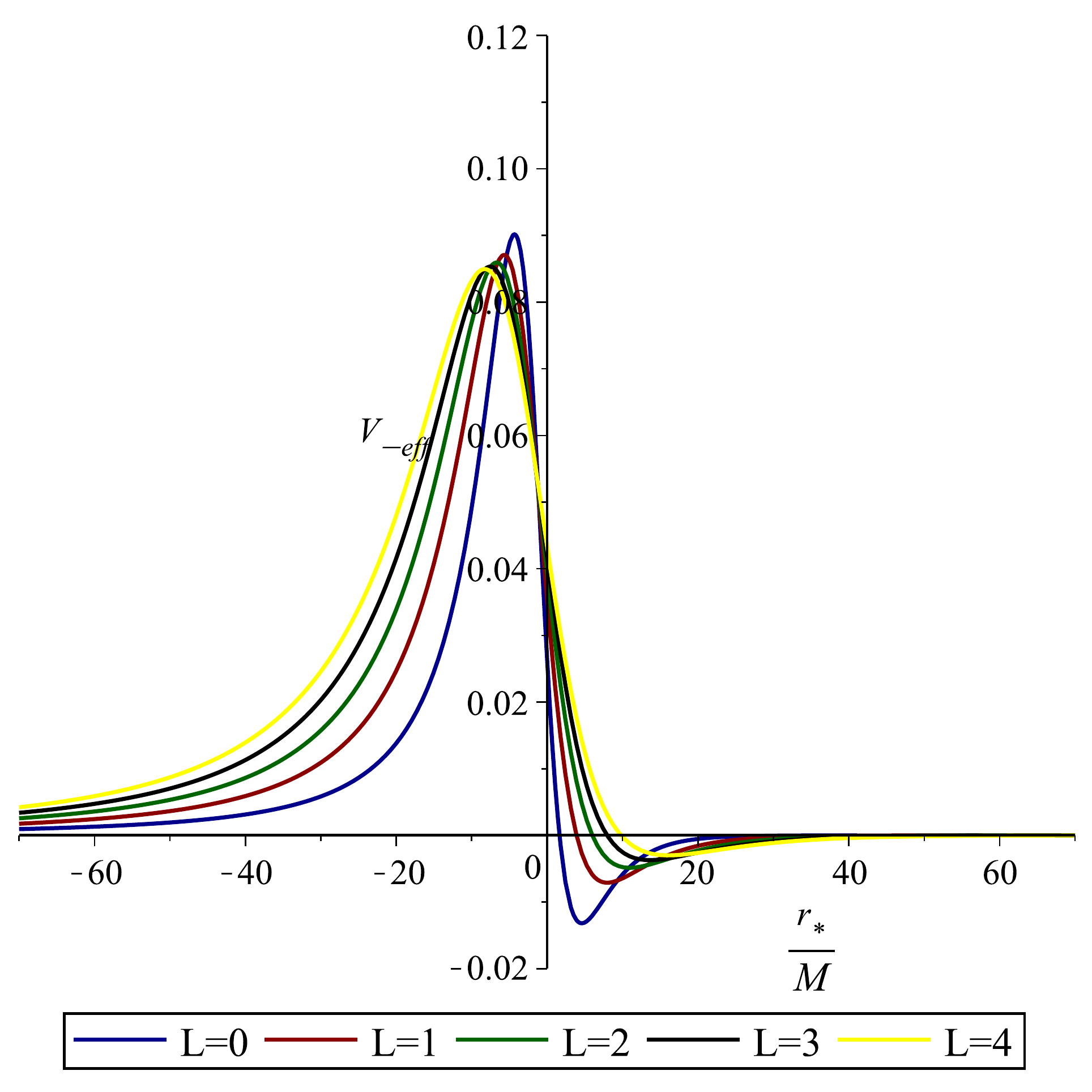}\caption{$V_{-}$ versus $\frac{r_{\star}}{M}$ graph (see Ref. \cite{SK27}).}%
\label{EffSBHBGM2}%
\end{figure}

For deriving the GFs of fermionic particles let us re-follow Ref. \cite{SK27} for SBHBGM \eqref{ww1}. In order to compute the $V_{eff}$ of the fermionic fields generating in the SBHBGM geometry, we consider the Chandrasekar-Dirac equations \cite{chandra}:
\begin{equation*}
\left( D+\epsilon -\rho \right) F_{1}+\left( \overline{\delta }+\pi -\alpha
\right) F_{2}=i\mu _{\ast }G_{1},
\end{equation*}%
\begin{equation*}
\left( \Delta +\mu -\gamma \right) F_{2}+\left( \delta +\beta -\tau \right)
F_{1}=i\mu _{\ast }G_{2},
\end{equation*}%
\begin{equation*}
\left( D+\overline{\epsilon }-\overline{\rho }\right) G_{2}-\left( \delta +%
\overline{\pi }-\overline{\alpha }\right) G_{1}=i\mu _{\ast }F_{2},
\end{equation*}%
\begin{equation}
\left( \Delta +\overline{\mu }-\overline{\gamma }\right) G_{1}-\left( 
\overline{\delta }+\overline{\beta }-\overline{\overline{\tau }}\right)
G_{2}=i\mu _{\ast }F_{1},  \label{a6}
\end{equation}
where $F_{1,2}$ and $G_{1,2}$ are known as the Dirac spinors which indicate the components of the wave functions. The spin coefficients are represented by the Greek letters, moreover, the complex conjugate of any parameter is denoted by a over-bar symbol. The spin coefficients which are non-zero are found to be

\begin{equation}
\epsilon=\gamma=\frac{\sqrt{2}f^\prime}{8\sqrt{f}\sqrt{1+L}}, \mu=\rho=\frac{\sqrt{2f}}{2r\sqrt{1+L}} , \beta=-\alpha=-\frac{\sqrt{2}cot\theta}{4r}.\label{SdW60}%
\end{equation}
Moreover, $D=\ell^{a}\partial_{a}, \Delta=n^{a}\partial_{a}$, and $\delta=m^{a}\partial_{a}$ are the corresponding directional derivatives \cite{SK27}. After a series of procedures whose details are given in  \cite{SK27}, the Dirac equations \eqref{a6} are decomposed into the radial and angular equations. The radial parts can be reduced to the 1-dimensional Schr\"{o}dinger equation with the following effective potential,
\begin{equation}
V_{\pm}=f\left[\left(\frac{2l+1}{2r}\pm i\mu\right)^{2}\mp\left(l+\frac{1}{2}\right)\frac{1}{\sqrt{1+L}}\frac{d}{dr}\left(-\frac{\sqrt{f}}{r}\mp\frac{2i\mu\sqrt{f}}{2l+1}\right)\right],\label{SdW61}%
\end{equation}
whose behaviors are depicted in Figs. \eqref{EffSBHBGM1} and \eqref{EffSBHBGM2}. After substituting the effective potential \eqref{SdW61} in the semi-analytic GF equation \eqref{is8}, one gets

\begin{equation}
 \sigma_{l}^f\left(\omega\right)\geq sech^{2}\left[\frac{\left(l+\frac{1}{2}\right)\sqrt{1+L}}{4M\omega}\left(l+\frac{1}{2}\pm\frac{1}{4\sqrt{1+L}}\right)\right].\label{SdW62}%
\end{equation}

Additionally, for the slowly rotating KlBH case \cite{Kanzi:2022vhp}, the expression for the fermionic GFs are defined as \cite{Kanzi:2022vhp}
\begin{equation}
\sigma_{l}\left(\omega\right)\geq\sec h^{2}\left\{ \frac{1}{2\omega} \int_{r_{h}}^{\infty}\left(\frac{\lambda^2}{r^2\sqrt{1+L}}\pm\lambda\frac{3M-r}{r^2\sqrt{r^2-2Mr}}\right)\right\},
\label{iz14}
\end{equation}
in which
\begin{equation}
\sigma_{l}\left(\omega\right)\geq\sec h^{2}\left\{ \frac{1}{2\omega} \left(\frac{\lambda^2}{r_{h}\sqrt{1+L}}\pm\lambda\left[\frac{-1}{r_h}+\frac{M}{r_{h}^2}+\frac{3M^2}{6r_{h}^2}+\frac{9M^3}{8r_{h}^4}\right]\right)\right\}.
\label{sk9}
\end{equation}

\begin{figure}[h]
\centering
\includegraphics[scale=0.5]{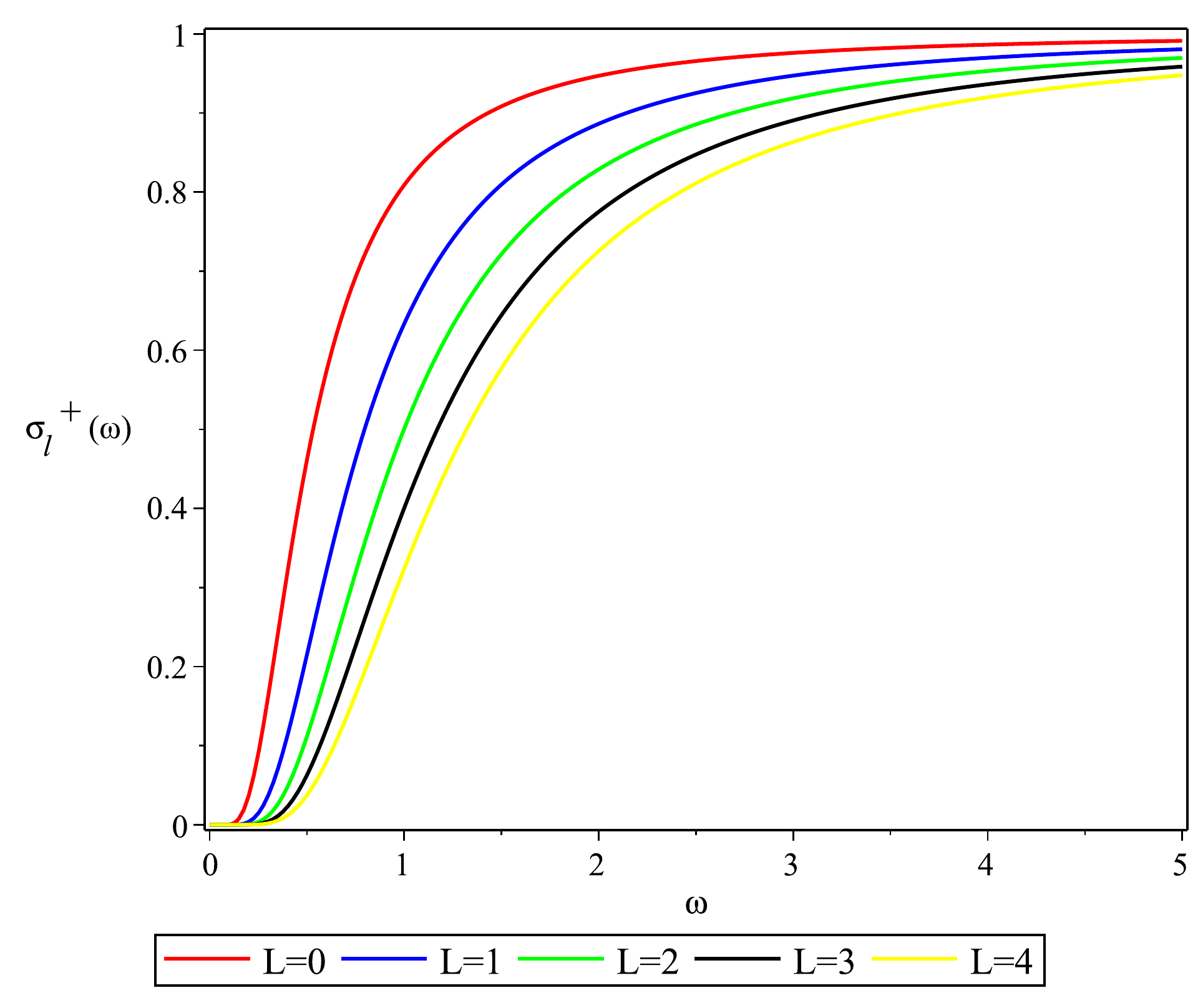}\caption{$\sigma_{l}^{+}\left(\omega\right)$ versus $\omega$ graph (see Ref. \cite{SK27}).}%
\label{GFSBHBGM1}%
\end{figure}

\begin{figure}[h]
\centering
\includegraphics[width=13cm,height=11cm]{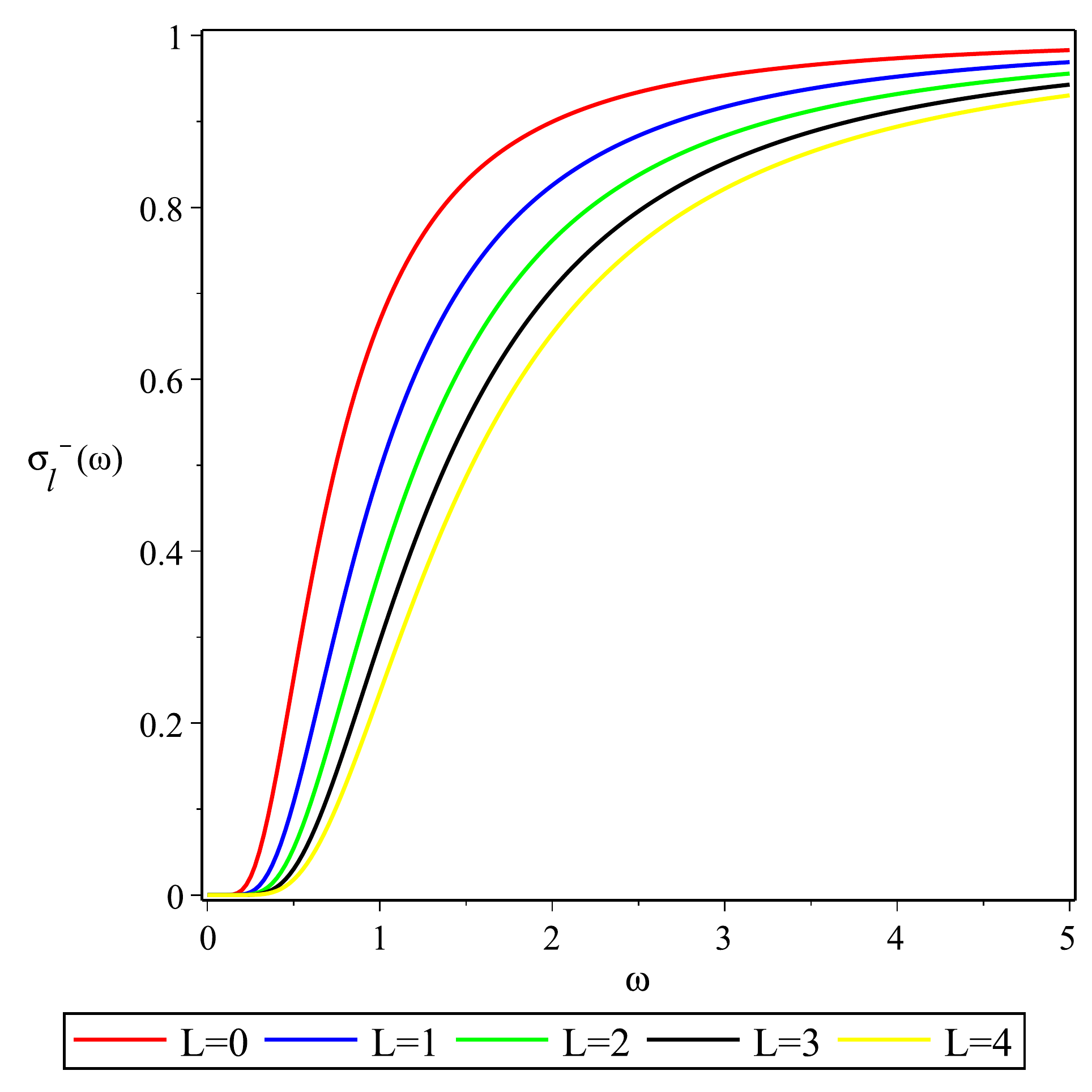}\caption{$\sigma_{l}^{-}\left(\omega\right)$ versus $\omega$ graph (see Ref. \cite{SK27}).}%
\label{GFSBHBGM2}%
\end{figure}

\begin{figure}[h]
\centering\includegraphics[scale=0.5]{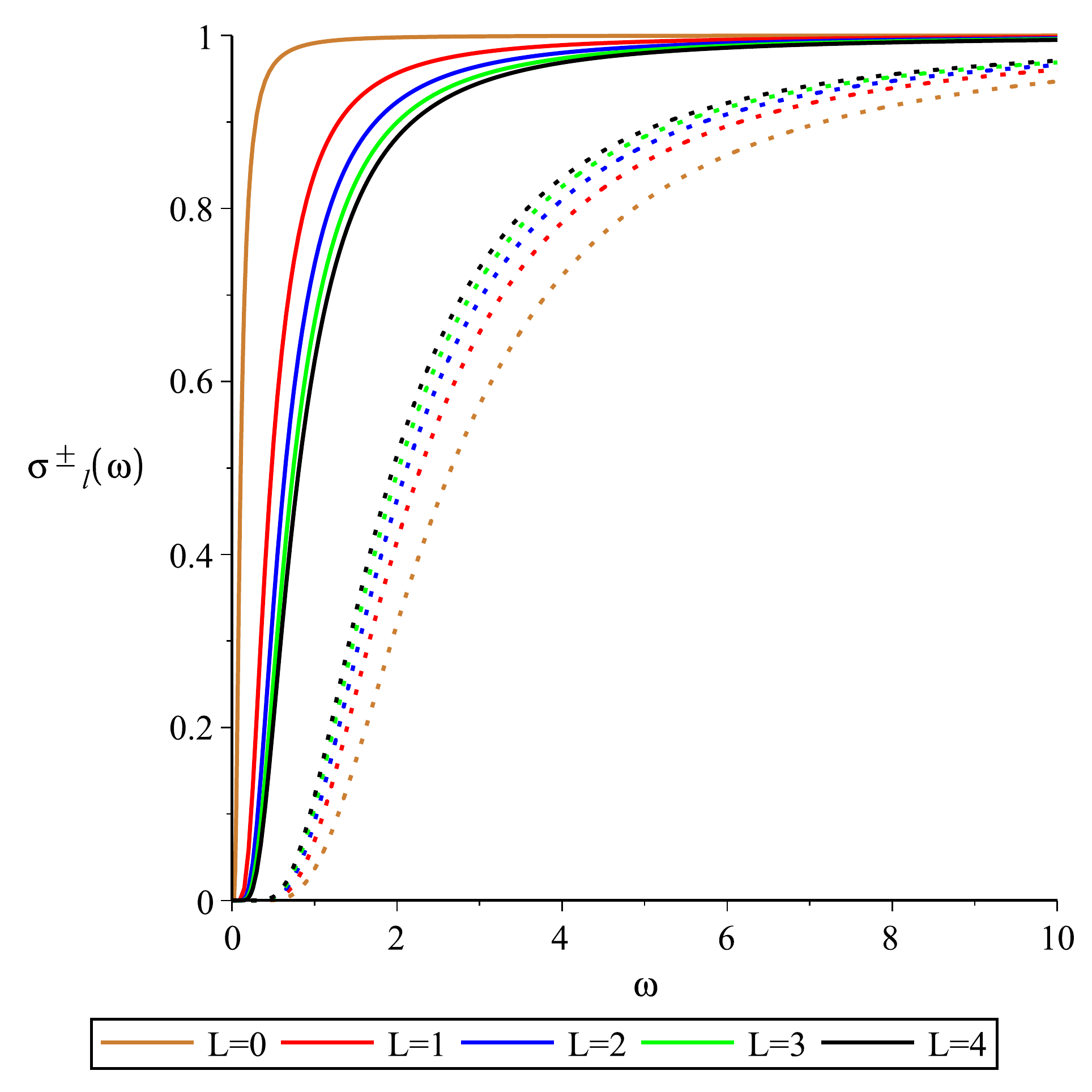}\caption{
$\sigma_{l}^{\pm}\left(  \omega\right)$ \eqref{isq42} plots for the spin-$\frac{\pm 1}{2}$ particles of slowly rotating KlBH. Solid lines stand for positive spin-half particles, however dotted lines represent negative spin-half particles. The physical quantities; $M=r=1$ and $\lambda=-1.5$.} \label{myfig41}
\end{figure}

The behaviors of the GFs of the fermions for various LSB parameters are shown in the Figs. \eqref{GFSBHBGM1} and \eqref{GFSBHBGM2} for SBHBGM, and for  the slowly rotating KlBH with Fig. \eqref{myfig41}.

Moreover, in line with the same thought, GFs of the CTNBHs have recently been studied in Refs. \cite{SK39} and \cite{SK40}, respectively. For the CTNBH \cite{SK40}, the charged Dirac equations \cite{Al-Badawi:2008ucc,Dariescu:2021zve} that govern the behavior of charged spin-$\frac{1}{2}$ particles are obtained with the NP formalism. Similarly, the charged KG equation has been employed in order to study the GF of bosons in the CTNBH geometry. For both type of charged particles, the corresponding Schr\"{o}dinger like wave equations with the relevant effective potentials have been derived \cite{SK40}.  Thus, the effect of the NUT parameter ($l$) on the charged bosonic and fermionic GFs and QNMs have been thoroughly analyzed, which was the main aim of \cite{SK40}. The result obtained in \cite{SK40} for the bosonic GFs emitted from the CTNBH is as follows
   
\begin{multline}
\sigma_{\ell}\left(  \omega\right)  \geq\sec h^{2}\left(  \frac{1}{2\omega
}\left[  -\frac{r_{h}}{4\left(  r_{h}^{2}+l^{2}\right)  }+\frac{Mr_{h}}{2l^{2}(l^{2}+r_{h}^{2})}-\frac{\arctan(\frac{r_{h}}{l})}{4l}\left(1-\frac{2M}{l^2}+\frac{3Q^2}{2l^2}+4\lambda\right)\right.  \right.
\\
\left. - \frac{2l^2r_{h}+q^2r_{h}+4M-2Ml^2+4l^2+2q^2}{4(r_{h}^2+l^2)}
-\frac{3rq^{2}}{8l^2(r_{h}^2+l^2)}-\right.  \\
\left.  \left.  -\frac{\tilde{q}^{2}q^{2}}{r_{h}}\left(1+\frac{M}{r_{h}}+\frac{4M^2+q^2}{3r_{h}^2}\right)  \right] \right), \label{isq40}
\end{multline}
where $M$, $l$, and $q$ denote the mass, the NUT parameter, and the electric parameter of the BH, respectively. Moreover, $\tilde{q}$ denotes the charge of the boson and $r_{h}=M\pm\sqrt{M^{2}+r^{2}+q^{2}}$ is the event horizon of the CTNBH.
\begin{figure}[h]
\centering
\includegraphics[scale=.5]{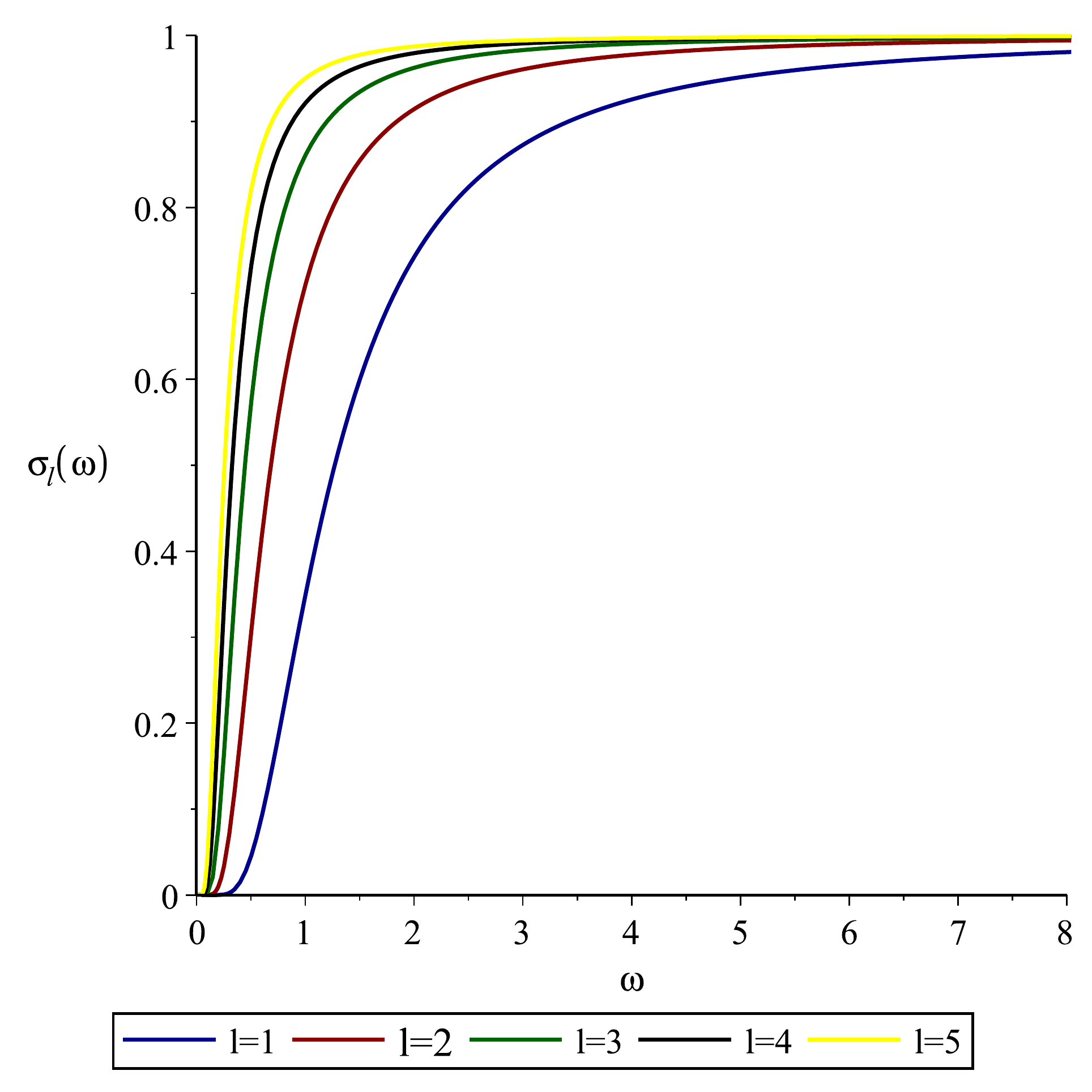} \caption{$\sigma_{\ell}$ plots (\ref{isq40})
for various values of the NUT parameter $l$. Physical quantities: $\lambda =  2, q =\tilde{q}= 0.5$, and $M = 1$.}
\label{Figure4}%
\end{figure}
Similarly, the fermionic GFs obtained from the Dirac perturbation of the CTNBH are as follows \cite{SK40}
\begin{multline}
\sigma_{l}^{\pm}\left(  \omega\right)\geq\sim\sec h^{2}\left(  \frac{\lambda}{2\omega
}\left[  \frac{2k^{2}\lambda l^{2}qe}{\left( qe r_{h}+kl^{2}\right)^{3}  }+\frac{k^{2}\lambda l^{4}}{3r_{h}^{2}(kl^{2}+qer_{h})^{2}}-\frac{k^{2}\lambda qer_{h}^3}{2(kl^2+qer_{h})^{3}(kl^{2}+kr_{h}^{2}+qer_{h})}\right.  \right.
\\
\left. - \frac{\arctan\left(\frac{kr_{h}}{\sqrt{(kl^2+qer_{h})k}}\right)k^{2}\lambda qer_{h}(r_{h}-2kl^2)}{(qer_{h}^3+kl^2)^{3}\sqrt{(qer_{h}^3+kl^2)k}}
\pm\left(\frac{3}{4r_{h}^{4}}+\frac{2(qe-Mk)}{5kr_{h}^{5}}\right. \right.  \\
\left.  \left. \left. +\frac{k(M^2+q^2+4l^2)+2qe}{12r_{h}^{6}}-\frac{qe(M+3qek)}{7kr_{h}^{7}} -\frac{qe}{8r_{h}^{8}}\left(\frac{3l^2}{k}+\frac{M^2+q^2}{2k}\right)\right) \right] \right), \label{isq42}
\end{multline}
in which $e$ represents the charge of the fermion.

\begin{figure}[h]
\centering
\includegraphics[scale=.6]{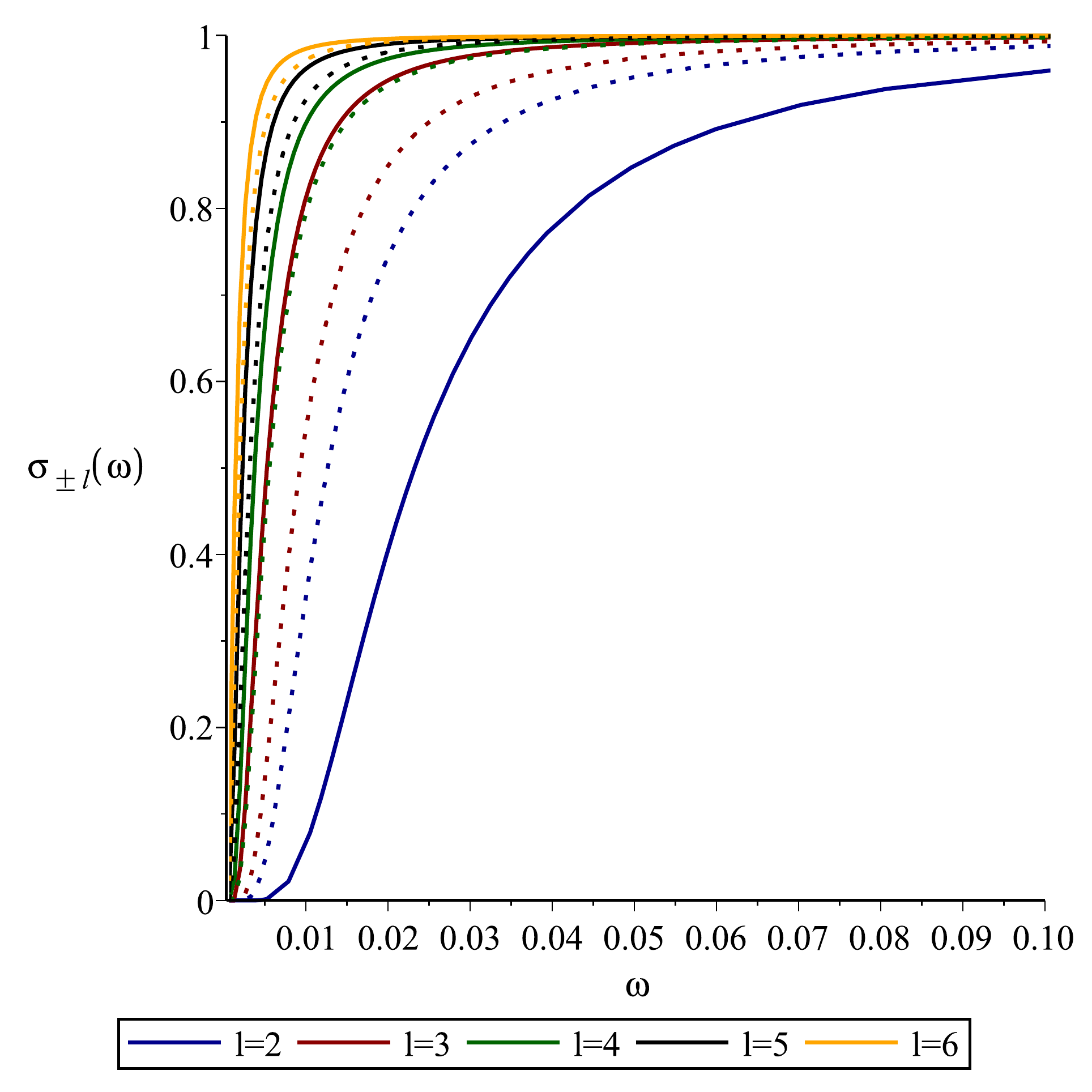} \caption{$\sigma_{\pm\ell}(\omega)$ plots \eqref{isq42}
for various values of the NUT parameter $l$. While the solid lines indicate the spin-$(+\frac{1}{2})$, dotted lines represent spin-$(-\frac{1}{2})$ particles.The physical quantities; $\lambda = -1.5, k = 0.2, q =0.1, e = 0.5$, and $M = 1$.}
\label{Figuresa5}%
\end{figure}
The impacts of the NUT parameter $l$ and the CTNBH charge $q$ on the GFs are illustrated in Figs. \eqref{Figure4} and \eqref{Figuresa5}. As can be easily seen from those figures, the GFs increase as the NUT parameter is increased. The latter remark is valid for the both bosonic and fermionic perturbations.

The influence of the NUT parameter $l$ on the QNMs is also investigated in \cite{SK40}. To this end, two sets of charge values are used: $q=\tilde{q}=0.1$ and $q=\tilde{q}=0.2$ for $n=0$ and $n=1$ when $l=1$. Table \ref{tab1} summarizes the findings, which show that increasing the NUT parameter reduces both oscillation and damping modes, in general. Moreover, for $n=1$ state although the real part remains intact, but the damping modes raises until $l=3.3$ then they start to decrease for both sets of the charges: $q=\tilde{q}=0.1$ and $q=\tilde{q}=0.2$. 

\begin{figure}[h]
\centering
\includegraphics[scale=.5]{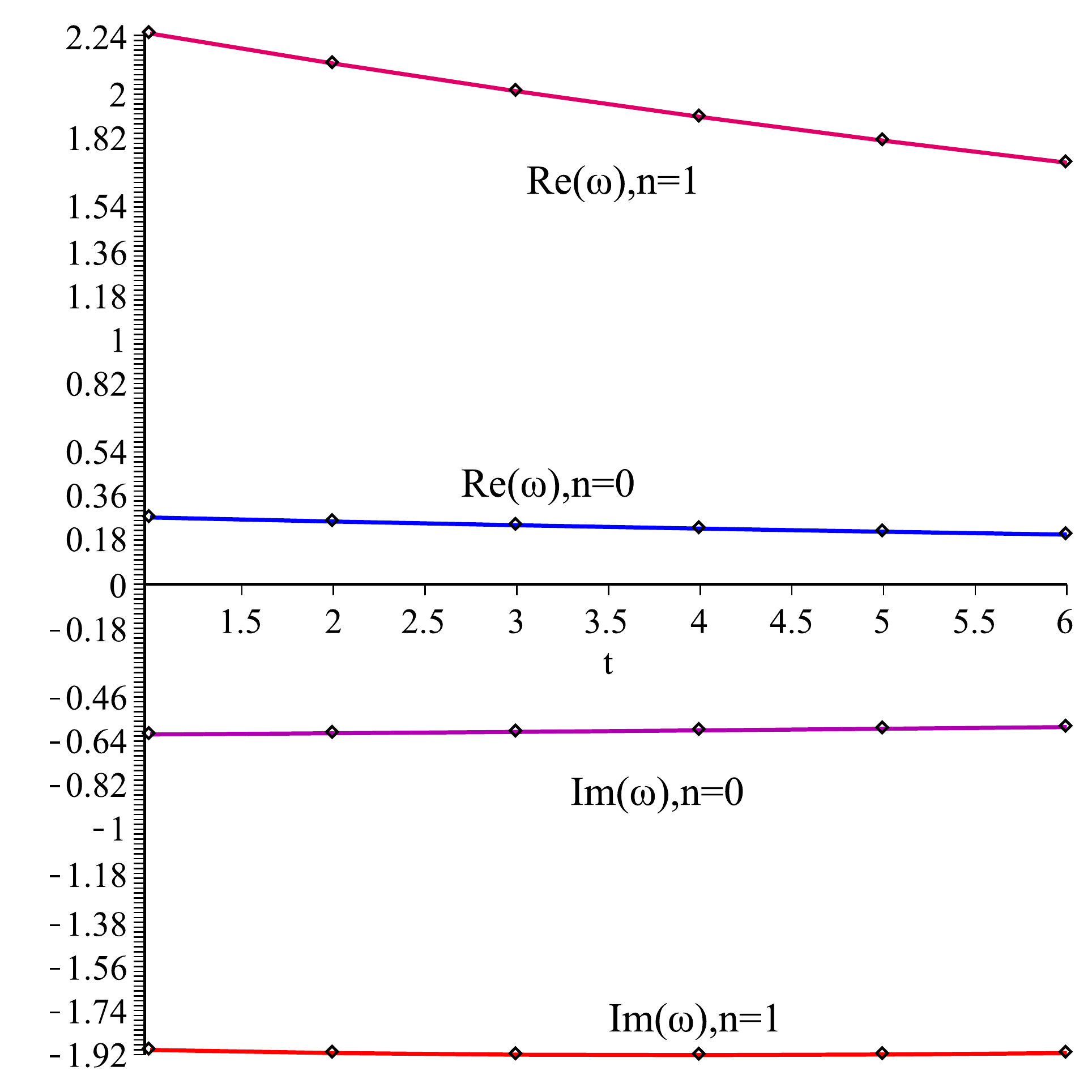} \caption{Scalar QNMs plots for various values of the NUT parameter $l$ based on Table \ref{tab1}. Here, the considered charge values are $q=\tilde{q}=0.2$
.}
\label{Figure6}%
\end{figure}

\begin{table}
\begin{center}
  \begin{tabular}{ |c|c|c|c|c|c|c|c|}
\hline
$\tilde{l}$ & $n$ & $q=\tilde{q}$ & $l$ & $\omega_{Bosons}$ & $q=\tilde{q}$& $l$ & $\omega_{Bosons}$\\
\hline\hline
1 & 0 & 0.2 & 3 & 0.2725502408-0.6131029148i & 0.1 & 3 & 0.2756084514-0.6136558658i\\
  &   &  & 3.1 & 0.2560591000-0.6082054931i& & 3.1 & 0.2587389727-0.6087803472i \\
  &   &  & 3.2 & 0.2409607665-0.6025973599i & & 3.2 & 0.2433163930-0.6031804616i\\
  &  & & 3.3 & 0.2271104777-0.5964385698i & & 3.3 & 0.2291872658-0.5970200571i \\
  &   & & 3.4 & 0.2143806411-0.5898593359i & & 3.4 & 0.2162166481-0.5904320921i \\
  &   & & 3.5 & 0.2026583779-0.5829655802i& & 3.5 & 0.2042858309-0.5835246352i \\
\hline
 & 1 & 0.2 & 3 & 2.248092173-1.900912954i& 0.1& 3 & 2.268320069-1.905126791i \\
  &   &  & 3.1 & 2.125469414-1.913465607i& & 3.1 & 2.143321140-1.917860042i\\
  &   & & 3.2 & 2.012248433-1.920073137i& & 3.2 & 2.028053175-1.924547860i \\
  &  & & 3.3 & 1.907546817-1.921767172i& & 3.3 & 1.921581800-1.926248930i\\
  &   & & 3.4 & 1.810574732-1.919400788i& & 3.4 & 1.823073953-1.923836247i\\
  &   & & 3.5 & 1.720624552-1.913681044i& & 3.5 & 1.731786697-1.918032284i \\
\hline
\end{tabular}
\end{center}
 \captionof{table}{Bosonic QNMs of the CTNBH for various combinations of charge and state values.} \label{tab1}
\end{table}
 To make the behaviors of bosonic QNMs more understandable and comprehensive, we have also plotted Fig. \ref{Figure6} for various NUT parameters $l$ and charge $q=\tilde{q}=0.2$. It is also worth noting that for other low charge values like $q=\tilde{q}=0.01$, the figures plotted are almost identical to Fig. \ref{Figure6} \big(that is why we have only depicted Fig. \ref{Figure6}\big).
 
\begin{figure}[h]
\centering
\includegraphics[scale=.5]{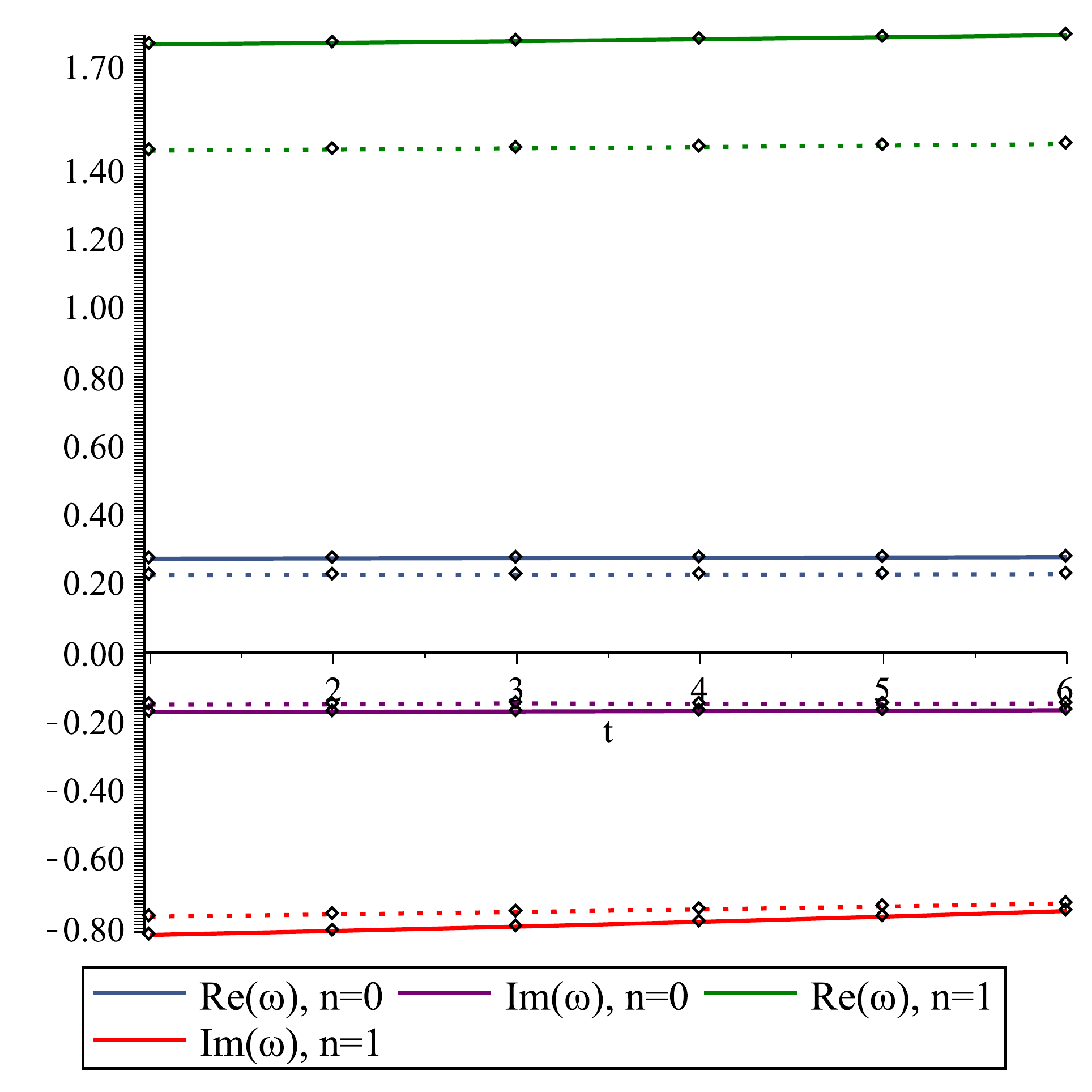} \caption{Dirac QNMs plots
for various values of the NUT parameter $l$ based on Table \ref{tab2}. As the solid lines are for $q=e=0.8$, the dotted lines stand for $q=e=0.9$.}
\label{Figure8}%
\end{figure}

For the Dirac perturbation of the CTNBH,  the results obtained for the charged fermionic QNMs \cite{SK40} are tabulated in Table \ref{tab2}. The latter table is formed with the values of $k=0.1$ and $M=1$ (for the other pairwise values of $k$ and $M$, one gets similar behaviors). Table \ref{tab2} indicates that the fermionic oscillation frequencies increase smoothly with the increasing NUT parameters $l$. However, such as in the bosonic case, this behavior is reversed for the damping modes. On the other hand, it is worth highlighting that the evolution (in the context of increase and decrease) of the QNM frequencies of the Dirac particles with the NUT parameter $l$ is in harmony with their GF characteristics. Finally, to reveal the behaviors of the imaginary and real frequencies,  we have depicted the Dirac QNMs in Fig. \ref{Figure8}. The both behaviors are in accordance not only with each other, but with the bosonic QNMs as well.

\begin{table}
\begin{center}
 \begin{tabular}{ |c|c|c|c|c|c|c|c|}
\hline
$\tilde{l}$ & $n$ & $q=e$ & $l$ & $\omega_{Fermions}$ & $q=e$& $l$ & $\omega_{Fermions}$\\
\hline\hline
1 & 0 & 0.9 & 1 & 0.2246744266-0.1507449260i & 0.8 & 1 & 0.2724250498-0.1725874216i\\
  &   &  & 1.1 & 0.2251419577-0.1502503846i& & 1.1 & 0.2731628029-0.1716878385i \\
  &   &  & 1.2 & 0.225654448-0.1497044313i & & 1.2 & 0.2739749307-0.1706919821i\\
  &  & & 1.3 & 0.2262187771-0.1491057793i & & 1.3 & 0.2748624250-0.1695967016i \\
  &   & & 1.4 & 0.2268297702-0.1484531124i & & 1.4 & 0.2758266950-0.1683984787i \\
  &   & & 1.5 & 0.2274901744-0.1477447054i& & 1.5 & 0.2768689571-0.1670932364i \\
\hline
 & 1 & 0.9 & 1 & 1.456007708-0.7655868096i& 0.8 & 1 & 1.763833077-0.8187154684i \\
  &   &  & 1.1 & 1.459193911-0.7592922722i& & 1.1 & 1.768465908-0.8073416560i\\
  &   & & 1.2 & 1.462685227-0.7523492681i& & 1.2 & 1.773520512-0.7947675850i \\
  &  & & 1.3 & 1.466482391-0.7447418782i& & 1.3 & 1.778990042-0.7809589163i\\
  &   & & 1.4 & 1.470585762-0.7364546462i& & 1.4 & 1.784867095-0.7658771666i\\
  &   & & 1.5 & 1.474995402-0.7274692890i& & 1.5 & 1.791142115-0.7494791876i \\
\hline
\end{tabular}
\end{center}
   \captionof{table}{Fermionic QNMs of the CTNBH for various combinations of charge and state values.} \label{tab2}

\end{table}

\section{Conclusion} \label{sec6}
This study serves as an introduction to the analysis of QNMs and GFs of numerous curved spacetimes examined in astrophysics, higher-dimensional gravity, and string theory. Although currently there are some great reviews on the QNMs including many valuable studies, which are supported by the observational characteristics of the recently discovered gravitational waves \cite{Nakamura:2016gri,Ghosh:2021mrv}, but the number of reviews written on the GFs is insignificant. Therefore, in this review article, we have mainly focused our efforts on the GFs rather than the QNMs. To this end,
 we have concentrated on certain approaches, topics, and computations that were either not reviewed or were briefly mentioned in the literature. As a result, instead of discussing general problems superficially, we have directed the reader to the more specialized literature. Thus, we  believe that this review article can pave the way for the reader to learn and follow the methodologies given in this study for signing new works.

Finally, we would like to state that to humanity's surprise, many theoretical studies that were previously assumed to be difficult to test empirically and observationally have been confirmed by scientists employing and developing unique setups. Some of the most recent theory-experiment/observation pairings are as follows:  1) Higgs boson found by the Compact Muon Solenoid \cite{CMS:2012qbp} and ATLAS \cite{ATLAS:2012yve} 2) Gravitational waves are observed \cite{LIGOScientific:2016aoc} 3)  First image of a BH \cite{EventHorizonTelescope:2019dse} 4) The first room-temperature superconductor identified \cite{superconductor} 5) Observation stationary Hawking radiation in an analog BH \cite{Kolobov:2019qfs}. All of those suggest that the GFs and QNMs may cease to be pure theoretical subjects and become measurable physical quantities much sooner than anticipated. In conclusion, the theoretical physics will continue to guide the experimental physics in this regard.

\section*{Acknowledgements}
\addcontentsline{toc}{section}{Acknowledgments}
We would like to express our gratitude to the following colleagues who contributed to the information presented in this review at various times and occasions: Mustafa Halilsoy, Ahmad Al-Badawi, Seyedhabibollah Mazharimousavi, Kimet Jusufi, Douglas Singleton, Roman
Konoplya, Eduardo Guendelman, Seyedeh Fatemeh Mirekhtiary, Ali \"{O}vg\"{u}n, G\"{u}lnihal Tokg\"{o}z Hyusein, Joel Saavedra, \"Ozay G\"{u}rtu\u{g}, Huriye G\"{u}rsel Mangut, Mert Mangut, and Halil Mutuk. We also thank Akhil Uniyal for clarifications
on the GF calculations. \.{I}. Sakall{\i}, who had Covid-19 at the time of writing this article, attributes this study to all Covid-19 victim scientists, which experienced the health, physical, and psychological difficulties during this pandemic.

\end{document}